\documentclass[structabstract]{aa}

\usepackage{txfonts}
\usepackage{natbib}
\usepackage{amsbsy}
\usepackage{graphicx}
\usepackage{longtable,lscape}
\usepackage{verbatim}
\usepackage[T1]{fontenc}

\bibpunct{(}{)}{;}{a}{}{,}     

\newcommand{\micron}{$\mu$m}
\newcommand{\el}[1]{\mathrm{#1}}

\newcommand{\ten}[1]{$10^{#1}$}
\newcommand{\scit}[2]{$#1\times10^{#2}$}
\newcommand{\scim}[2]{#1\times10^{#2}}

\newcommand{\ps}{s$^{-1}$}
\newcommand{\pcm}{cm$^{-1}$}
\newcommand{\pcs}{cm$^{-2}$}
\newcommand{\pcc}{cm$^{-3}$}
\newcommand{\eq}[1]{Eq.\ (\ref{eq:#1})}
\newcommand{\eqq}[1]{Equation (\ref{eq:#1})}
\newcommand{\fig}[1]{Fig.\ \ref{fig:#1}}

\newcommand{\tb}[1]{Table \ref{tb:#1}}
\newcommand{\rx}[1]{Reaction (\ref{eq:#1})}
\newcommand{\mh}{H$_2$}
\newcommand{\co}[2]{$^{#1}$C$^{#2}$O}
\newcommand{\com}[2]{^{#1}\el{C}^{#2}\el{O}}
\newcommand{\coc}[1]{$^{#1}$CO}
\newcommand{\coo}[1]{C$^{#1}$O}
\newcommand{\cocm}[1]{^{#1}\el{CO}}
\newcommand{\coom}[1]{\el{C}^{#1}\el{O}}
\newcommand{\cop}{CO$^{+}$}
\newcommand{\cdco}{N(\el{CO})}

\newcommand{\cdmh}{N(\el{H}_2)}
\newcommand{\cdh}{N(\el{H})}

\newcommand{\osc}{f_{\el{v}'0}}
\newcommand{\bvu}{B_\el{v}'}

\newcommand{\dvu}{D_\el{v}'}

\newcommand{\qvu}{q_\el{v}'}
\newcommand{\pvu}{p_\el{v}'}
\newcommand{\weu}{\omega_\el{e}'}
\newcommand{\wexeu}{\omega_\el{e}x_\el{e}'}
\newcommand{\weyeu}{\omega_\el{e}y_\el{e}'}
\newcommand{\wezeu}{\omega_\el{e}z_\el{e}'}
\newcommand{\zoph}{$\zeta$ Oph}
\newcommand{\av}{A_\el{V}}
\newcommand{\bco}{b(\el{CO})}
\newcommand{\bmh}{b(\el{H}_2)}
\newcommand{\texco}{T_\el{ex}(\el{CO})}
\newcommand{\texmh}{T_\el{ex}(\el{H}_2)}
\newcommand{\mc}[1]{\multicolumn{2}{c}{#1}}

\defcitealias{vandishoeck88a}{vDB88}
\newcommand{\vdb}{\citetalias{vandishoeck88a}}
\defcitealias{eidelsberg90a}{ER90}
\newcommand{\er}{\citetalias{eidelsberg90a}}

\begin{document}

\title{The photodissociation and chemistry of CO isotopologues: applications to interstellar clouds and circumstellar disks}

\author{
R. Visser \inst{1}
 \and E.F. van Dishoeck \inst{1,2}
 \and J.H. Black \inst{3}
}

\institute{Leiden Observatory, Leiden University, P.O. Box 9513, 2300 RA Leiden, The Netherlands \\
           \email{ruvisser@strw.leidenuniv.nl}
 \and Max-Planck-Institut f\"ur Extraterrestrische Physik, Giessenbachstrasse 1, 85748 Garching, Germany
 \and Onsala Space Observatory, Chalmers University of Technology, 43992 Onsala, Sweden
}

\titlerunning{The photodissociation and chemistry of interstellar CO isotopologues}
\authorrunning{R. Visser et al.}

\date{Received $<$date$>$ / Accepted $<$date$>$}


\abstract
{} 
{Photodissociation by UV light is an important destruction mechanism for carbon monoxide (CO) in many astrophysical environments, ranging from interstellar clouds to protoplanetary disks. The aim of this work is to gain a better understanding of the depth dependence and isotope-selective nature of this process.} 
{We present a photodissociation model based on recent spectroscopic data from the literature, which allows us to compute depth-dependent and isotope-selective photodissociation rates at higher accuracy than in previous work. The model includes self-shielding, mutual shielding and shielding by atomic and molecular hydrogen, and it is the first such model to include the rare isotopologues \coo{17} and \co{13}{17}. We couple it to a simple chemical network to analyse CO abundances in diffuse and translucent clouds, photon-dominated regions, and circumstellar disks.} 
{The photodissociation rate in the unattenuated interstellar radiation field is \scit{2.6}{-10} \ps, 30\% higher than currently adopted values. Increasing the excitation temperature or the Doppler width can reduce the photodissociation rates and the isotopic selectivity by as much as a factor of three for temperatures above 100 K. The model reproduces column densities observed towards diffuse clouds and PDRs, and it offers an explanation for both the enhanced and the reduced $N(\cocm{12})/N(\cocm{13})$ ratios seen in diffuse clouds. The photodissociation of \coo{17} and \co{13}{17} shows almost exactly the same depth dependence as that of \coo{18} and \co{13}{18}, respectively, so \element[][17]{O} and \element[][18]{O} are equally fractionated with respect to \element[][16]{O}. This supports the recent hypothesis that CO photodissociation in the solar nebula is responsible for the anomalous \element[][17]{O} and \element[][18]{O} abundances in meteorites. Grain growth in circumstellar disks can enhance the $N(\cocm{12})/N(\coom{17})$ and $N(\cocm{12})/N(\coom{18})$ ratios by a factor of ten relative to the initial isotopic abundances.}   
{} 


\maketitle



\section{Introduction}
\label{sec:intro}
Carbon monoxide (CO) is one of the most important molecules in astronomy. It is second in abundance only to molecular hydrogen (\mh) and it is the main gas-phase reservoir of interstellar carbon. Because it is readily detectable and chemically stable, CO and its less abundant isotopologues are the main tracers of the gas properties, structure and kinematics in a wide variety of astrophysical environments \citep[for recent examples, see][]{dame01a,najita03a,wilson05a,greve05a,leroy05a,huggins05a,bayet06a,oka07a,narayanan08a}. In particular, the pure rotational lines at millimetre wavelengths are often used to determine the total gas mass. This requires knowledge of the CO--\mh{} abundance ratio, which may differ by several orders of magnitude from one object to the next \citep{lacy94a,burgh07a,panic08a}. If isotopologue lines are used, the isotopic ratio enters as an additional unknown.

CO also controls much of the chemistry in the gas phase and on grain surfaces, and is a precursor to more complex molecules. In photon-dominated regions (PDRs), dark cores and shells around evolved stars, the amount of carbon locked up in CO compared with that in atomic C and C$^+$ determines the abundances of small and large carbon-chain molecules \citep[e.g.,][]{millar87a,jansen95a,aikawa99a,brown03a,teyssier04a,cernicharo04a,morata08a}. CO ice on the surfaces of grains can be hydrogenated to more complex saturated molecules such as CH$_3$OH \citep[e.g.,][]{charnley95a,watanabe02a}, so the partioning of CO between the gas and grains is important for the overall chemical composition as well \citep[e.g.,][]{caselli93a,rodgers03a,doty04a,garrod06a}.

A key process in controlling the gas-phase abundance of \coc{12} and its isotopologues is photodissociation by ultraviolet (UV) photons. This is governed entirely by discrete absorptions into predissociative excited states; any possible contributions from continuum channels are negligible \citep{hudson71a,fock80a,letzelter87a,cooper87a}. Spectroscopic measurements in the laboratory at increasingly higher spectral resolution have made it possible for detailed photodissociation models to be constructed (\citealt{solomon72a}; \citealt{bally82a}; \citealt{glassgold85a}; \citealt{vandishoeck86a}; \citealt{viala88b}; \citealt{vandishoeck88a}, hereafter vDB88; \citealt{warin96a}; \citealt{lee96a}). The currently adopted photodissociation rate in the
unattenuated interstellar radiation field is \scit{2}{-10} \ps.

Because the photodissociation of CO is a line process, it is subject to self-shielding: the lines become saturated at a \coc{12} column depth of about \ten{15} \pcs, and the photodissociation rate strongly decreases (\vdb; \citealt{lee96a}). \citet{bally82a} realised this is an isotope-selective effect. Due to their lower abundance, isotopologues other than \coc{12} are not self-shielded until much deeper into a cloud or other object. This results in a zone where the abundances of these isotopologues are reduced with respect to \coc{12}, and the abundances of atomic \element[][13]{C}, \element[][17]{O} and \element[][18]{O} are enhanced with respect to \element[][12]{C} and \element[][16]{O}. For example, the \coo{17}--\coc{12} and \coo{18}--\coc{12} column density ratios towards \object{X Per} are a factor of five lower than the elemental oxygen isotope ratios \citep{sheffer02a}. The \coc{13}--\coc{12} ratio along the same line of sight is unchanged from the elemental carbon isotope ratio, indicating that \coc{13} is replenished through low-temperature isotope-exchange reactions. A much larger sample of sources shows $N(\cocm{13})/N(\cocm{12})$ column density ratios both enhanced and reduced by up to a factor of two relative to the elemental isotopic ratio \citep{sonnentrucker07a,burgh07a,sheffer07a}. The reduced ratios have so far defied explanation, as all models predict that isotope-exchange reactions prevail over selective photodissociation in translucent clouds.

CO self-shielding has been suggested as an explanation for the anomalous \element[][17]{O}--\element[][18]{O} abundance ratio found in meteorites \citep{clayton73a,clayton02a,lyons05a,lee08a}. In cold environments, molecules such as water (H$_2$O) may be enhanced in heavy isotopes. This so-called isotope fractionation process is due to the difference in vibrational energies of H$_2^{16}$O, H$_2^{17}$O and H$_2^{18}$O, and is therefore mass-dependent. It results in \element[][18]{O} being about twice as fractionated as \element[][17]{O}. However, \element[][17]{O} and \element[][18]{O} are nearly equally fractionated in the most refractory phases in meteorites (calcium-aluminium-rich inclusions, or CAIs), hinting at a mass-independent fractionation mechanism. Isotope-selective photodissociation of CO in the surface of the early circumsolar disk is such a mechanism, because it depends on the relative abundances of the isotopologues and the mutual overlap of absorption lines, rather than on the mass of the isotopologues. The enhanced amounts of \element[][17]{O} and \element[][18]{O} are subsequently transported to the planet- and comet-forming zones and eventually incorporated into CAIs. Recent observations of \coc{12}, \coo{17} and \coo{18} in two young stellar objects support the hypothesis of CO photodissociation as the cause of the anomalous oxygen isotope ratios in CAIs \citep{smith09a}. A crucial point in the \citeauthor{lyons05a} model is the assumption that the photodissociation rates of \coo{17} and \coo{18} are equal. Our model can test this at least partially.

Detailed descriptions of the CO photodissociation process are also important in other astronomical contexts. The circumstellar envelopes of evolved stars are widely observed through CO emission lines. The measurable sizes of these envelopes are limited primarily by the photodissociation of CO in the radiation field of background starlight \citep{mamon88a}. Finally, proper treatment of the line-by-line contributions to the photodissociation of CO may affect the analysis of CO photochemistry in the upper atmospheres of planets \citep{fox89a}.

In this paper, we present an updated version of the photodissociation model from \vdb, based on laboratory experiments performed in the past twenty years (Sect.\ \ref{sec:moldata}). We expand the model to include \coo{17} and \co{13}{17} and we cover a broader range of CO excitation temperatures and Doppler widths (Sects.\ \ref{sec:photdis} and \ref{sec:temp}). We rederive the shielding functions from \vdb{} and extend these also to higher excitation temperatures and larger Doppler widths (Sect.\ \ref{sec:shield}). Finally, we couple the model to a chemical network and discuss the implications for translucent clouds, PDRs and circumstellar disks, with a special focus on the meteoritic \element[][18]{O} anomaly (Sect.\ \ref{sec:implic}).


\section{Molecular data}
\label{sec:moldata}
The photodissociation of CO by interstellar radiation occurs through discrete absorptions into predissociated bound states, as first suggested by \citet{hudson71a} and later confirmed by \citet{fock80a}. Any possible contributions from continuum channels are negligible at wavelengths longer than the Lyman limit of atomic hydrogen \citep{letzelter87a,cooper87a}.

Ground-state CO has a dissociation energy of 11.09 eV and the general interstellar radiation field is cut off at 13.6 eV, so knowledge of all absorption lines within that range (911.75--1117.80 \AA) is required to compute the photodissociation rate. These data were only partially available in 1988, but ongoing laboratory work has filled in a lot of gaps. Measurements have also been extended to include CO isotopologues, providing more accurate values than can be obtained from theoretical isotopic relations. \tb{moldata-ls} lists the values we adopt for \coc{12}.

\addtocounter{table}{1}

\subsection{Band positions and identifications}
\label{subsec:bandpos}
\citet[hereafter ER90]{eidelsberg90a} and \citet{eidelsberg92a} redid the experiments of \citet{letzelter87a} at higher spectral resolution and higher accuracy, and also for \coc{13}, \coo{18} and \co{13}{18}. They reported 46 predissociative absorption bands between 11.09 and 13.6 eV, many of which were rotationally resolved. Nine of these have a cross section too low to contribute significantly to the overall dissociation rate. The remaining 37 bands are largely the same as the 33 bands of \vdb; bands 1 and 2 of the latter are resolved into four and two individual bands, respectively. Throughout this work, band numbers refer to our numbering scheme (\tb{moldata-ls}), unless noted otherwise.
 
Thanks to the higher resolution and the isotopologue data, \er{} could identify the electronic and vibrational character of the upper states more reliably than Stark et al. in \vdb. The vibrational levels are required to compute the positions for those isotopologue bands that have not been measured directly. Nine of the \vdb{} bands (not counting the previously unresolved bands 1 and 2) have a revised $v'$ value.

The \citet{eidelsberg92a} positions ($\nu_0$ or $\lambda_0$) are the best available for most bands, with an estimated accuracy of 0.1--0.5 \pcm{}. Seven of their \coc{12} bands were too weak or diffuse for a reliable analysis, so their positions are accurate only to within 5 \pcm. Nevertheless, we adopt the \citeauthor{eidelsberg92a} positions for three of these: bands 2A, 6 and 14. The former was blended with band 2B in \vdb, and the other two show a better match with the isotopologue band positions if we take the \citeauthor{eidelsberg92a} values. For the other four weak or diffuse bands, Nos. 4, 15, 19 and 28, we keep the \vdb{} positions. \citet{ubachs94a} further improved the experiments, obtaining an accuracy of about 0.01--0.1 \pcm, so we adopt their band positions where available. Finally, we adopt the even more accurate positions (0.003 \pcm{} or better) available for the $C1$, $E0$, $E1$ and $L0$ bands \citep{ubachs00a,cacciani01a,cacciani02a,cacciani04a}.\footnote{All transitions in our model arise from the $v''$=0 level of the electronic ground state. We use a shorthand that only identifies the upper state, with $C1$ indicating the $C\,^1\Sigma^+$ $v'$=1 state, etc.}

Band positions for isotopologues other than \coc{12} are still scarce, although many more are currently known from experiments than in 1988. The $C1$ and $E1$ bands have been measured for all six natural isotopologues, and the $E0$ band for all but \co{13}{17}, at an accuracy of 0.003 \pcm{} \citep{cacciani95a,ubachs00a,cacciani01a,cacciani04a}. The positions of the $E0$ band are especially important because of its key role in the isotope-selective nature of the CO photodissociation (Sect.\ \ref{subsec:depth}). Positions are known at lower accuracy (0.003--0.5 \pcm) for an additional 25 \coo{18}, 30 \coc{13} and 9 \co{13}{18} bands \citep{eidelsberg92a,ubachs94a,cacciani02a}; these are included throughout.

{\setlength\arraycolsep{2pt}
We compute the remaining band positions from theoretical isotopic relations. For band $b$ of isotopologue $i$, the position is
\begin{eqnarray}
\label{eq:isohead}
\nu_0(b,i) & = & \nu_0(b,{}\!\cocm{12}) + \left[E_\el{v}'(b,i)-E_\el{v}''(X,i)\right] - \nonumber\\
& & \left[E_\el{v}'(b,{}\!\cocm{12})-E_\el{v}''(X,{}\!\cocm{12})\right] \,,
\end{eqnarray}
with $E_\el{v}'(b,i)$ and $E_\el{v}''(X,i)$ the vibrational energy of the excited and ground states, respectively. Hence, we need the vibrational constants ($\weu$, $\wexeu$, \ldots) for all the excited states other than the $C\,^1\Sigma^+$. These have only been determined experimentally for the $E\,^1\Pi$ state \citep{kepa88a}. For the other states, we employ this scheme:
\begin{itemize}
\item[$\bullet$] if it is part of a vibrational series (such as band 30, for which the corresponding $v'$=1 band is No. 27), we can derive $\weu$ from the difference in $\nu_0$;
\item[$\bullet$] else, if it is part of a Rydberg series converging to the $X\,^2\Sigma^+$ or $A\,^2\Pi$ state of \cop, we take those constants \citep{haridass00a};
\item[$\bullet$] else, we take the constants of ground-state CO \citep{guelachvili83a}.
\end{itemize}
The choice for each band and the values of the constants are given in \tb{moldata-ls}.
}


\subsection{Rotational constants}
\label{subsec:rotcon}
The rotational constants ($\bvu$ and $\dvu$) for each excited state are needed to compute the positions of the individual absorption lines. \er{} provided $\bvu$ values for most bands, at an estimated accuracy of better than 1\%. Their $\dvu$ values are less well constrained and may be off by more than a factor of two. However, this is of little importance for the low-$J$ lines typically involved in the photodissociation of CO. More accurate values ($\bvu$ to better than 0.1\%, $\dvu$ to 10\% or better) are available for 12 states from higher-resolution experiments \citep{eikema94a,ubachs94a,ubachs00a,cacciani01a,cacciani02a,cacciani04a}. Again, the data for isotopologues other than \coc{12} are generally scarce, so we have to compute their constants from theoretical isotopic relations. This increases the uncertainty in $\bvu$ to a few per cent. In case of bands 12, 15, 19 and 28, \er{} reported constants for \coo{18} but not for \coc{12}, so we employ the theoretical relations for the latter. We adopt the \vdb{} constants for bands 4 and 14, because they are more accurate than those of \er. No constants are available for Rydberg bands 2A and 6, so we adopt the constants of the associated \cop{} states ($A\,^2\Pi$ and $X\,^2\Sigma^+$, respectively).

In seven cases, the rotational constants of the P and R branch ($e$ parity) were found to differ from those of the Q branch \citep[$f$ parity;][]{ubachs94a,ubachs00a,cacciani02a,cacciani04a}. For these bands, the $f$ parity values are given in \tb{moldata-ls}. \tb{parbc} lists the difference between the $f$ and $e$ values, defined as $q_{\el{v}}'=B_{\el{v},e}'-B_{\el{v},f}'$ and $p_{\el{v}}'=D_{\el{v},e}'-D_{\el{v},f}'$. The uncertainties in $q_{\el{v}}'$ and $p_{\el{v}}'$ are on the order of 1 and 10\%, respectively.

\begin{table}
\caption{Parity-dependent rotational constants for \coc{12}.}
\label{tb:parbc}
\centering
\begin{tabular}{c r@{.}l r@{.}l c}
\hline\hline
Band & \mc{$\qvu$} & \mc{$\pvu{}^{\mathrm{a}}$} & Refs.$^{\mathrm{b}}$ \\
\#   & \mc{(\pcm)} & \mc{(\pcm)} & \\
\hline
16 & $-2$&7(-3) & \mc{---} & 1 \\
22 & 2&212(-2) & 7&9(-6) & 2 \\
23 & 3&0(-2) & \mc{---} & 1 \\
25 & $-7$&0(-4) & \mc{---} & 1 \\
26 & $-1$&11(-2) & \mc{---} & 1 \\
31 & 1&14(-2) & 3&0(-8) & 3 \\
33 & 1&196(-2) & 2&1(-7) & 4 \\
\hline
\end{tabular}
\begin{list}{}{}
\item[$^{\mathrm{a}}$] Dashes indicate that no measurement is available, so we adopt a value of zero.
\item[$^{\mathrm{b}}$] (1) \citet{ubachs94a}; (2) \citet{cacciani02a}; (3) \citet{ubachs00a}; (4) \citet{cacciani04a}
\end{list}
\end{table}


\subsection{Oscillator strengths}
\label{subsec:osc}
\er{} measured the integrated absorption cross sections ($\sigma_\el{int}$) for all their bands to a typical accuracy of 20\%, but they cautioned that some values, especially for mutually overlapping bands, may be off by up to a factor of two. The oscillator strengths ($\osc$) derived from these data are different from \vdb{} for most bands, sometimes by even more than a factor of two. In addition, there are differences of up to an order of magnitude between the cross sections of \coc{12} and those of the other isotopologues for many bands shortwards of 990 \AA. The isotopic differences are likely due in part to the difficulty in determining individual cross sections for strongly overlapping bands, but isotope-selective oscillator strengths in general are not unexpected. For example, they were also observed recently in high-resolution measurements of N$_2$ (G.\ Stark, priv.\ comm.). For CO, the oscillator strengths depend on the details of the interactions between the $J'$, $v'$ levels of the excited states and other rovibronic levels. These interactions, in turn, depend on the energy levels, which are different between the isotopologues. We adopt the isotope-selective oscillator strengths where available. In case of transitions where no isotopic data exist, we choose to take the value of the isotopologue nearest in mass. This gives the closest match in energy levels and should, in general, also give the closest match in oscillator strengths, which to first order are determined by the Franck-Condon factors.

For ten of our bands, we adopt oscillator strenghts from studies that aimed specifically at measuring this parameter \citep{federman01a,eidelsberg04a,eidelsberg06a}. Their esimated accuracy is 5--15\%. The oscillator strength for the $E0$ band from \citet{federman01a} is almost twice as large as that of \vdb{} and \er, which appears to be due to an inadequate treatment of saturation effects in the older work. The 2001 value corresponds well to other values derived since 1990. \citeauthor{federman01a} also measured the oscillator strength of the weaker $C1$ band and found it to be the same, within the error margins, as that of \vdb{} and \er. Recent measurements of the $K0$, $L0$ and $E1$ transitions and the four $W$ transitions show larger oscillator strengths than those of \vdb{} and \er{} \citep{eidelsberg04a,eidelsberg06a}. The new values correspond closely to those of \citet{sheffer03a}, who derived oscillator strengths for eight bands by fitting a synthetic spectrum to Far Ultraviolet Spectroscopic Explorer (FUSE) data taken towards the star \object{HD 203374A}. Lastly, the \citet{eidelsberg04a} value for the $L'1$ band is 33\% lower than that of \vdb, but very similar to those of \er{} and \citet{sheffer03a}, so we adopt it as well.

We compute the oscillator strengths of individual lines as the product of the appropriate H\"onl-London factor and the oscillator strength of the corresponding band \citep{morton94a}. Significant departures from H\"onl-London patterns have been reported for many N$_2$ lines, sometimes even for the lowest rotational levels \citep{stark05a,stark08a}. The oscillator strength measured in one particular N$_2$ band for the P(22) line was twenty times stronger than that for the P(2) line, due to strong mixing of the upper state with a nearby Rydberg state. For other bands where deviations from H\"onl-London factors were observed, the effect was generally less than 50\% at $J'$=10. Similar deviations are likely to occur for CO, but a lack of experimental data prevents us from including this in our model. Note, however, that large deviations are only expected for specific levels that happen to be strongly interacting with another state. Many hundreds of levels contribute to the photodissociation rate, so the effect of some erroneous individual line oscillator strengths is small.


\subsection{Lifetimes and predissociation probabilities}
\label{subsec:eta}
Upon excitation, there is competition between dissociative and radiative decay. A band's predissociation probability ($\eta$) can be computed if the upper state's total and radiative lifetimes are known: $\eta=1-\tau_\el{tot}/\tau_\el{rad}$. \er{} reported total lifetimes (1/$A_\el{tot}$) for all their bands, but many of these are no more than order-of-magnitude estimates. Higher-resolution experiments have since produced more accurate values for 17 of our bands \citep{ubachs94a,ubachs00a,cacciani98a,cacciani01a,cacciani02a,eidelsberg06a}. In several cases, values that differ by up to a factor of three are reported for different isotopologues. Where available, we take isotope-specific values. Otherwise, we follow the same procedure as for the oscillator strengths, and take the value of the isotopologue nearest in mass.

The total lifetimes of some upper states have been shown to depend on the rotational level \citep{drabbels93b,ubachs00a,eidelsberg06a}. In case of $^1\Pi$ states, a dependence on parity was sometimes observed as well. We include these effects for the five bands in which they have been measured (\tb{parrota}).

\begin{table}
\caption{Parity- and rotation-dependent inverse lifetimes for \coc{12}.}
\label{tb:parrota}
\centering
\begin{tabular}{c r@{.}l r@{.}l c}
\hline\hline
Band & \mc{$A_{\el{tot},f}{}^{\mathrm{a}}$} & \mc{$A_{\el{tot},e}{}^{\mathrm{a}}$} & Refs.$^{\mathrm{b}}$ \\
\#   & \mc{(\ps)}                           & \mc{(\ps)}                           & \\
\hline
\phantom{0}8 & 3&6(11)+4.0(9)$x$ & 1&6(11)+1.3(10)$x$ & 1 \\
\phantom{$^{\mathrm{c}}$}13$^{\mathrm{c}}$ & 3&4(10)+7.3(10)$x$ & \mc{---}
 & 2 \\
16 & 1&0(11)+1.8(9)$x$ & 1&0(11)+3.4(9)$x$ & 1 \\
22 & 1&83(9) & 1&91(9)+1.20(9)$x$ & 3 \\
25 & 1&2(10) & 1&2(10)+2.4(9)$x$ & 1 \\
\hline
\end{tabular}
\begin{list}{}{}
\item[$^{\mathrm{a}}$] $x$ stands for $J'(J'+1)$.
\item[$^{\mathrm{b}}$] (1) \citet{eidelsberg06a}; (2) \citet{ubachs94a}; (3) \citet{drabbels93b}.
\item[$^{\mathrm{c}}$] This is a $^1\Sigma^+$ upper state, so there is no distinction between $e$ and $f$ parity.
\end{list}
\end{table}

Recent experiments by \citet{chakraborty08a} suggest isotope-dependent photodissociation rates for the $E0$, $E1$, $K0$ and $W2$ bands. These have been interpreted to imply different predissociation probabilities of individual lines of the various isotopologues due to a near-resonance accidental predissociation process. Similar effects have been reported for ClO$_2$ and CO$_2$ \citep{lim99a,bhattacharya00a}. In this process, the bound-state levels into which the UV absorption takes place do not couple directly with the continuum of a dissociative state. Instead, they first transfer population to another bound state, whose levels happen to lie close in energy. For the CO $E1$ state, this process was rotationally resolved by \citet{ubachs00a} for all naturally occurring isotopologues and shown to be due to spin-orbit interaction with the $k\,^3\Pi$ $v'$=6 state, which in turn couples with a repulsive state. The predissociation rates of the $E1$ state are found to increase significantly due to this process, but only for specific $J'$ levels that accidentally overlap. For example, interaction occurs at $J'$=7, 9 and 12 for \coc{12}, but at $J'$=1 and 6 for \co{13}{18}. Since the dissociation probabilities for the $E1$ state due to direct predissociation were already high, $\eta=0.96$, this increase in dissociation rate only has a very minor effect \citep{cacciani98a}. Moreover, under astrophysical conditions a range of $J'$ values are populated, so that the effect of individual levels is diluted. Since \citeauthor{chakraborty08a} did not derive line-by-line molecular parameters, we cannot easily incorporate their results into our model. In Sect.\ \ref{subsec:disk}, we show that our results do not change significantly when we include the proposed effects in an ad-hoc way.

The radiative lifetime of an excited state is a sum over the decay into all lower-lying levels, including the $A\,^1\Pi$ and $B\,^1\Sigma^+$ electronic states and the $v'$$\neq$0 
levels of the ground state. The total decay rate to the ground state is obtained by summing the oscillator strengths from \tb{moldata-ls} for each vibrational series \citep{morton94a,cacciani98a}. Theoretical work by \citet{kirby89a} showed that transitions to electronic states other than the ground state contribute about 1\% of the overall radiative decay rate for the $C$ state and about 8\% for the $E$ state. No data are available for the excited states at higher energies. Fortunately, the radiative decay rate from these higher states to the ground state is small compared to the dissociation rate, so an uncertainty of $\sim$10\% will not affect the $\eta$ values.

The predissociation probabilities thus computed are nearly identical to those of \vdb: the largest difference is a 10\% decrease for the $C1$ band. This is due to the larger oscillator strength now adopted.

There have been suggestions that the $C0$ state can also contribute to the photodissociation rate. \citet{cacciani01a} measured upper-state lifetimes in the $C0$ and $C1$ states for several CO isotopologues. For the $v'$=0 state of \coc{13} they found a total lifetime of 1770 ps, consistent with a value of 1780 ps for \coc{12}, but different from the lifetime of 1500 ps in \co{13}{18}. Although the three values agree within the mutual uncertainties of 10--15\%
on each measurement, \citeauthor{cacciani01a} suggested that the heaviest species, \co{13}{18}, has a predissociation yield of $\eta=0.17$ rather than zero if the measurements are taken at face value and if the radiative lifetime of the $C0$ state is presumed to be 1780 ps for all three species. The accurate absorption oscillator strength measured by \citet{federman01a} for the $C0$ band, $0.123\pm0.016$, implies a radiative lifetime that can be no longer than 1658 ps at the lower bound of measurement uncertainty in $\osc$. Taken together, the lifetime measurements of \citeauthor{cacciani01a} and the absorption measurements of \citeauthor{federman01a} favour a conservative conclusion that the dissociation yield is zero for each of these three isotopologues. We assume the $C0$ band is also non-dissociative in \coo{17}, \coo{18} and \co{13}{17}; this is consistent with earlier studies (e.g., \vdb; \er; \citealt{morton94a}).


\subsection{Atomic and molecular hydrogen}
\label{subsec:hydrogen}
Lines of atomic and molecular hydrogen (H and \mh) form an important contribution to the overall shielding of CO. As in \vdb, we include H Lyman lines up to $n$=50 and \mh{} Lyman and Werner lines (transitions to the $B\,^1\Sigma_u^+$ and $C\,^1\Pi_u$ states) from the $v''$=0, $J''$=0--7 levels of the electronic ground state. We adopt the line positions, oscillator strengths and lifetimes from \citet{abgrall93a,abgrall93b}, as compiled for the freely available \textsc{Meudon} PDR code \citep{lebourlot93a,lepetit02a,lepetit06a}.\footnote{http://aristote.obspm.fr/MIS/pdr/pdr1.html} Ground-state rotational constants, required to compute the level populations, come from \citet{jennings84a}.


\section{Depth-dependent photodissociation}
\label{sec:photdis}


\subsection{Default model parameters}
\label{subsec:default}
The simplest way of modelling the depth-dependent photodissociation involves dividing a one-dimensional model of an astrophysical object, irradiated only from one side, into small steps, in which the photodissociation rates can be assumed constant. We compute the abundances from the edge inwards, so that at each step we know the columns of CO, H, \mh{} and dust shielding the unattenuated radiation field.

Following \vdb, \citet{lebourlot93a}, \citet{lee96a} and \citet{lepetit06a}, we treat the line and continuum attenuation separately. For each of our 37 CO bands, we include all lines originating from the first ten rotational levels ($J''$=0--9) of the $v''$=0 level of the electronic ground state. That results in 855 lines per isotopologue. In addition, we have 48 H lines and 444 \mh{} lines, for a total of 5622. We use an adaptive wavelength grid that resolves all lines without wasting computational time on empty regions. For typical model parameters, the wavelength range from 911.75 to 1117.80 \AA{} is divided into $\sim$47,000 steps.

We characterise the population distribution of CO over the rotational levels by a single temperature, $\texco$. The \mh{} population requires a more detailed treatment, because UV pumping plays a large role for the $J''$>4 levels \citep{vandishoeck86a}. We populate the $J''$=0--3 levels according to a single temperature, $\texmh$, and adopt fixed columns of \scit{4}{15}, \scit{1}{15}, \scit{2}{14} and \scit{1}{14} \pcs{} for $J''$=4--7. This reproduces observed translucent cloud column densities to within a factor of two (\citealt{vandishoeck86a} and references therein). The $J''$>4 population scales with the UV intensity, so we re-evaluate this point for PDRs and circumstellar disks in Sects.\ \ref{subsec:pdr} and \ref{subsec:disk}.

The line profiles of CO, \mh{} and H are taken to be Voigt functions, with default Doppler widths ($b$) of 0.3, 3.0 and 5.0 km \ps{}, respectively. We adopt \citet{draine78a} as our standard unattenuated interstellar radiation field.


\subsection{Unshielded photodissociation rates}
\label{subsec:urate}
We obtain an unshielded CO photodissociation rate of \scit{2.6}{-10} \ps. This rate is 30\% higher than that of \vdb, due to the generally larger oscillator strengths in our data set. The new data for bands 33 and 24 (the $E0$ and $K0$ transitions) have the largest effect: they account for 63 and 21\% of the overall increase. Clearly, the rate depends on the choice of radiation field. If we adopt \citet{habing68a}, \citet{gondhalekar80a} or \citet{mathis83a} instead of \citet{draine78a}, the photodissociation rate becomes 2.0, 2.0 or \scit{2.3}{-10} \ps, respectively. \vdb{} reported the same relative differences between these three fields.


\begin{table*}
\caption{Relative and absolute shielding effects for the ten most important bands in the \object{\zoph} cloud.$^{\mathrm{a}}$}
\label{tb:zophrate}
\centering
\begin{tabular}{lccccccccccc}
\hline\hline
Band & 33 & 28 & 24 & 23 & 22 & 20 & 19 & 16 & 15 & 13 & Total$^{\mathrm{c}}$ \\
$\lambda_0$ (\AA)$^{\mathrm{b}}$ & 1076.1 & 985.6 & 970.4 & 968.9 & 968.3 & 956.2 & 950.0 & 941.2 & 940.0 & 933.1 & \\
\hline
\multicolumn{12}{c}{\rule{0pt}{1em}\coc{12}} \\
\hline
Edge (\%)$^{\mathrm{d}}$ & 32.2\phantom{000} & \phantom{0}4.7\phantom{000} & \phantom{0}8.7\phantom{000} & \phantom{0}3.0\phantom{000} & \phantom{0}3.6\phantom{000} & \phantom{0}3.4\phantom{000} & \phantom{0}4.3\phantom{000} & \phantom{0}5.1\phantom{000} & \phantom{0}3.5\phantom{000} & \phantom{0}3.1\phantom{000} & 7.8(-10) \\
Centre (\%)$^{\mathrm{d}}$ & \phantom{0}2.8\phantom{000} & \phantom{0}0.4\phantom{000} & \phantom{0}5.7\phantom{000} & \phantom{0}9.3\phantom{000} & \phantom{0}0.8\phantom{000} & 12.1\phantom{000} & 13.8\phantom{000} & \phantom{0}6.1\phantom{000} & 11.0\phantom{000} & \phantom{0}7.0\phantom{000} & 7.5(-11) \\
Shielding$^{\mathrm{e}}$ & \phantom{0}0.0084 & \phantom{0}0.0072 & \phantom{0}0.063\phantom{0} & \phantom{0}0.30\phantom{00} & \phantom{0}0.022\phantom{0} & \phantom{0}0.34\phantom{00} & \phantom{0}0.31\phantom{00} & \phantom{0}0.12\phantom{00} & \phantom{0}0.30\phantom{00} & \phantom{0}0.22\phantom{00} & 0.10\phantom{(-0)} \\
\hline
\multicolumn{12}{c}{\rule{0pt}{1em}\coo{17}} \\
\hline
Edge (\%) & 32.0\phantom{000} & \phantom{0}4.7\phantom{000} & \phantom{0}8.7\phantom{000} & \phantom{0}3.0\phantom{000} & \phantom{0}3.6\phantom{000} & \phantom{0}3.4\phantom{000} & \phantom{0}4.3\phantom{000} & \phantom{0}5.1\phantom{000} & \phantom{0}3.5\phantom{000} & \phantom{0}3.1\phantom{000} & 7.8(-10) \\
Centre (\%) & 22.0\phantom{000} & \phantom{0}0.1\phantom{000} & \phantom{0}5.3\phantom{000} & \phantom{0}3.5\phantom{000} & \phantom{0}0.8\phantom{000} & \phantom{0}8.7\phantom{000} & 10.3\phantom{000} & 14.2\phantom{000} & \phantom{0}9.1\phantom{000} & \phantom{0}2.4\phantom{000} & 2.4(-10) \\
Shielding & \phantom{0}0.21\phantom{00} & \phantom{0}0.0067 & \phantom{0}0.19\phantom{00} & \phantom{0}0.36\phantom{00} & \phantom{0}0.065\phantom{0} & \phantom{0}0.79\phantom{00} & \phantom{0}0.74\phantom{00} & \phantom{0}0.85\phantom{00} & \phantom{0}0.81\phantom{00} & \phantom{0}0.23\phantom{00} & 0.31\phantom{(-0)} \\
\hline
\multicolumn{12}{c}{\rule{0pt}{1em}\coo{18}} \\
\hline
Edge (\%) & 31.5\phantom{000} & \phantom{0}5.1\phantom{000} & \phantom{0}6.2\phantom{000} & \phantom{0}1.5\phantom{000} & \phantom{0}4.9\phantom{000} & \phantom{0}3.7\phantom{000} & \phantom{0}4.7\phantom{000} & \phantom{0}5.2\phantom{000} & \phantom{0}3.8\phantom{000} & \phantom{0}3.4\phantom{000} & 7.2(-10) \\
Centre (\%) & 35.3\phantom{000} & \phantom{0}0.1\phantom{000} & \phantom{0}4.7\phantom{000} & \phantom{0}2.6\phantom{000} & \phantom{0}7.8\phantom{000} & \phantom{0}3.3\phantom{000} & \phantom{0}7.7\phantom{000} & \phantom{0}8.1\phantom{000} & \phantom{0}5.9\phantom{000} & \phantom{0}3.0\phantom{000} & 3.7(-10) \\
Shielding & \phantom{0}0.58\phantom{00} & \phantom{0}0.0067 & \phantom{0}0.39\phantom{00} & \phantom{0}0.88\phantom{00} & \phantom{0}0.82\phantom{00} & \phantom{0}0.46\phantom{00} & \phantom{0}0.83\phantom{00} & \phantom{0}0.80\phantom{00} & \phantom{0}0.81\phantom{00} & \phantom{0}0.45\phantom{00} & 0.52\phantom{(-0)} \\
\hline
\multicolumn{12}{c}{\rule{0pt}{1em}\coc{13}} \\
\hline
Edge (\%) & 28.1\phantom{000} & \phantom{0}6.5\phantom{000} & \phantom{0}8.3\phantom{000} & \phantom{0}1.8\phantom{000} & \phantom{0}5.2\phantom{000} & \phantom{0}3.4\phantom{000} & \phantom{0}4.3\phantom{000} & \phantom{0}4.9\phantom{000} & \phantom{0}3.9\phantom{000} & \phantom{0}2.4\phantom{000} & 7.8(-10) \\
Centre (\%) & 18.5\phantom{000} & \phantom{0}0.0\phantom{000} & \phantom{0}5.3\phantom{000} & \phantom{0}4.4\phantom{000} & 11.2\phantom{000} & \phantom{0}4.1\phantom{000} & 10.7\phantom{000} & \phantom{0}4.4\phantom{000} & \phantom{0}9.1\phantom{000} & \phantom{0}0.8\phantom{000} & 2.8(-10) \\
Shielding & \phantom{0}0.23\phantom{00} & \phantom{0}0.0007 & \phantom{0}0.23\phantom{00} & \phantom{0}0.87\phantom{00} & \phantom{0}0.76\phantom{00} & \phantom{0}0.42\phantom{00} & \phantom{0}0.87\phantom{00} & \phantom{0}0.32\phantom{00} & \phantom{0}0.83\phantom{00} & \phantom{0}0.13\phantom{00} & 0.35\phantom{(-0)} \\
\hline
\multicolumn{12}{c}{\rule{0pt}{1em}\co{13}{17}} \\
\hline
Edge (\%) & 28.0\phantom{000} & \phantom{0}6.3\phantom{000} & \phantom{0}8.3\phantom{000} & \phantom{0}1.8\phantom{000} & \phantom{0}5.2\phantom{000} & \phantom{0}3.4\phantom{000} & \phantom{0}4.3\phantom{000} & \phantom{0}4.9\phantom{000} & \phantom{0}3.9\phantom{000} & \phantom{0}2.4\phantom{000} & 7.8(-10) \\
Centre (\%) & 32.5\phantom{000} & \phantom{0}0.0\phantom{000} & \phantom{0}7.4\phantom{000} & \phantom{0}2.9\phantom{000} & \phantom{0}7.5\phantom{000} & \phantom{0}5.0\phantom{000} & \phantom{0}7.2\phantom{000} & \phantom{0}8.5\phantom{000} & \phantom{0}6.1\phantom{000} & \phantom{0}1.0\phantom{000} & 4.2(-10) \\
Shielding & \phantom{0}0.63\phantom{00} & \phantom{0}0.0007 & \phantom{0}0.48\phantom{00} & \phantom{0}0.88\phantom{00} & \phantom{0}0.79\phantom{00} & \phantom{0}0.78\phantom{00} & \phantom{0}0.91\phantom{00} & \phantom{0}0.94\phantom{00} & \phantom{0}0.85\phantom{00} & \phantom{0}0.24\phantom{00} & 0.54\phantom{(-0)} \\
\hline
\multicolumn{12}{c}{\rule{0pt}{1em}\co{13}{18}} \\
\hline
Edge (\%) & 30.1\phantom{000} & \phantom{0}7.0\phantom{000} & \phantom{0}5.9\phantom{000} & \phantom{0}1.5\phantom{000} & \phantom{0}4.7\phantom{000} & \phantom{0}3.6\phantom{000} & \phantom{0}4.4\phantom{000} & \phantom{0}5.0\phantom{000} & \phantom{0}4.0\phantom{000} & \phantom{0}2.4\phantom{000} & 7.5(-10) \\
Centre (\%) & 39.0\phantom{000} & \phantom{0}0.0\phantom{000} & \phantom{0}5.8\phantom{000} & \phantom{0}2.4\phantom{000} & \phantom{0}5.7\phantom{000} & \phantom{0}1.7\phantom{000} & \phantom{0}7.5\phantom{000} & \phantom{0}8.5\phantom{000} & \phantom{0}6.3\phantom{000} & \phantom{0}1.7\phantom{000} & 4.1(-10) \\
Shielding & \phantom{0}0.71\phantom{00} & \phantom{0}0.0007 & \phantom{0}0.54\phantom{00} & \phantom{0}0.87\phantom{00} & \phantom{0}0.67\phantom{00} & \phantom{0}0.26\phantom{00} & \phantom{0}0.92\phantom{00} & \phantom{0}0.94\phantom{00} & \phantom{0}0.85\phantom{00} & \phantom{0}0.38\phantom{00} & 0.55\phantom{(-0)} \\
\hline
\end{tabular}
\begin{list}{}{}
\item[$^{\mathrm{a}}$] See the text for the adopted column densities, Doppler widths, excitation temperatures and radiation field.
\item[$^{\mathrm{b}}$] \coc{12} band head position.
\item[$^{\mathrm{c}}$] Total photodissociation rate in \ps{} at the edge and the centre, and the shielding factor at the centre.
\item[$^{\mathrm{d}}$] Relative contribution per band to the overall photodissociation rate at the edge and the centre of the cloud.
\item[$^{\mathrm{e}}$] Shielding factor per band: the absolute contribution at the centre divided by the absolute contribution at the edge.
\end{list}
\end{table*}

\subsection{Shielding by CO, \mh{} and H}
\label{subsec:depth}
Self-shielding, shielding by H, \mh{} and the other CO isotopologues, and continuum shielding by dust all reduce the photodissociation rates inside a cloud or other environment relative to the unshielded rates. For a given combination of column depths ($N$) and visual extinction ($\av$), the photodissociation rate for isotopologue $i$ is
\begin{equation}
\label{eq:pdrate}
k_i = \chi k_{0,i} \Theta_i \exp(-\gamma \av)\,,
\end{equation}
with $\chi$ a scaling factor for the UV intensity and $k_{0,i}$ the unattenuated rate in a given radiation field. The shielding function $\Theta_i$ accounts for self-shielding and shielding by H, \mh{} and the other CO isotopologues; tabulated values for some typical model parameters are presented in Tables \ref{tb:shield}--\ref{tb:shield4}. The dust extinction term, $\exp(-\gamma\av)$, is discussed in Sect.\ \ref{subsec:dust}. \eqq{pdrate} assumes the radiation is coming from all directions. If this is not the case, such as for a cloud irradiated only from one side, $k_{0,i}$ should be reduced accordingly.

For now, we ignore shielding by dust and compute the depth-dependent photodissociation rates due to line shielding only. Our test case is the centre of the diffuse cloud towards the star \zoph. The observed column densities of H, \mh, \coc{12} and \coc{13} are \scit{5.2}{20}, \scit{4.2}{20}, \scit{2.5}{15} and \scit{1.5}{13} \pcs{} \citep{vandishoeck86a,lambert94a}, and we assume \coo{17}, \coo{18}, \co{13}{17} and \co{13}{18} column densities of \scit{4.1}{11}, \scit{1.6}{12}, \scit{5.9}{9} and \scit{2.3}{10} \pcs, consistent with observational constraints. For the centre of the cloud, we adopt half of these values. We set $\bco=0.48$ km \ps{} and $\texco=4.2$ K \citep{sheffer92a}, and we populate the \mh{} rotational levels explicitly according to the observed distribution. The cloud is illuminated by three times the \citeauthor{draine78a} field ($\chi=3$).

\tb{zophrate} lists the relative contribution of the most important bands at the edge and centre for each isotopologue, as well as the overall photodissociation rate at each point. The column densities are small, but isotope-selective shielding already occurs: the \coc{12} rate at the centre is lower than that of the other isotopologues by factors of three to six.

The $E0$ band is the most important contributor at the edge. This was also found by \vdb, but the higher oscillator strength now adopted makes it even stronger. Going to the centre, it saturates rapidly for \coc{12}: its absolute contribition to the total dissociation rate decreases by two orders of magnitude, and it goes from the strongest band to the 12th strongest band. The five strongest bands at the centre are the same as in \vdb: Nos. 13, 15, 19, 20 and 23.

\begin{figure}
\resizebox{\hsize}{!}{\includegraphics{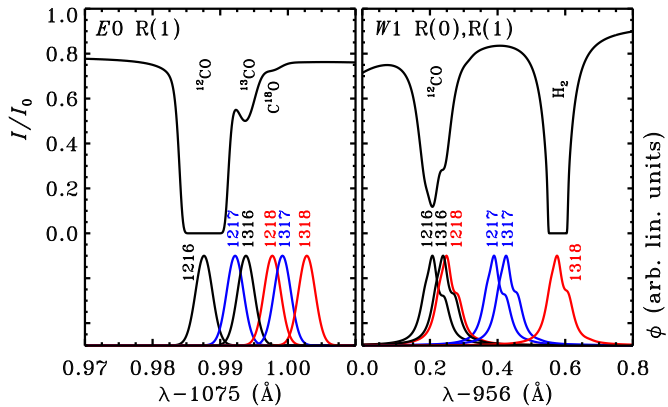}}
\caption{Relative intensity of the radiation field ($I/I_0$) and intrinsic line profiles for the six CO isotopologues ($\phi$, in arbitrary linear units) at the centre of the \zoph{} cloud in two wavelength ranges.}
\label{fig:zoph4c}
\end{figure}

Figure \ref{fig:zoph4c} illustrates the isotope-selective shielding. The left panel is centred on the R(1) line of the $E0$ band (No. 33 from \tb{moldata-ls}). This line is fully saturated in \coc{12} and the relative intensity of the radiation field ($I/I_0$, with $I_0$ the intensity at the edge of the cloud) goes to zero. \coc{13} and \coo{18} also visibly reduce the intensity, to $I/I_0=0.50$ and 0.72, but the other three isotopologues are not abundant enough to do so. Consequently, these three are not self-shielded in the \zoph{} cloud, but they are shielded by \coc{12}, \coc{13} and \coo{18}. The weaker shielding of \co{13}{17} and \co{13}{18} in the $E0$ band compared to \coo{17} is due to their lines having less overlap with the \coc{12} lines.

The right panel of \fig{zoph4c} contains the R(0) line of the $W1$ band (No. 20), with the R(1) line present as a shoulder on the red wing. Also visible is the saturated $B13$ R(2) line of \mh{} at 956.58 \AA. The $W1$ band is weaker than the $E0$ band, so \coc{12} is the only isotopologue to cause any appreciable reduction in the radiation field and to be (partially) self-shielded. The shielding of the other five isotopologues is dominated by overlap with the \coc{12} and \mh{} lines. This figure also shows the need for accurate line positions: if the \co{13}{18} line were shifted by 0.1 \AA{} in either direction, it would no longer overlap with the \mh{} line and be less strongly shielded. Note that the position of the $W1$ band has only been measured for \coc{12}, \coc{13} and \coo{18}, so we have to compute the position for the other isotopologues from theoretical isotopic relations. This causes the \coo{17} line to appear longwards of the \coo{18} line.


\subsection{Continuum shielding by dust}
\label{subsec:dust}
Dust can provide a very strong attenuation of the radiation field. This effect is largely independent of wavelength for the 912--1118 \AA{} radiation available to dissociate CO, so it affects all isotopologues to the same extent. It can be expressed as an exponential function of the visual extinction, as expressed in \eq{pdrate}. For typical interstellar dust grains (radius of 0.1 \micron{} and optical properties from \citealt{roberge91a}), the extinction coefficient $\gamma$ is 3.53 for CO \citep{vandishoeck06a}. Larger grains have less opacity in the UV and do not shield CO as strongly. For ice-coated grains with a mean radius of 1 \micron, appropriate for circumstellar disks \citep{jonkheid06a}, the extinction coefficient is only 0.6. The effects of dust shielding are discussed more fully in Sect.\ \ref{sec:implic}.

Photodissociation of CO may still take place even in highly extincted regions. Cosmic rays or energetic electrons generated by cosmic rays can excite \mh, allowing it to emit in a multitude of bands, including the Lyman and Werner systems \citep{prasad83a}. The resulting UV photons can dissociate CO at a rate of about \ten{-15} \ps{} \citep{gredel87a}, independent of depth. That is enough to increase the atomic C abundance by some three orders of magnitude compared to a situation where the photodissociation rate is absolutely zero. The cosmic-ray-induced photodissociation rate is sensitive to the spectroscopic constants of CO, especially where it concerns the overlap between CO and \mh{} lines, so it would be interesting to redo the calculations of \citeauthor{gredel87a} with the new data from \tb{moldata-ls}. However, that is beyond the scope of this paper.


\subsection{Uncertainties}
\label{subsec:unc}
The uncertainties in the molecular data are echoed in the model results. When coupled to a chemical network, as in Sect.\ \ref{sec:implic}, the main observables produced by the model are the column densities of the CO isotopologues for a given astrophysical environment. The accuracy of the photodissociation rates is only relevant in a specific range of depths; in the average interstellar UV field, this range runs from an $\av$ of $\sim$0.2 to $\sim$2 mag. Photoprocesses are so dominant at lower extinctions and so slow at higher extinctions that the exact rate does not matter. In the intermediate regime, both the absolute photodissociation rates and the differences between the rates for individual isotopologues are important. The oscillator strengths are the key variable in both cases and these are generally known rather accurately. Taking account of the experimental uncertainties in the band oscillator strengths and of the theoretical uncertainties in computing the properties for individual lines, and identifying which bands are important contributors (\tb{zophrate}), we estimate the absolute photodissociation rates to be accurate to about 20\%. This error margin carries over into the absolute CO abundances and column densities for the $\av\approx0.2$--2 mag range when the rates are put into a chemical model. The accuracy on the rates and abundances of the isotopologues relative to each other is estimated to be about 10\% when summed over all states, even when we allow for the kind of isotope-specific predissociation probabilities suggested by \citet{chakraborty08a}.


\section{Excitation temperature and Doppler width}
\label{sec:temp}
The calculations of \vdb{} were only done for low excitation temperatures of CO and \mh{}. Here, we extend this work to higher temperatures, as required for PDRs and disks, and we re-examine the effect of the Doppler widths of CO, \mh{} and H on the photodissociation rates. We first treat four cases separately, increasing either $\texco$, $\bco$, $\texmh$ or $\bmh$. At the end of this section we combine these effects in a grid of excitation temperatures and Doppler widths. As a template model we take the centre of the \zoph{} cloud, with column densities and other parameters as described in Sect.\ \ref{subsec:depth}.


\subsection{Increasing $\texco$}
\label{subsec:inc-texco}
\begin{figure*}
\resizebox{\hsize}{!}{\includegraphics{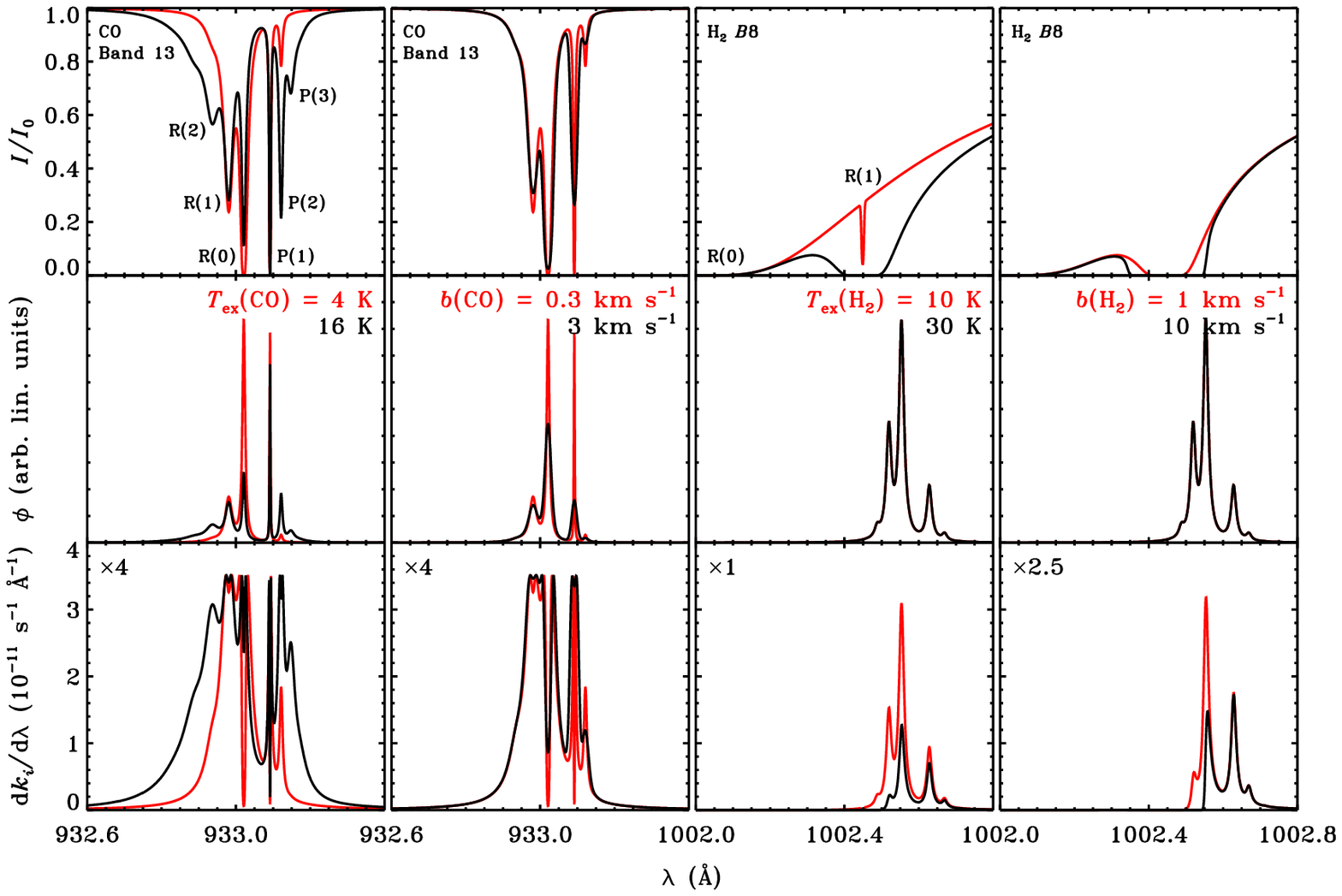}}
\caption{Illustration of the effect of increasing $\texco$, $\bco$, $\texmh$ or $\bmh$ (\emph{from left to right}) in our \zoph{} cloud model (Sect.\ \ref{subsec:depth}). \emph{Top:} relative intensity of the radiation field, including absorption by only \coc{12} (\emph{left two panels}) or \mh{} (\emph{right two panels}). \emph{Middle:} intrinsic line profile for band 13 (933.1 \AA) or 30 (1002.6 \AA) of \coc{12}. \emph{Bottom:} photodissociation rate per unit wavelength, multiplied by a constant as indicated.}
\label{fig:btexincdec}
\end{figure*}

As the excitation temperature of CO increases, additional rotational levels are populated and photodissociation is spread across more lines. Figure \ref{fig:btexincdec} visualises this for band 13 of \coc{12}. At 4 K, only four lines are active: the R(0), R(1), P(1) and P(2) lines at 933.02, 932.98, 933.09 and 933.12 \AA. The R(0) and P(1) lines are both fully self-shielded at the line centre. Going to 16 K, the R(0) line loses about 70\%
of its intrinsic intensity and ceases to be self-shielding. In addition, the R(2), P(3) and higher-$J$ lines start to absorb. The combination of less saturated low-$J$ lines and more active higher-$J$ lines yields a 39\% higher photodissociation rate at 16 K compared to 4 K.

A higher CO excitation temperature has the same favourable effect for \coc{13}, which is partially self-shielded at the centre of the \zoph{} cloud. Its photodissociation rates increase by 16\% when going from 4 to 16 K. \coo{18} is also partially self-shielded, but less so than \coc{13}, so the favourable effect is smaller. At the same time, it suffers from increased overlap by \coc{12}. The net result is a small increase in the photodissociation rate of 0.2\%.

The two heaviest isotopologues, \co{13}{17} and \co{13}{18}, are not abundant enough to be self-shielded. Their $J''$<2 lines generally have little overlap with the corresponding \coc{12} lines, especially in the $E0$ band near 1076 \AA. This band, whose lines are amongst the narrowest in our data set, is the strongest contributor to the photodissociation rate at the centre of the cloud for \co{13}{17} and \co{13}{18} (\tb{zophrate}). In fact, its narrow lines are part of the reason it is the strongest contributor. The $J''$=3 and 4 lines that become active at 16 K do have some overlap with \coc{12}. Without the favourable effect of less self-shielding, this causes the photodissociation rate for \co{13}{17} and \co{13}{18} to decrease for higher excitation temperatures. The change is only small, though: 0.4\% for \co{13}{17} and 2\% for \co{13}{18}.

Finally, \coo{17} experiences an increase of 18\% in its photodissociation rate. Its lines lie closer to those of \coc{12} than do the \co{13}{17} and \co{13}{18} lines, so it is generally more strongly shielded. At 4 K, most of the shielding is due to the saturated R(0) lines of \coc{12}. These become partially unsaturated at higher $\texco$, so the corresponding R(0) lines of \coo{17} become a stronger contributor to the photodissociation rate, even though the shift towards higher-$J$ lines make them intrinsically weaker. Overall, increasing $\texco$ from 4 to 16 K thus results in a higher \coo{17} photodissociation rate.


\subsection{Increasing $\bco$}
\label{subsec:inc-bco}
The width of the absorption lines is due to Doppler broadening and natural (or lifetime) broadening. The integrated intensity in each line remains the same when $\bco$ increases, so a larger width is accompanied by a lower peak intensity. The resulting reduction in self-shielding then causes a higher \coc{12} photodissociation rate, as shown in \fig{btexincdec} for band 13. However, the effect is rather small because the Doppler width is smaller than the natural width for most lines at typical $b$ values. Natural broadening is the dominant broadening mechanism up to $\bco \approx \scim{6}{-12}A_\el{tot}$, with both parameters in their normal units. The R(0) line of band 13 has an inverse lifetime of \scit{1.8}{11} \ps{} (Tables \ref{tb:moldata-ls} and \ref{tb:parrota}), so Doppler broadening becomes important at about 1 km \ps. From 0.3 to 3 km \ps, as in \fig{btexincdec}, the line width only increases by a factor of 1.9. Integrated over all lines, the \coc{12} photodissociation rate becomes 26\% higher.

The rates of the other five isotopologues decrease along this $\bco$ interval due to increased shielding by the $E0$ lines of \coc{12}. With an inverse lifetime of only \scit{1}{9} \ps, Doppler broadening is this band's dominant broadening mechanism in the regime of interest. A tenfold increase in the Doppler parameter from 0.3 to 3 km \ps{} results in a nearly tenfold increase in the line widths. At 0.3 km \ps, the $E0$ lines of \coc{12} are still sufficiently narrow that they do not strongly shield the lines of the other isotopologues. This is no longer the case at 3 km \ps. \coc{13} still benefits somewhat from reduced self-shielding in other bands, but it is not enough to overcome the reduced strength of the $E0$ band, and its photodissociation rates decrease by 2\%. The decrease is 13\% for \coo{17} and 26--28\% for the remaining three isotopologues. The relatively small decrease for \coo{17} is due to its $E0$ band being already partially shielded by \coc{12} at 0.3 km \ps, so the stronger shielding at 3 km \ps{} has less of an effect.


\subsection{Increasing $\texmh$ or $\bmh$}
\label{subsec:inc-texmh}
Increasing the excitation temperature of \mh, while keeping the CO parameters constant, results in a decreased photodissociation rate for all six isotopologues. The cause, as again illustrated in \fig{btexincdec}, is the activation of more \mh{} lines. At $\texmh=10$ K, the R(1) line of the $B8$ band at 1002.45 \AA{} is very narrow and does not shield the $F0$ band (No. 30) of the CO isotopologues. (The continuum-like shielding visible in \fig{btexincdec} is due to the strongly saturated $B8$ R(0) line at 1001.82 \AA.) It becomes much more intense at 30 K and widens due to being saturated, thereby shielding part of the $F0$ band. The same thing happens to other CO bands, resulting in an overall rate decrease of 0.6--2.5\%. There is no particular trend visible amongst the isotopologues; the magnitude of the rate change depends purely on the chance that a given CO band overlaps with an \mh{} line.

Similar decreases of one or two percent in the CO photodissociation rates are seen when the \mh{} Doppler width is changed from 1 to 10 km \ps. As the \mh{} lines become broader, the amount of overlap with CO increases across the entire wavelength range. As an example, \fig{btexincdec} shows again the 1002.0--1002.8 \AA{} region, where the $B8$ R(1) line of \mh{} further reduces the contribution of the $F0$ band to the \coc{12} photodissociation rate.


\subsection{Grid of $T_\el{ex}$ and $b$}
\label{subsec:btexgrid}
We now combine the four individual cases into a grid of excitation temperatures and Doppler widths to see how they influence each other. $\texco$ is raised from 4 to 512 K in steps of factors of two. The $v''$=1 vibrational level of \coc{12} lies at 2143 \pcm{} above the $v''$=0 level, so it starts to be thermally populated at $\sim$500 K. No data are available on dissociative transitions out of this level, so we choose not to go to higher excitation temperatures. We increase the range of rotational levels up to $J''$=39, at 2984 \pcm{} above the $J''$=0 level for \coc{12}. At $\texco=512$ K, the normalised population distribution peaks at $J''$=9 and decreases to \scit{9.7}{-5} at $J''$=39. The \mh{} excitation temperature is set to $[\texco]^{1.5}$ to take account of the fact that its critical densities for thermalisation are lower than those of CO. Where necessary, absorption by rotational levels above our normal limit of $J''$=7 and by non-zero vibrational levels is taken into account \citep{dabrowski84a,abgrall93a,abgrall93b}. All \mh{} rovibrational levels are strictly thermally populated; no UV pumping is included. The grid is run for CO Doppler widths of 0.1, 0.3, 1.0 and 3.2 km \ps; we set $\bmh=\sqrt{14}\bco$ and $b(\el{H})=\sqrt{28}\bco$, corresponding to the differences appropriate for thermal broadening.

\begin{figure*}
\resizebox{\hsize}{!}{\includegraphics{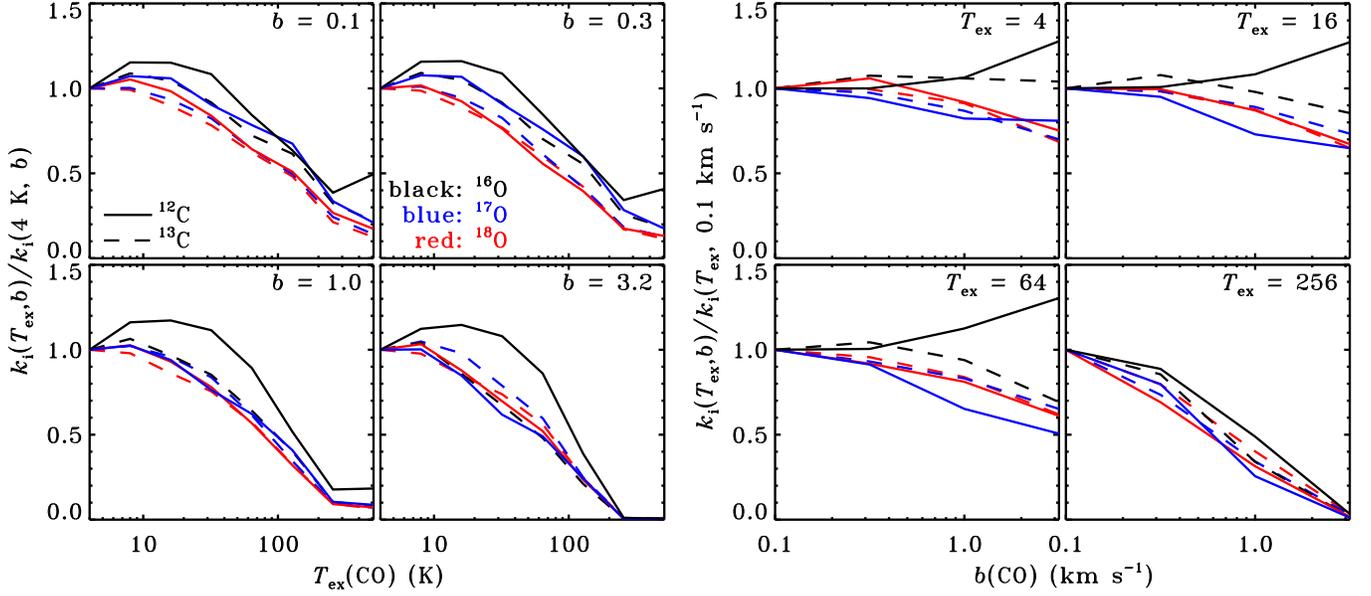}}
\caption{\emph{Left:} photodissociation rate of the CO isotopologues as a function of excitation temperature, normalised to the rate at 4 K, at four different Doppler widths, for parameters corresponding to the centre of the \zoph{} diffuse cloud. \emph{Right:} photodissociation rates as a function of Doppler width, normalised to the rate at 0.1 km \ps, at four different excitation temperatures. The other model parameters are as in Sect.\ \ref{subsec:depth}, except $\bmh=\sqrt{14}\bco$, $b(\el{H})=\sqrt{28}\bco$ and $\texmh=[\texco]^{1.5}$.}
\label{fig:tex1}
\end{figure*}

\begin{figure*}
\resizebox{\hsize}{!}{\includegraphics{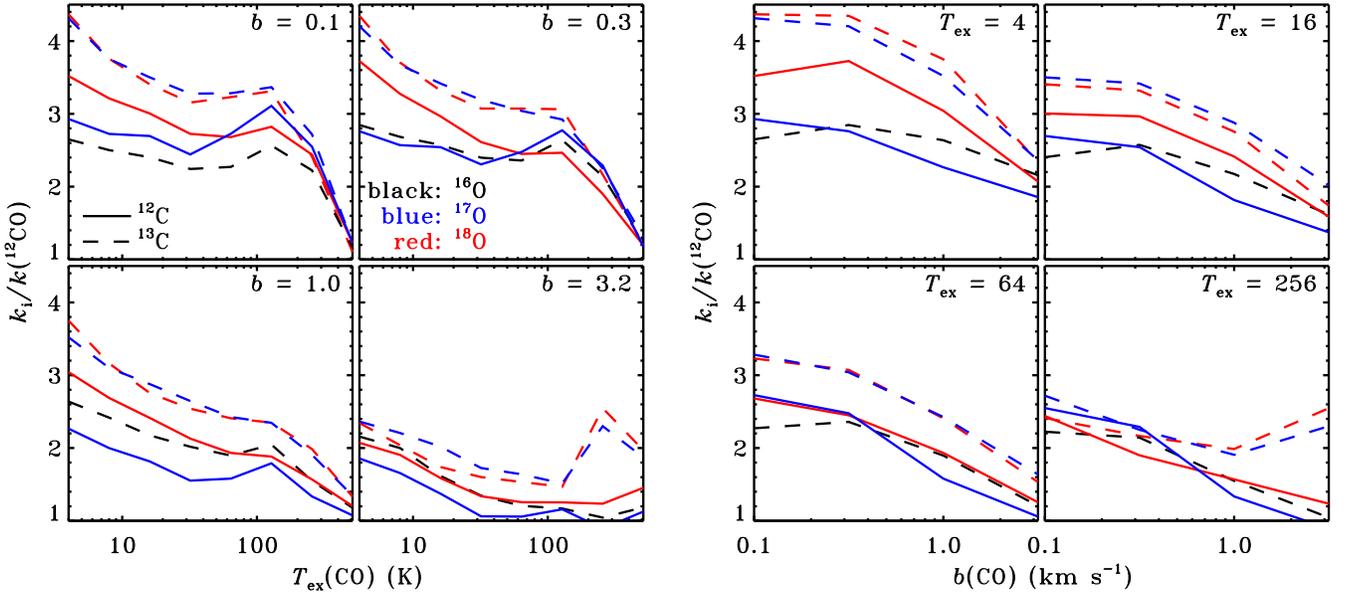}}
\caption{As \fig{tex1}, but with all rates normalised to the \coc{12} rate.}
\label{fig:tex2}
\end{figure*}

The left set of panels in \fig{tex1} shows the photodissociation rate of the six isotopologues at the centre of the \zoph{} cloud as a function of excitation temperature for the different Doppler widths. The rates are normalised to the rate at 4 K. The \coc{12} rate increases from 4 to 16 K, as described in Sect.\ \ref{subsec:inc-texco}. At higher temperatures the increased overlap with \mh{} lines takes over and the rate goes down. As long as the CO excitation temperature is less than $\sim$100 K, the \coc{12} rate remains constant up to $\bco=0.3$ km \ps{} and increases as $\bco$ goes from 0.3 to 3.2 km \ps (right side of \fig{tex1}). At higher temperatures there is so much shielding by \mh{} that reduced self-shielding in the CO lines has no discernable effect on the rate. Instead, the rate goes down with $b$ due to stronger shielding by the broadened \mh{} lines.

The \coc{13} rate also increases initially with $\texco$ and then goes down as \mh{} shielding takes over. The rate increases from $\bco=0.1$ to 0.3 km \ps, but decreases for higher values as described in Sect.\ \ref{subsec:inc-bco}. For the remaining four isotopologues, the plotted curves likewise result from a combination of weaker shielding by \coc{12} and stronger shielding by \mh. At CO excitation temperatures between 4 and 8 K, the rates typically change by a few per cent either way. Going to higher temperatures, all rates decrease monotonically. Likewise, the rates generally decrease towards higher $\bco$ values.

A change in behaviour is seen when increasing $\texco$ from 256 to 512 K. It is at this point that the $v''$>0 levels of \mh{} become populated. Less energy is now needed to excite \mh{} to the $B$ and $C$ states, so absorption shifts towards longer wavelengths. This causes even stronger shielding in the heavy CO isotopologues, for whom the $E0$ band at 1076 \AA{} is still an important contributor to the photodissociation rate, at least as long as $\bco$ does not exceed 0.3 km \ps. The $E0$ band is strongly self-shielded in \coc{12} (\tb{zophrate}), so the shift of the \mh{} absorption to longer wavelengths does not reduce its contribution by much. In fact, the weaker \mh{} absorption at shorter wavelengths allows for an increased contribution of bands like Nos. 13 and 16 at 933 and 941 \AA, causing a net increase in the \coc{12} photodissociation rate from 256 to 512 K. The situation changes somewhat when the CO Doppler width increases to 1.0 km \ps{} or more. The $E0$ band of the heavy isotopologues is now much less of a contributor, because it is shielded by the broader \coc{12} lines. The shifting \mh{} absorption does not cause any additional shielding, so the rates remain almost the same. Furthermore, the \mh{} lines are also broader and continue to shield the \coc{12} bands at shorter wavelengths, preventing its photodissociation rate from increasing like it does in the low-$b$ cases.

The main astrophysical consequence becomes clear when we look at the photodissociation rates of the five heavier isotopologues with respect to that of \coc{12}. Figure \ref{fig:tex2} shows this $k_i/k(\cocm{12})$ ratio as a function of $\texco$ and $\bco$. In these plots, the \coc{12} rate shows as a horizontal line with a value of unity. The ratios generally decrease with both parameters: a higher excitation temperature and a larger Doppler width both cause less self-shielding in \coc{12}, so the rate differences between the isotopologues become smaller. Shielding by \mh{} increases at the same time, further reducing the differences between the isotopologues. This means that photodissociation of CO is more strongly isotope-selective in cold sources than in hot sources.


\section{Shielding function approximations}
\label{sec:shield}
It is unpractical for many astrophysical applications to do the full integration of all 5622 lines in our model every time a photodissociation rate is required. Therefore, we present approximations to the shielding functions $\Theta$ introduced in \eq{pdrate}. The approximations are derived for several sets of model parameters and are valid across a wide range of astrophysical environments (Sect.\ \ref{subsec:shcomp}).


\subsection{Shielding functions on a two-dimensional grid of $\cdco$ and $\cdmh$}
\label{subsec:shield2d}
The transition from atomic to molecular hydrogen occurs much closer to the edge of the cloud than the \element[+]{C}--C--CO transition, so the column density of atomic H is roughly constant at the depths where shielding of CO is important. In addition, H shields CO by only a few per cent. Therefore, it is a good approximation to compute the shielding functions on a grid of CO and \mh{} column densities, while taking a constant column of H. It is sufficient to express the shielding of all CO isotopologues as a function of $N(\cocm{12})$, because self-shielding of the heavier CO isotopologues is a small effect compared to shielding by \coc{12}.

\begin{table*}
\caption{Two-dimensional shielding functions $\Theta[N(\cocm{12}),\cdmh]$ for $\bco=0.3$ km \ps{}, $\texco=5$ K and $N(\cocm{12})/N(\cocm{13})=69$.$^{\mathrm{a}}$}
\label{tb:shield}
\centering
\begin{tabular}{ccccccccc}
\hline\hline
\rule{0pt}{1em}$\log \cdmh$ & \multicolumn{8}{c}{$\log{} \cocm{12}$ (\pcs)} \\
\cline{2-9}
\rule{0pt}{1em}(\pcs) & 0 & 13 & 14 & 15 & 16 & 17 & 18 & 19 \\
\hline
\multicolumn{9}{c}{\rule{0pt}{1em}\coc{12}: unattenuated rate $k_{0,i}=\scim{2.592}{-10}$ \ps} \\
\hline
\phantom{0}0 & 1.000\phantom{(-0)} & 8.080(-1) & 5.250(-1) & 2.434(-1) & 5.467(-2) & 1.362(-2) & 3.378(-3) & 5.240(-4) \\
19 & 8.176(-1) & 6.347(-1) & 3.891(-1) & 1.787(-1) & 4.297(-2) & 1.152(-2) & 2.922(-3) & 4.662(-4) \\
20 & 7.223(-1) & 5.624(-1) & 3.434(-1) & 1.540(-1) & 3.515(-2) & 9.231(-3) & 2.388(-3) & 3.899(-4) \\
21 & 3.260(-1) & 2.810(-1) & 1.953(-1) & 8.726(-2) & 1.907(-2) & 4.768(-3) & 1.150(-3) & 1.941(-4) \\
22 & 1.108(-2) & 1.081(-2) & 9.033(-3) & 4.441(-3) & 1.102(-3) & 2.644(-4) & 7.329(-5) & 1.437(-5) \\
23 & 3.938(-7) & 3.938(-7) & 3.936(-7) & 3.923(-7) & 3.901(-7) & 3.893(-7) & 3.890(-7) & 3.875(-7) \\
\hline
\multicolumn{9}{c}{\rule{0pt}{1em}\coo{17}: unattenuated rate $k_{0,i}=\scim{2.607}{-10}$ \ps} \\
\hline
\phantom{0}0 & 1.000\phantom{(-0)} & 9.823(-1) & 8.911(-1) & 6.149(-1) & 3.924(-1) & 2.169(-1) & 4.167(-2) & 2.150(-3) \\
19 & 8.459(-1) & 8.298(-1) & 7.490(-1) & 5.009(-1) & 3.196(-1) & 1.850(-1) & 3.509(-2) & 1.984(-3) \\
20 & 7.337(-1) & 7.195(-1) & 6.481(-1) & 4.306(-1) & 2.741(-1) & 1.556(-1) & 2.645(-2) & 1.411(-3) \\
21 & 3.335(-1) & 3.290(-1) & 3.039(-1) & 2.293(-1) & 1.685(-1) & 9.464(-2) & 1.460(-2) & 6.823(-4) \\
22 & 1.193(-2) & 1.191(-2) & 1.172(-2) & 1.095(-2) & 9.395(-3) & 5.644(-3) & 1.183(-3) & 2.835(-5) \\
23 & 3.959(-7) & 3.959(-7) & 3.959(-7) & 3.959(-7) & 3.958(-7) & 3.954(-7) & 3.924(-7) & 3.873(-7) \\
\hline
\multicolumn{9}{c}{\rule{0pt}{1em}\coo{18}: unattenuated rate $k_{0,i}=\scim{2.392}{-10}$ \ps} \\
\hline
\phantom{0}0 & 1.000\phantom{(-0)} & 9.974(-1) & 9.777(-1) & 8.519(-1) & 5.060(-1) & 1.959(-1) & 2.764(-2) & 1.742(-3) \\
19 & 8.571(-1) & 8.547(-1) & 8.368(-1) & 7.219(-1) & 4.095(-1) & 1.581(-1) & 2.224(-2) & 1.618(-3) \\
20 & 7.554(-1) & 7.532(-1) & 7.371(-1) & 6.336(-1) & 3.572(-1) & 1.372(-1) & 1.889(-2) & 1.383(-3) \\
21 & 3.559(-1) & 3.549(-1) & 3.477(-1) & 3.035(-1) & 1.948(-1) & 7.701(-2) & 1.071(-2) & 6.863(-4) \\
22 & 1.214(-2) & 1.212(-2) & 1.199(-2) & 1.105(-2) & 8.233(-3) & 3.324(-3) & 6.148(-4) & 3.225(-5) \\
23 & 4.251(-7) & 4.251(-7) & 4.251(-7) & 4.251(-7) & 4.249(-7) & 4.233(-7) & 4.180(-7) & 4.142(-7) \\
\hline
\multicolumn{9}{c}{\rule{0pt}{1em}\coc{13}: unattenuated rate $k_{0,i}=\scim{2.595}{-10}$ \ps} \\
\hline
\phantom{0}0 & 1.000\phantom{(-0)} & 9.824(-1) & 9.019(-1) & 6.462(-1) & 3.547(-1) & 9.907(-2) & 1.131(-2) & 7.591(-4) \\
19 & 8.447(-1) & 8.276(-1) & 7.502(-1) & 5.113(-1) & 2.745(-1) & 7.652(-2) & 8.635(-3) & 6.747(-4) \\
20 & 7.415(-1) & 7.266(-1) & 6.581(-1) & 4.451(-1) & 2.360(-1) & 6.574(-2) & 7.187(-3) & 5.429(-4) \\
21 & 3.546(-1) & 3.502(-1) & 3.270(-1) & 2.452(-1) & 1.398(-1) & 3.750(-2) & 3.973(-3) & 2.703(-4) \\
22 & 1.180(-2) & 1.177(-2) & 1.153(-2) & 1.023(-2) & 6.728(-3) & 1.955(-3) & 2.665(-4) & 1.471(-5) \\
23 & 2.385(-7) & 2.385(-7) & 2.385(-7) & 2.384(-7) & 2.379(-7) & 2.348(-7) & 2.310(-7) & 2.292(-7) \\
\hline
\multicolumn{9}{c}{\rule{0pt}{1em}\co{13}{17}: unattenuated rate $k_{0,i}=\scim{2.598}{-10}$ \ps} \\
\hline
\phantom{0}0 & 1.000\phantom{(-0)} & 9.979(-1) & 9.820(-1) & 8.832(-1) & 5.942(-1) & 3.177(-1) & 1.523(-1) & 3.885(-2) \\
19 & 8.540(-1) & 8.520(-1) & 8.374(-1) & 7.469(-1) & 4.901(-1) & 2.677(-1) & 1.302(-1) & 3.135(-2) \\
20 & 7.405(-1) & 7.387(-1) & 7.254(-1) & 6.439(-1) & 4.198(-1) & 2.333(-1) & 1.142(-1) & 2.607(-2) \\
21 & 3.502(-1) & 3.494(-1) & 3.434(-1) & 3.076(-1) & 2.214(-1) & 1.386(-1) & 6.941(-2) & 1.195(-2) \\
22 & 1.279(-2) & 1.278(-2) & 1.267(-2) & 1.198(-2) & 1.045(-2) & 7.743(-3) & 4.088(-3) & 4.581(-4) \\
23 & 2.370(-7) & 2.370(-7) & 2.370(-7) & 2.370(-7) & 2.369(-7) & 2.368(-7) & 2.359(-7) & 2.312(-7) \\
\hline
\multicolumn{9}{c}{\rule{0pt}{1em}\co{13}{18}: unattenuated rate $k_{0,i}=\scim{2.503}{-10}$ \ps} \\
\hline
\phantom{0}0 & 1.000\phantom{(-0)} & 9.988(-1) & 9.900(-1) & 9.329(-1) & 7.253(-1) & 3.856(-1) & 1.524(-1) & 2.664(-2) \\
19 & 8.744(-1) & 8.734(-1) & 8.656(-1) & 8.164(-1) & 6.403(-1) & 3.441(-1) & 1.347(-1) & 2.491(-2) \\
20 & 7.572(-1) & 7.562(-1) & 7.492(-1) & 7.047(-1) & 5.518(-1) & 3.006(-1) & 1.185(-1) & 2.224(-2) \\
21 & 3.546(-1) & 3.542(-1) & 3.506(-1) & 3.283(-1) & 2.638(-1) & 1.666(-1) & 6.887(-2) & 1.149(-2) \\
22 & 1.561(-2) & 1.560(-2) & 1.550(-2) & 1.475(-2) & 1.235(-2) & 7.850(-3) & 3.416(-3) & 5.290(-4) \\
23 & 2.490(-7) & 2.490(-7) & 2.490(-7) & 2.489(-7) & 2.487(-7) & 2.482(-7) & 2.471(-7) & 2.421(-7) \\
\hline
\end{tabular}
\begin{list}{}{}
\item[$^{\mathrm{a}}$] These shielding functions were computed for the \citet{draine78a} radiation field ($\chi=1$) and the following set of parameters: $\bco=0.3$ km \ps, $\bmh=3.0$ km \ps{} and $b(\el{H})=5.0$ km \ps; $\texco=5$ K and $\texmh=5^{1.5}$ K; $\cdh=\scim{5}{20}$ \pcs; $N(\cocm{12})/N(\cocm{13})=69$, $N(\cocm{12})/N(\coom{18})=N(\cocm{13})/N(\com{13}{18})=557$ and $N(\coom{18})/N(\coom{17})=N(\com{13}{18})/N(\com{13}{17})=3.6$. Self-shielding is mostly negligible for the heavier isotopologues, so all shielding functions are expressed as a function of the \coc{12} column density. Continuum attenuation by dust is not included in this table (see \eq{pdrate}).
\end{list}
\end{table*}

\onltab{6}{
\begin{table*}
\caption{As \tb{shield}, for $\bco=0.3$ km \ps{}, $\texco=50$ K and $N(\cocm{12})/N(\cocm{13})=69$.$^{\mathrm{a}}$}
\label{tb:shield2}
\centering
\begin{tabular}{ccccccccc}
\hline\hline
\rule{0pt}{1em}$\log \cdmh$ & \multicolumn{8}{c}{$\log{} \cocm{12}$ (\pcs)} \\
\cline{2-9}
\rule{0pt}{1em}(\pcs) & 0 & 13 & 14 & 15 & 16 & 17 & 18 & 19 \\
\hline
\multicolumn{9}{c}{\rule{0pt}{1em}\coc{12}: unattenuated rate $k_{0,i}=\scim{2.590}{-10}$ \ps} \\
\hline
\phantom{0}0 & 1.000\phantom{(-0)} & 9.405(-1) & 7.046(-1) & 4.015(-1) & 9.964(-2) & 1.567(-2) & 3.162(-3) & 4.839(-4) \\
19 & 7.546(-1) & 6.979(-1) & 4.817(-1) & 2.577(-1) & 6.505(-2) & 1.135(-2) & 2.369(-3) & 3.924(-4) \\
20 & 5.752(-1) & 5.228(-1) & 3.279(-1) & 1.559(-1) & 3.559(-2) & 6.443(-3) & 1.526(-3) & 2.751(-4) \\
21 & 2.493(-1) & 2.196(-1) & 1.135(-1) & 4.062(-2) & 7.864(-3) & 1.516(-3) & 4.448(-4) & 9.367(-5) \\
22 & 1.550(-3) & 1.370(-3) & 6.801(-4) & 2.127(-4) & 5.051(-5) & 1.198(-5) & 6.553(-6) & 3.937(-6) \\
23 & 8.492(-8) & 8.492(-8) & 8.492(-8) & 8.492(-8) & 8.492(-8) & 8.492(-8) & 8.488(-8) & 8.453(-8) \\
\hline
\multicolumn{9}{c}{\rule{0pt}{1em}\coo{17}: unattenuated rate $k_{0,i}=\scim{2.607}{-10}$ \ps} \\
\hline
\phantom{0}0 & 1.000\phantom{(-0)} & 9.756(-1) & 8.826(-1) & 7.413(-1) & 4.507(-1) & 1.508(-1) & 2.533(-2) & 1.684(-3) \\
19 & 7.266(-1) & 7.036(-1) & 6.209(-1) & 5.209(-1) & 3.212(-1) & 1.074(-1) & 1.832(-2) & 1.436(-3) \\
20 & 5.418(-1) & 5.206(-1) & 4.469(-1) & 3.695(-1) & 2.222(-1) & 7.000(-2) & 1.197(-2) & 9.425(-4) \\
21 & 2.208(-1) & 2.089(-1) & 1.687(-1) & 1.352(-1) & 8.073(-2) & 2.082(-2) & 3.360(-3) & 2.492(-4) \\
22 & 1.401(-3) & 1.352(-3) & 1.190(-3) & 1.043(-3) & 7.022(-4) & 1.687(-4) & 2.525(-5) & 4.211(-6) \\
23 & 8.509(-8) & 8.509(-8) & 8.509(-8) & 8.509(-8) & 8.509(-8) & 8.509(-8) & 8.505(-8) & 8.470(-8) \\
\hline
\multicolumn{9}{c}{\rule{0pt}{1em}\coo{18}: unattenuated rate $k_{0,i}=\scim{2.396}{-10}$ \ps} \\
\hline
\phantom{0}0 & 1.000\phantom{(-0)} & 9.822(-1) & 9.163(-1) & 8.067(-1) & 5.498(-1) & 2.188(-1) & 3.412(-2) & 1.992(-3) \\
19 & 8.007(-1) & 7.833(-1) & 7.206(-1) & 6.272(-1) & 4.226(-1) & 1.597(-1) & 2.469(-2) & 1.630(-3) \\
20 & 5.848(-1) & 5.688(-1) & 5.123(-1) & 4.363(-1) & 2.816(-1) & 9.826(-2) & 1.653(-2) & 1.158(-3) \\
21 & 2.277(-1) & 2.187(-1) & 1.881(-1) & 1.546(-1) & 9.811(-2) & 3.114(-2) & 5.506(-3) & 3.655(-4) \\
22 & 1.411(-3) & 1.375(-3) & 1.256(-3) & 1.126(-3) & 8.200(-4) & 2.912(-4) & 5.310(-5) & 5.241(-6) \\
23 & 8.937(-8) & 8.937(-8) & 8.937(-8) & 8.937(-8) & 8.937(-8) & 8.936(-8) & 8.933(-8) & 8.896(-8) \\
\hline
\multicolumn{9}{c}{\rule{0pt}{1em}\coc{13}: unattenuated rate $k_{0,i}=\scim{2.597}{-10}$ \ps} \\
\hline
\phantom{0}0 & 1.000\phantom{(-0)} & 9.765(-1) & 8.965(-1) & 7.701(-1) & 4.459(-1) & 1.415(-1) & 1.748(-2) & 8.346(-4) \\
19 & 7.780(-1) & 7.550(-1) & 6.789(-1) & 5.753(-1) & 3.262(-1) & 1.032(-1) & 1.183(-2) & 6.569(-4) \\
20 & 5.523(-1) & 5.309(-1) & 4.617(-1) & 3.793(-1) & 1.958(-1) & 5.888(-2) & 7.515(-3) & 4.382(-4) \\
21 & 2.100(-1) & 1.979(-1) & 1.601(-1) & 1.244(-1) & 5.373(-2) & 1.573(-2) & 2.535(-3) & 1.361(-4) \\
22 & 1.318(-3) & 1.268(-3) & 1.107(-3) & 9.017(-4) & 3.658(-4) & 1.114(-4) & 2.414(-5) & 2.608(-6) \\
23 & 4.511(-8) & 4.511(-8) & 4.511(-8) & 4.511(-8) & 4.511(-8) & 4.511(-8) & 4.509(-8) & 4.490(-8) \\
\hline
\multicolumn{9}{c}{\rule{0pt}{1em}\co{13}{17}: unattenuated rate $k_{0,i}=\scim{2.605}{-10}$ \ps} \\
\hline
\phantom{0}0 & 1.000\phantom{(-0)} & 9.853(-1) & 9.344(-1) & 8.453(-1) & 5.978(-1) & 3.075(-1) & 1.180(-1) & 3.205(-2) \\
19 & 8.097(-1) & 7.954(-1) & 7.474(-1) & 6.755(-1) & 4.947(-1) & 2.711(-1) & 1.065(-1) & 2.898(-2) \\
20 & 5.925(-1) & 5.792(-1) & 5.363(-1) & 4.815(-1) & 3.576(-1) & 2.062(-1) & 8.289(-2) & 1.967(-2) \\
21 & 2.389(-1) & 2.315(-1) & 2.083(-1) & 1.848(-1) & 1.410(-1) & 8.627(-2) & 3.185(-2) & 3.489(-3) \\
22 & 1.937(-3) & 1.907(-3) & 1.812(-3) & 1.711(-3) & 1.474(-3) & 1.042(-3) & 3.690(-4) & 1.704(-5) \\
23 & 4.523(-8) & 4.523(-8) & 4.523(-8) & 4.523(-8) & 4.523(-8) & 4.523(-8) & 4.521(-8) & 4.502(-8) \\
\hline
\multicolumn{9}{c}{\rule{0pt}{1em}\co{13}{18}: unattenuated rate $k_{0,i}=\scim{2.510}{-10}$ \ps} \\
\hline
\phantom{0}0 & 1.000\phantom{(-0)} & 9.873(-1) & 9.424(-1) & 8.744(-1) & 6.740(-1) & 3.885(-1) & 1.268(-1) & 2.980(-2) \\
19 & 7.975(-1) & 7.852(-1) & 7.424(-1) & 6.894(-1) & 5.447(-1) & 3.236(-1) & 1.036(-1) & 2.722(-2) \\
20 & 5.869(-1) & 5.754(-1) & 5.364(-1) & 4.953(-1) & 3.932(-1) & 2.445(-1) & 8.070(-2) & 1.894(-2) \\
21 & 2.465(-1) & 2.401(-1) & 2.187(-1) & 2.011(-1) & 1.628(-1) & 1.079(-1) & 3.267(-2) & 4.406(-3) \\
22 & 2.164(-3) & 2.138(-3) & 2.050(-3) & 1.971(-3) & 1.772(-3) & 1.421(-3) & 5.147(-4) & 4.066(-5) \\
23 & 4.454(-8) & 4.454(-8) & 4.454(-8) & 4.454(-8) & 4.454(-8) & 4.454(-8) & 4.452(-8) & 4.434(-8) \\
\hline
\end{tabular}
\begin{list}{}{}
\item[$^{\mathrm{a}}$] These shielding functions were computed for the same parameters as in \tb{shield}, except $\texco=50$ K and $\texmh=50^{1.5}$ K. Additional rotational lines of CO and \mh{} were included as described in Sect.\ \ref{subsec:btexgrid}.
\end{list}
\end{table*}
}

\onltab{7}{
\begin{table*}
\caption{As \tb{shield}, for $\bco=3.0$ km \ps{}, $\texco=5$ K and $N(\cocm{12})/N(\cocm{13})=69$.$^{\mathrm{a}}$}
\label{tb:shield3}
\centering
\begin{tabular}{ccccccccc}
\hline\hline
\rule{0pt}{1em}$\log \cdmh$ & \multicolumn{8}{c}{$\log{} \cocm{12}$ (\pcs)} \\
\cline{2-9}
\rule{0pt}{1em}(\pcs) & 0 & 13 & 14 & 15 & 16 & 17 & 18 & 19 \\
\hline
\multicolumn{9}{c}{\rule{0pt}{1em}\coc{12}: unattenuated rate $k_{0,i}=\scim{2.588}{-10}$ \ps} \\
\hline
\phantom{0}0 & 1.000\phantom{(-0)} & 9.639(-1) & 7.443(-1) & 3.079(-1) & 5.733(-2) & 1.321(-2) & 3.322(-3) & 5.193(-4) \\
19 & 7.928(-1) & 7.594(-1) & 5.606(-1) & 2.145(-1) & 4.172(-2) & 1.039(-2) & 2.696(-3) & 4.426(-4) \\
20 & 7.037(-1) & 6.743(-1) & 4.984(-1) & 1.861(-1) & 3.410(-2) & 8.272(-3) & 2.212(-3) & 3.716(-4) \\
21 & 3.176(-1) & 3.082(-1) & 2.477(-1) & 1.007(-1) & 1.833(-2) & 4.252(-3) & 1.066(-3) & 1.850(-4) \\
22 & 1.083(-2) & 1.067(-2) & 9.413(-3) & 4.528(-3) & 1.054(-3) & 2.353(-4) & 6.789(-5) & 1.378(-5) \\
23 & 3.931(-7) & 3.931(-7) & 3.930(-7) & 3.922(-7) & 3.896(-7) & 3.886(-7) & 3.883(-7) & 3.869(-7) \\
\hline
\multicolumn{9}{c}{\rule{0pt}{1em}\coo{17}: unattenuated rate $k_{0,i}=\scim{2.604}{-10}$ \ps} \\
\hline
\phantom{0}0 & 1.000\phantom{(-0)} & 9.703(-1) & 7.959(-1) & 5.192(-1) & 3.667(-1) & 2.247(-1) & 6.359(-2) & 6.394(-3) \\
19 & 8.049(-1) & 7.764(-1) & 6.108(-1) & 3.734(-1) & 2.733(-1) & 1.692(-1) & 4.882(-2) & 5.903(-3) \\
20 & 7.008(-1) & 6.759(-1) & 5.311(-1) & 3.212(-1) & 2.348(-1) & 1.424(-1) & 3.533(-2) & 3.847(-3) \\
21 & 3.125(-1) & 3.056(-1) & 2.638(-1) & 1.885(-1) & 1.474(-1) & 8.575(-2) & 1.807(-2) & 1.839(-3) \\
22 & 1.077(-2) & 1.076(-2) & 1.062(-2) & 9.872(-3) & 8.416(-3) & 5.126(-3) & 1.119(-3) & 4.151(-5) \\
23 & 3.950(-7) & 3.950(-7) & 3.950(-7) & 3.950(-7) & 3.949(-7) & 3.945(-7) & 3.919(-7) & 3.864(-7) \\
\hline
\multicolumn{9}{c}{\rule{0pt}{1em}\coo{18}: unattenuated rate $k_{0,i}=\scim{2.394}{-10}$ \ps} \\
\hline
\phantom{0}0 & 1.000\phantom{(-0)} & 9.767(-1) & 8.397(-1) & 6.181(-1) & 4.521(-1) & 1.999(-1) & 3.383(-2) & 1.883(-3) \\
19 & 8.114(-1) & 7.886(-1) & 6.554(-1) & 4.491(-1) & 3.245(-1) & 1.475(-1) & 2.665(-2) & 1.723(-3) \\
20 & 7.177(-1) & 6.979(-1) & 5.816(-1) & 3.981(-1) & 2.851(-1) & 1.284(-1) & 2.244(-2) & 1.476(-3) \\
21 & 3.313(-1) & 3.260(-1) & 2.936(-1) & 2.274(-1) & 1.646(-1) & 7.081(-2) & 1.188(-2) & 6.966(-4) \\
22 & 1.130(-2) & 1.129(-2) & 1.113(-2) & 1.016(-2) & 7.643(-3) & 3.131(-3) & 6.377(-4) & 3.195(-5) \\
23 & 4.233(-7) & 4.233(-7) & 4.233(-7) & 4.233(-7) & 4.231(-7) & 4.217(-7) & 4.167(-7) & 4.126(-7) \\
\hline
\multicolumn{9}{c}{\rule{0pt}{1em}\coc{13}: unattenuated rate $k_{0,i}=\scim{2.595}{-10}$ \ps} \\
\hline
\phantom{0}0 & 1.000\phantom{(-0)} & 9.750(-1) & 8.292(-1) & 6.083(-1) & 4.222(-1) & 1.249(-1) & 1.250(-2) & 6.838(-4) \\
19 & 7.899(-1) & 7.656(-1) & 6.246(-1) & 4.244(-1) & 2.974(-1) & 9.238(-2) & 9.445(-3) & 5.851(-4) \\
20 & 6.933(-1) & 6.720(-1) & 5.484(-1) & 3.694(-1) & 2.561(-1) & 7.843(-2) & 7.799(-3) & 4.678(-4) \\
21 & 3.278(-1) & 3.219(-1) & 2.858(-1) & 2.162(-1) & 1.496(-1) & 4.297(-2) & 4.117(-3) & 2.360(-4) \\
22 & 1.108(-2) & 1.106(-2) & 1.089(-2) & 9.813(-3) & 6.942(-3) & 2.057(-3) & 2.618(-4) & 1.306(-5) \\
23 & 2.378(-7) & 2.378(-7) & 2.378(-7) & 2.378(-7) & 2.375(-7) & 2.354(-7) & 2.304(-7) & 2.285(-7) \\
\hline
\multicolumn{9}{c}{\rule{0pt}{1em}\co{13}{17}: unattenuated rate $k_{0,i}=\scim{2.604}{-10}$ \ps} \\
\hline
\phantom{0}0 & 1.000\phantom{(-0)} & 9.800(-1) & 8.605(-1) & 6.553(-1) & 5.067(-1) & 2.970(-1) & 1.491(-1) & 4.514(-2) \\
19 & 8.134(-1) & 7.939(-1) & 6.781(-1) & 4.891(-1) & 3.826(-1) & 2.401(-1) & 1.228(-1) & 3.233(-2) \\
20 & 7.049(-1) & 6.879(-1) & 5.867(-1) & 4.192(-1) & 3.270(-1) & 2.096(-1) & 1.079(-1) & 2.669(-2) \\
21 & 3.302(-1) & 3.256(-1) & 2.970(-1) & 2.382(-1) & 1.931(-1) & 1.292(-1) & 6.607(-2) & 1.136(-2) \\
22 & 1.174(-2) & 1.173(-2) & 1.161(-2) & 1.093(-2) & 9.517(-3) & 7.132(-3) & 3.652(-3) & 3.436(-4) \\
23 & 2.312(-7) & 2.312(-7) & 2.312(-7) & 2.312(-7) & 2.311(-7) & 2.311(-7) & 2.308(-7) & 2.290(-7) \\
\hline
\multicolumn{9}{c}{\rule{0pt}{1em}\co{13}{18}: unattenuated rate $k_{0,i}=\scim{2.509}{-10}$ \ps} \\
\hline
\phantom{0}0 & 1.000\phantom{(-0)} & 9.826(-1) & 8.767(-1) & 6.711(-1) & 5.145(-1) & 3.329(-1) & 1.592(-1) & 4.547(-2) \\
19 & 8.157(-1) & 7.988(-1) & 6.964(-1) & 5.074(-1) & 3.964(-1) & 2.777(-1) & 1.330(-1) & 4.212(-2) \\
20 & 7.096(-1) & 6.950(-1) & 6.057(-1) & 4.394(-1) & 3.440(-1) & 2.477(-1) & 1.197(-1) & 3.796(-2) \\
21 & 3.192(-1) & 3.153(-1) & 2.911(-1) & 2.365(-1) & 1.943(-1) & 1.424(-1) & 6.786(-2) & 1.888(-2) \\
22 & 1.077(-2) & 1.076(-2) & 1.065(-2) & 9.980(-3) & 8.532(-3) & 6.289(-3) & 2.998(-3) & 6.921(-4) \\
23 & 2.410(-7) & 2.410(-7) & 2.410(-7) & 2.410(-7) & 2.410(-7) & 2.409(-7) & 2.407(-7) & 2.395(-7) \\
\hline
\end{tabular}
\begin{list}{}{}
\item[$^{\mathrm{a}}$] These shielding functions were computed for the same parameters as in \tb{shield}, except $\bco=3.0$ km \ps, $\bmh=11.2$ km \ps{} and $b(\el{H})=15.9$ km \ps.
\end{list}
\end{table*}
}

\onltab{8}{
\begin{table*}
\caption{As \tb{shield}, for $\bco=0.3$ km \ps{}, $\texco=5$ K and $N(\cocm{12})/N(\cocm{13})=35$.$^{\mathrm{a}}$}
\label{tb:shield4}
\centering
\begin{tabular}{ccccccccc}
\hline\hline
\rule{0pt}{1em}$\log \cdmh$ & \multicolumn{8}{c}{$\log{} \cocm{12}$ (\pcs)} \\
\cline{2-9}
\rule{0pt}{1em}(\pcs) & 0 & 13 & 14 & 15 & 16 & 17 & 18 & 19 \\
\hline
\multicolumn{9}{c}{\rule{0pt}{1em}\coc{12}: unattenuated rate $k_{0,i}=\scim{2.592}{-10}$ \ps} \\
\hline
\phantom{0}0 & 1.000\phantom{(-0)} & 8.078(-1) & 5.245(-1) & 2.427(-1) & 5.428(-2) & 1.342(-2) & 3.321(-3) & 5.157(-4) \\
19 & 8.176(-1) & 6.345(-1) & 3.886(-1) & 1.781(-1) & 4.271(-2) & 1.136(-2) & 2.875(-3) & 4.591(-4) \\
20 & 7.223(-1) & 5.623(-1) & 3.430(-1) & 1.534(-1) & 3.493(-2) & 9.095(-3) & 2.347(-3) & 3.842(-4) \\
21 & 3.260(-1) & 2.810(-1) & 1.951(-1) & 8.696(-2) & 1.894(-2) & 4.691(-3) & 1.127(-3) & 1.913(-4) \\
22 & 1.108(-2) & 1.081(-2) & 9.030(-3) & 4.434(-3) & 1.096(-3) & 2.606(-4) & 7.208(-5) & 1.420(-5) \\
23 & 3.938(-7) & 3.938(-7) & 3.936(-7) & 3.923(-7) & 3.901(-7) & 3.893(-7) & 3.890(-7) & 3.875(-7) \\
\hline
\multicolumn{9}{c}{\rule{0pt}{1em}\coo{17}: unattenuated rate $k_{0,i}=\scim{2.607}{-10}$ \ps} \\
\hline
\phantom{0}0 & 1.000\phantom{(-0)} & 9.791(-1) & 8.675(-1) & 5.829(-1) & 3.848(-1) & 2.122(-1) & 4.089(-2) & 2.102(-3) \\
19 & 8.459(-1) & 8.266(-1) & 7.258(-1) & 4.699(-1) & 3.139(-1) & 1.810(-1) & 3.440(-2) & 1.941(-3) \\
20 & 7.337(-1) & 7.168(-1) & 6.280(-1) & 4.036(-1) & 2.691(-1) & 1.518(-1) & 2.583(-2) & 1.378(-3) \\
21 & 3.335(-1) & 3.283(-1) & 2.988(-1) & 2.218(-1) & 1.655(-1) & 9.193(-2) & 1.417(-2) & 6.647(-4) \\
22 & 1.193(-2) & 1.191(-2) & 1.171(-2) & 1.091(-2) & 9.219(-3) & 5.456(-3) & 1.149(-3) & 2.666(-5) \\
23 & 3.959(-7) & 3.959(-7) & 3.959(-7) & 3.959(-7) & 3.958(-7) & 3.954(-7) & 3.923(-7) & 3.873(-7) \\
\hline
\multicolumn{9}{c}{\rule{0pt}{1em}\coo{18}: unattenuated rate $k_{0,i}=\scim{2.392}{-10}$ \ps} \\
\hline
\phantom{0}0 & 1.000\phantom{(-0)} & 9.971(-1) & 9.757(-1) & 8.410(-1) & 4.805(-1) & 1.677(-1) & 2.310(-2) & 1.529(-3) \\
19 & 8.571(-1) & 8.544(-1) & 8.349(-1) & 7.120(-1) & 3.881(-1) & 1.349(-1) & 1.861(-2) & 1.421(-3) \\
20 & 7.554(-1) & 7.531(-1) & 7.353(-1) & 6.247(-1) & 3.379(-1) & 1.166(-1) & 1.562(-2) & 1.212(-3) \\
21 & 3.559(-1) & 3.549(-1) & 3.470(-1) & 2.995(-1) & 1.827(-1) & 6.493(-2) & 8.710(-3) & 5.900(-4) \\
22 & 1.214(-2) & 1.212(-2) & 1.197(-2) & 1.091(-2) & 7.560(-3) & 2.781(-3) & 4.965(-4) & 2.789(-5) \\
23 & 4.251(-7) & 4.251(-7) & 4.251(-7) & 4.251(-7) & 4.248(-7) & 4.228(-7) & 4.175(-7) & 4.141(-7) \\
\hline
\multicolumn{9}{c}{\rule{0pt}{1em}\coc{13}: unattenuated rate $k_{0,i}=\scim{2.595}{-10}$ \ps} \\
\hline
\phantom{0}0 & 1.000\phantom{(-0)} & 9.788(-1) & 8.746(-1) & 5.882(-1) & 2.889(-1) & 6.582(-2) & 7.841(-3) & 6.070(-4) \\
19 & 8.447(-1) & 8.241(-1) & 7.241(-1) & 4.601(-1) & 2.237(-1) & 5.052(-2) & 6.312(-3) & 5.444(-4) \\
20 & 7.415(-1) & 7.235(-1) & 6.353(-1) & 4.000(-1) & 1.921(-1) & 4.324(-2) & 5.236(-3) & 4.396(-4) \\
21 & 3.546(-1) & 3.494(-1) & 3.205(-1) & 2.277(-1) & 1.138(-1) & 2.454(-2) & 2.807(-3) & 2.164(-4) \\
22 & 1.180(-2) & 1.176(-2) & 1.148(-2) & 9.874(-3) & 5.564(-3) & 1.339(-3) & 1.818(-4) & 1.270(-5) \\
23 & 2.385(-7) & 2.385(-7) & 2.384(-7) & 2.383(-7) & 2.374(-7) & 2.335(-7) & 2.306(-7) & 2.292(-7) \\
\hline
\multicolumn{9}{c}{\rule{0pt}{1em}\co{13}{17}: unattenuated rate $k_{0,i}=\scim{2.598}{-10}$ \ps} \\
\hline
\phantom{0}0 & 1.000\phantom{(-0)} & 9.977(-1) & 9.803(-1) & 8.723(-1) & 5.661(-1) & 2.965(-1) & 1.436(-1) & 3.191(-2) \\
19 & 8.540(-1) & 8.519(-1) & 8.359(-1) & 7.377(-1) & 4.699(-1) & 2.518(-1) & 1.236(-1) & 2.638(-2) \\
20 & 7.405(-1) & 7.385(-1) & 7.241(-1) & 6.360(-1) & 4.035(-1) & 2.204(-1) & 1.084(-1) & 2.199(-2) \\
21 & 3.502(-1) & 3.494(-1) & 3.428(-1) & 3.038(-1) & 2.130(-1) & 1.330(-1) & 6.616(-2) & 1.008(-2) \\
22 & 1.279(-2) & 1.278(-2) & 1.267(-2) & 1.193(-2) & 1.021(-2) & 7.576(-3) & 3.904(-3) & 3.924(-4) \\
23 & 2.370(-7) & 2.370(-7) & 2.370(-7) & 2.370(-7) & 2.369(-7) & 2.368(-7) & 2.358(-7) & 2.311(-7) \\
\hline
\multicolumn{9}{c}{\rule{0pt}{1em}\co{13}{18}: unattenuated rate $k_{0,i}=\scim{2.503}{-10}$ \ps} \\
\hline
\phantom{0}0 & 1.000\phantom{(-0)} & 9.988(-1) & 9.893(-1) & 9.277(-1) & 7.041(-1) & 3.637(-1) & 1.353(-1) & 1.838(-2) \\
19 & 8.744(-1) & 8.734(-1) & 8.652(-1) & 8.132(-1) & 6.262(-1) & 3.264(-1) & 1.205(-1) & 1.727(-2) \\
20 & 7.572(-1) & 7.562(-1) & 7.488(-1) & 7.022(-1) & 5.411(-1) & 2.868(-1) & 1.064(-1) & 1.546(-2) \\
21 & 3.546(-1) & 3.541(-1) & 3.504(-1) & 3.272(-1) & 2.592(-1) & 1.608(-1) & 6.195(-2) & 7.985(-3) \\
22 & 1.561(-2) & 1.560(-2) & 1.550(-2) & 1.471(-2) & 1.214(-2) & 7.623(-3) & 3.112(-3) & 3.736(-4) \\
23 & 2.490(-7) & 2.490(-7) & 2.490(-7) & 2.489(-7) & 2.487(-7) & 2.482(-7) & 2.470(-7) & 2.419(-7) \\
\hline
\end{tabular}
\begin{list}{}{}
\item[$^{\mathrm{a}}$] These shielding functions were computed for the same parameters as in \tb{shield}, except $N(\cocm{12})/N(\cocm{13})=35$.
\end{list}
\end{table*}
}

\tb{shield} presents the shielding functions in the same manner as \vdb{} did, but for somewhat different model parameters: $\bco=0.3$ instead of 1.0 km \ps, $\texco=5$ instead of 10 K, and $\texmh=5^{1.5}$ instead of $10^{1.5}$ K. The present parameters correspond more closely to what is observed in diffuse and translucent clouds. The column density ratios for the six isotopologues are kept constant at the elemental isotope ratios from \citet{wilson99a}: $[\element[][12]{C}]/[\element[][13]{C}]=69$, $[\element[][16]{O}]/[\element[][18]{O}]=557$ and $[\element[][18]{O}]/[\element[][17]{O}]=3.6$. A small column of \scit{5.2}{15} \pcs{} of \mh{} at $J''$=4--7 is included throughout (except at $\log \cdmh=0$) to account for UV pumping (Sect.\ \ref{subsec:default}).\footnote{Although not mentioned explicitly by \vdb, their tabulated shielding functions also include this extra column of $J''$=4--7 \mh.} Shielding functions for larger values of $b$ and $T_\el{ex}$ and for other isotope ratios are given in Tables \ref{tb:shield2}--\ref{tb:shield4} in the online appendix to this paper. For ease of use, we have also set up a webpage\footnote{http://www.strw.leidenuniv.nl/$\sim$ewine/photo} where the shielding functions can be downloaded in plain text format. This webpage offers shielding functions for a wider variety of parameters than is possible to include in this paper. In addition, it uses a grid of $N(\cocm{12}$) and $\cdmh$ values that is five times finer than the grid in Tables \ref{tb:shield}--\ref{tb:shield4}, allowing for more accurate interpolation.

\begin{figure}
\resizebox{\hsize}{!}{\includegraphics{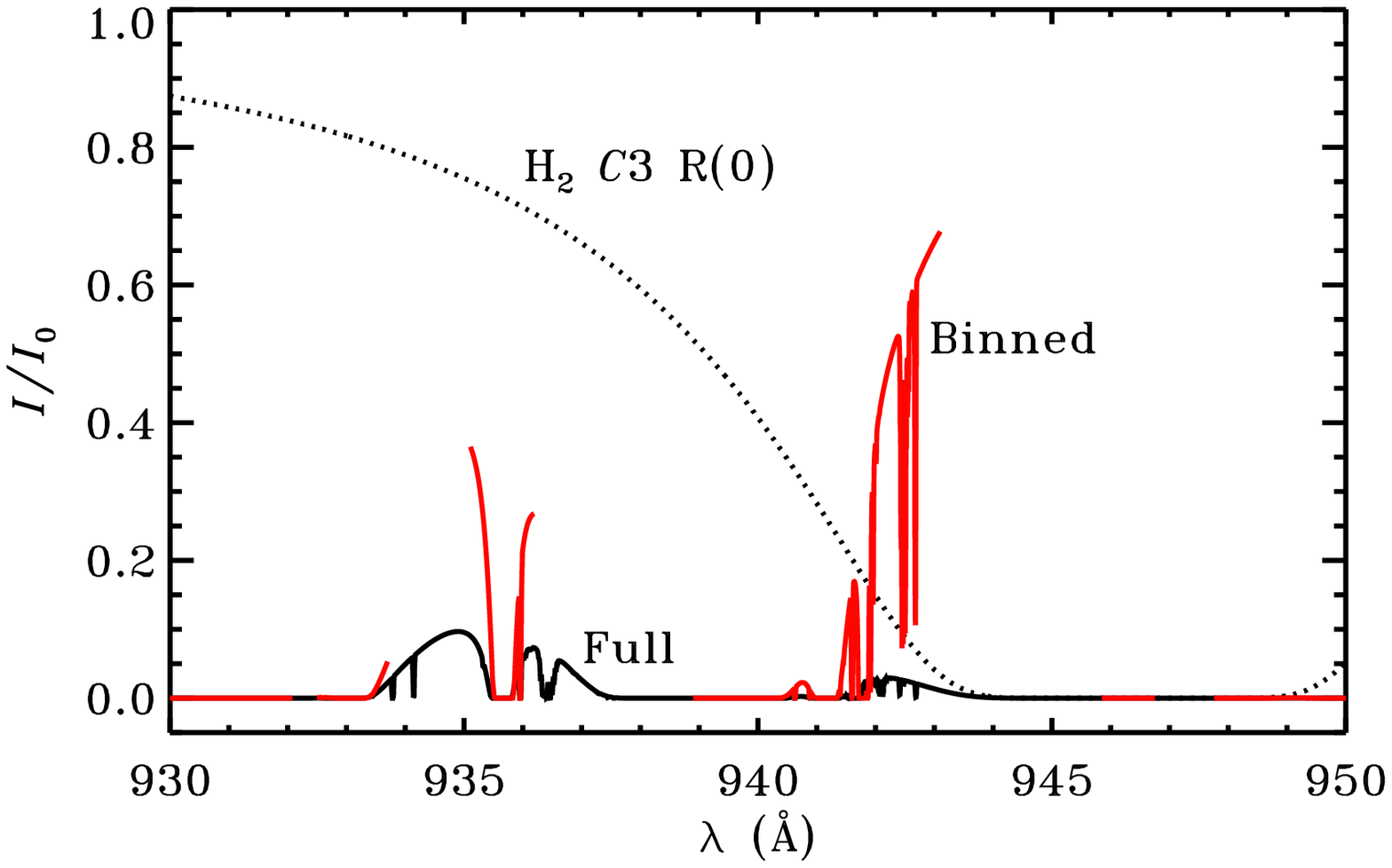}}
\caption{Illustration of the effect of not integrating strongly saturated lines over the entire wavelength regime. Full black curve: relative intensity of the radiation field at $\cdco=10^{18}$ and $\cdmh=10^{22}$ \pcs{} for the parameters of \tb{shield}, computed with a full integration of all lines over all wavelengths. Red curve: the same, but computed with the binned integration method of \vdb. Dotted curve: attenuation due to the $C3$ R(0) line of \mh, centred on  946.42 \AA. This line was not included by \vdb{} when computing the shielding in the 938.9--943.1 \AA{} bin, thus overestimating the photodissociation rate in the $W2$ band at 941.1 \AA.}
\label{fig:binfull}
\end{figure}

For column densities of up to \ten{17} \pcs{} of CO and \ten{21} \pcs{} of \mh, our shielding functions are generally within a few per cent of the \vdb{} values when corrected for the difference in $b$ and $T_\el{ex}$. Larger differences occur for larger columns: we predict the shielding to be about five times weaker at $\cdco=10^{19}$ \pcs{} and more than a hundred times stronger at $\cdmh=10^{23}$ \pcs. The 912-1118 \AA{} wavelength range was divided into 23 bins by \vdb, and most lines were included only in one bin to speed up the computation. We integrate all lines over all wavelengths. As the lines get strongly saturated at high column depths, absorption in the line wings becomes important. Thus, \mh{} lines can cause substantial shielding over a range of more than 10 \AA, while CO lines may still absorb several \AA{} away from the line centre. The binned integration method of \vdb{} did not take these effects into account, so they underpredicted shielding at large \mh{} columns and overpredicted shielding at large CO columns (\fig{binfull}). It should be noted, however, that photodissociation at these depths is typically already so slow a process that it is no longer the dominant destruction pathway for CO. In addition, a large CO column is usually accompanied by a large \mh{} column, so the two effects partially cancel each other.

\subsection{Comparison between the full model and the approximations}
\label{subsec:shcomp}
Despite being computed for a limited number of model parameters, the shielding functions from Tables \ref{tb:shield}--\ref{tb:shield4} provide a good approximation to the rates from the full model for a wide range of astrophysical environments. Section \ref{subsec:trans} presents a grid of translucent cloud models, where the photodissociation model is coupled to a chemical network and CO is traced as a function of depth. This presents a large range of column densities, with the ratios between the isotopologues deviating from the fixed values adopted for Tables \ref{tb:shield}--\ref{tb:shield4}. The grid covers gas densities from 100 to 1000 \pcc{} and gas temperatures from 15 to 100 K, while keeping the excitation temperatures and Doppler widths constant at the values used for \tb{shield}. Altogether, the grid contains 2880 points per isotopologue where the photodissociation rate is computed. Here, we compare the photodissociaton rates from the approximate method to the full integration for each of these points. The sensitivity of $k(\cocm{13})$ to the $N(\cocm{12})/N(\cocm{13})$ ratio can be corrected for in a simple manner: we use the shielding functions from \tb{shield4} when the ratio is closer to 35 than to 69, and those from \tb{shield} otherwise.

\begin{figure}
\resizebox{\hsize}{!}{\includegraphics{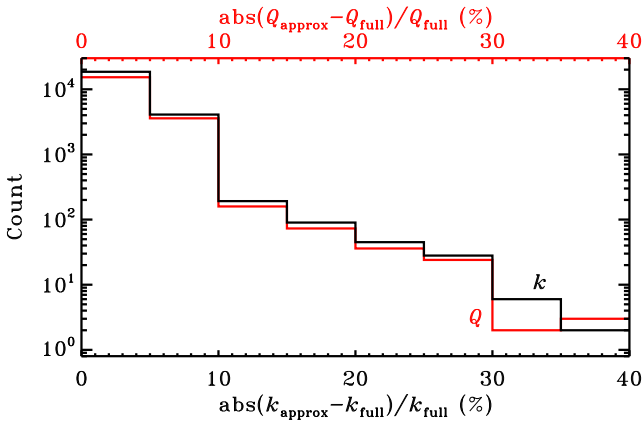}}
\caption{Histogram of the absolute relative difference between the approximate method (Tables \ref{tb:shield} and \ref{tb:shield4}; see text for details) and the full computation for the absolute photodissociation rates ($k$, black line) and the isotopologue rate ratios ($Q$, red line) in a grid of translucent cloud models. The data for all six isotopologues are taken together, but the $Q$ values for \coc{12} (which are unity regardless of the method) are omitted.}
\label{fig:transhistcomb}
\end{figure}

The rate from our approximate method is within 10\% of the ``real'' rate in 98.3\% of all points (\fig{transhistcomb}). In no cases is the difference between the approximate rates and the full model more than 40\%. Perhaps even more important than the absolute photodissociation rates are the ratios, $Q$, between the rates of \coc{12} and the other five isotopologues \co{x}{y}:
\begin{equation}
\label{eq:qratio}
Q_i = \frac{k(\com{x}{y})}{k(\cocm{12})}\,.
\end{equation}
The shielding functions from Tables \ref{tb:shield} and \ref{tb:shield4} together reproduce the ratios from the full method to the same accuracy as the absolute rates. At 98.4\% of all points in the grid of translucent clouds, the ratios are off by less than 10\% (\fig{transhistcomb}). Similar scores can be obtained for models of PDRs or other environments if one uses shielding functions computed for the right combination of parameters. Otherwise, the accuracy goes down. For example, the shielding functions from \tb{shield} ($\texco=5$ K) can easily give photodissociation rates off by a factor of two when applied to a high-density, high-temperature PDR.

\section{Chemistry of CO: astrophysical implications}
\label{sec:implic}
Photodissociation is an important destruction mechanism for CO in many environments. In this section, we couple the photodissociation model to a small chemical network in order to explore abundances and column densities. Specifically, we model the CO chemistry in translucent clouds, PDRs and circumstellar disks, and we compare our results to observations of such objects. The photodissociation rates are computed with our full model throughout this section.


\subsection{Translucent clouds}
\label{subsec:trans}


\subsubsection{Model setup}
\label{subsec:transmod}
Translucent clouds, with visual extinctions between 1 and 5 mag, form an excellent test case for our CO photodissociation model. Two recent studies of several dozen lines of sight through diffuse ($\av<1$ mag) and translucent clouds provide a set of CO and \mh{} column densities for comparison \citep{sonnentrucker07a,sheffer08a}. These surveys show a clear correlation between $\cdmh$ and $N(\cocm{12})$, with distinctly different slopes for \mh{} columns of less and more than \scit{2.5}{20} \pcs. \citeauthor{sheffer08a} attributed this break to a change in the formation mechanism of CO. Their models show that the two-step conversion from \element[+]{C} to CH$^+$ and CO$^+$,
\begin{equation}
\label{eq:cp+h2}
\element[+]{C}\ +\ \el{H}_2\ \to\ \el{CH}^+\ +\ \element{H}\,,
\end{equation}
\begin{equation}
\label{eq:chp+o}
\el{CH}^+\ +\ \element{O}\ \to\ \element{CO}^+\ +\ \element{H}\,,
\end{equation}
followed by reaction with H (forming CO directly) or \mh{} (forming HCO$^+$, which then recombines with an electron to give CO) is the dominant pathway at low column densities. However, the highly endothermic Reaction (\ref{eq:cp+h2}) is not fast enough at gas kinetic temperatures typical for these environments to explain the observed abundances of CH$^+$ and CO. Suprathermal chemistry has been suggested as a solution to this problem. \citeauthor{sheffer08a} followed the approach of \citet{federman96a}, who argued that Alfv\'en waves entering the cloud from the outside result in non-thermal motions between ions and neutrals. Other mechanisms have been suggested by \citet{joulain98a} and \citet{pety00a}. The effect of the Alfv\'en waves can be incorporated into a chemical model by replacing the kinetic temperature in the rate equation for Reaction (\ref{eq:cp+h2}) and all other ion-neutral reactions by an effective temperature:
\begin{equation}
\label{eq:teff}
T_\el{eff} = T_\el{gas} + \frac{\mu v_\el{A}^2}{3k_\el{B}}\,.
\end{equation}
Here, $k_\el{B}$ is the Boltzmann constant, $\mu$ is the reduced mass of the reactants and $v_\el{A}$ is the Alfv\'en speed. The Alfv\'en waves reach a depth of a few \ten{20} \pcs{} of \mh{}, corresponding to an $\av$ of a few tenths of a magnitude, beyond which suprathermal chemistry ceases to be important. CO can therefore no longer be formed efficiently through Reactions (\ref{eq:cp+h2}) and (\ref{eq:chp+o}), and the reaction between \element[+]{C} and OH (producing CO either directly or via a CO$^+$ intermediate) takes over as the key route to CO. The identification of these two different chemistry regimes supports the conclusion of \citet{zsargo03a} that suprathermal chemistry is required to explain observed CO abundances in diffuse environments. Suprathermal chemistry also drives up the HCO$^+$ abundance, confirming the conclusion of \citet{liszt94a} and \citet{liszt07a} that HCO$^+$ is the dominant precursor to CO in diffuse clouds.

We present here a grid of translucent cloud models to see how well the new photodissociation results match the observations. We set the Alfv\'en speed to 3.3 km \ps{} for $\cdmh<\scim{4}{20}$ \pcs{} and to zero for larger column densities \citep{sheffer08a}. The grid comprises densities ($n_\el{H}=n(\el{H})+2n(\el{H}_2)$) of 100, 300, 500, 700 and 1000 \pcc{}, gas temperatures of 15, 30, 50 and 100 K, and relative UV intensities ($\chi$) of 1, 3 and 10. The dust temperature is assumed to stay low for all models ($\sim15$ K), so the \mh{} formation rate does not change. The ionisation rate of H due to cosmic rays is set to a constant value of \scit{1.3}{-17} \ps. Attenuation by 0.1 \micron{} dust grains (Sect.\ \ref{subsec:dust}) is taken into account. The Doppler widths and level populations are as described in Sect.\ \ref{subsec:default}, with CO and \mh{} excitation temperatures of 5 and $5^{1.5}$ K. Taking other $b$ or $T_\el{ex}$ values plausible for these environments does not alter our results significantly. All models are run to an $\av$ of 5 mag; results are also presented for a range of smaller extinctions.

The models require a chemical network to compute the abundances at each depth step. Since we are only interested in CO, the number of relevant species and reactions is limited. We adopt the network from a recent PDR benchmark study \citep{rollig07a}, which includes only 31 species consisting of H, He, C and O. We duplicate all C- and O-containing species and reactions for \element[][13]{C}, \element[][17]{O} and \element[][18]{O}. Freeze-out and thermal evaporation are added for all neutral species, but no grain-surface reactions are included other than \mh{} formation according to \citet{black87a}. We add ion-molecule exchange reactions such as
\begin{equation}
\label{eq:imex1}
\cocm{12}\ +\ \element[+][13]{C}\ \rightleftarrows\ \element[+][12]{C}\ +\ \cocm{13}\ +\ 35\ \el{K}\,,
\end{equation}
\begin{equation}
\label{eq:imex2}
\el{H}^{12}\el{CO}^+\ +\ \cocm{13}\ \rightleftarrows\ \cocm{12}\ +\ \el{H}^{13}\el{CO}^+\ +\ 9\ \el{K}\,,
\end{equation}
which can enhance the abundances of the heavy isotopologues of CO and HCO$^+$ \citep{watson76a,smith80a,langer84a}. The temperature-dependence of the rate of these two reactions was fitted by \citet{liszt07a}; the alternative equations from \citet{woods09a} give the same results. The effective temperature from \eq{teff} is used instead of the kinetic temperature for all ion-neutral reactions, including Reactions (\ref{eq:imex1}) and (\ref{eq:imex2}). Altogether, the network contains 118 species and 1723 reactions. We adopt the elemental abundances of \citet{cardelli96a} and isotope ratios appropriate for the local ISM ($[\element[][12]{C}]/[\element[][13]{C}]=69$, $[\element[][16]{O}]/[\element[][18]{O}]=557$ and $[\element[][18]{O}]/[\element[][17]{O}]=3.6$; \citealt{wilson99a}); the complete list of elemental abundances is given in \tb{elabun}. Chemical steady state is reached at all depths after $\sim$1 Myr, regardless of whether the gas starts in atomic or molecular form.

\begin{table}
\caption{Elemental abundances.}
\label{tb:elabun}
\centering
\begin{tabular}{cc}
\hline\hline
Element           & Abundance relative to $n_\el{H}$ \\
\hline
\rule{0pt}{1em}He & \scit{1.00}{-1} \\
\element[][12]{C} & \scit{1.40}{-4} \\
\element[][13]{C} & \scit{2.03}{-6} \\
\element[][16]{O} & \scit{3.10}{-4} \\
\element[][17]{O} & \scit{1.55}{-7} \\
\element[][18]{O} & \scit{5.57}{-7} \\
\hline
\end{tabular}
\end{table}


\subsubsection{\coc{12}}
\label{subsec:trans12co}
\begin{figure*}
\resizebox{\hsize}{!}{\includegraphics{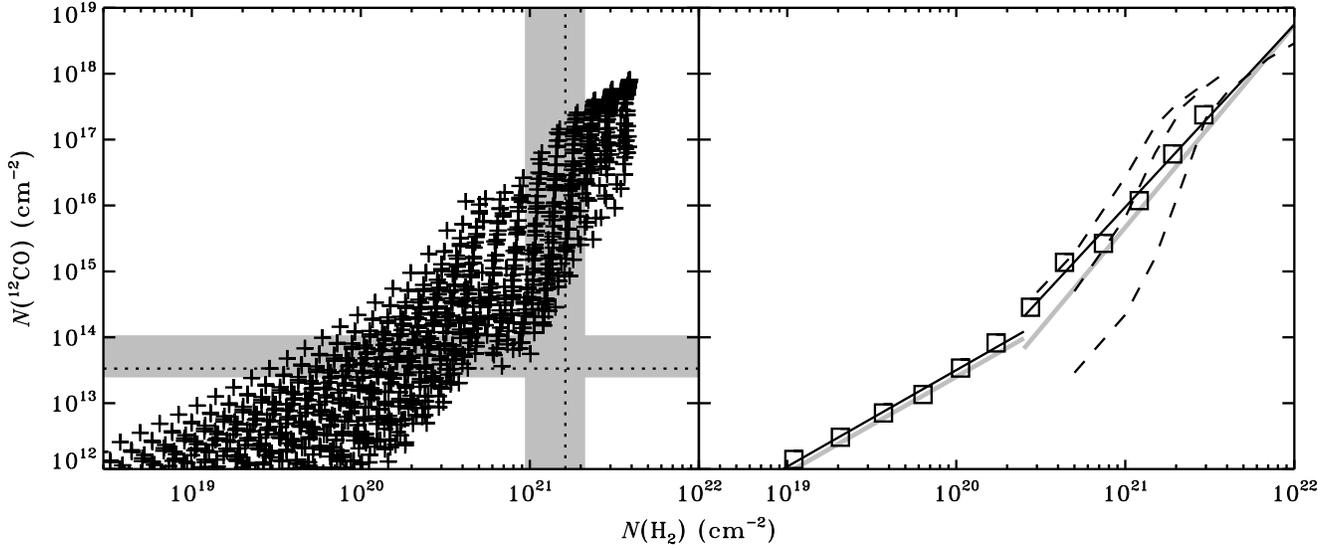}}
\caption{Column densities of \coc{12} versus those of \mh. \emph{Left:} data for all points with $\av\leq5$ mag from our grid of translucent cloud models, covering $n_\el{H}=100$--1000 \pcc, $T_\el{gas}=15$--100 K and $\chi=1$--10. The vertical grey band indicates the range over which photodissociation ceases to be the dominant destruction mechanism of \coc{12}; the dotted line is the median. Likewise, the horizontal band indicates the range over which \coc{12} becomes self-shielding. \emph{Right:} 100-point means of our model results (black squares), fitted with two straight lines. Also shown are the fits to the observations by \citet[grey lines]{sheffer08a} and the results from the translucent cloud models of \vdb{} (dashed lines) for $\chi=0.5$, 1 and 10 (\emph{left to right}).}
\label{fig:co-vs-h2}
\end{figure*}

The left panel of \fig{co-vs-h2} shows $N(\cocm{12})$ versus $\cdmh$ for all depth steps in our grid. These and other column densities are also listed for a set of selected $\av$ values in Tables \ref{tb:trans1}--\ref{tb:trans4} in the online appendix to this paper. The scatter in the data is due to the different physical parameters. For a given $\cdmh$, $N(\cocm{12})$ is about two times larger at $T_\el{gas}=100$ K than at 15 K. The formation rate of \mh{} increases with temperature; through the chain of reactions starting with \rx{cp+h2}, that results in a larger CO abundance and column density. Up to $\cdmh=10^{21}$ \pcs, increasing the gas density by a factor of ten increases $N(\cocm{12})/\cdmh$ also by about a factor of ten. This is due to the photodissociation rate being mostly independent of density, while the rates of the two-body reactions forming CO are not. Photodissociation ceases to be the main destruction mechanism for CO deeper into the cloud, so increasing $n_\el{H}$ has a smaller effect there. Increasing the UV intensity from $\chi=1$ to 10 has the simple effect of decreasing $N(\cocm{12})/\cdmh$ roughly tenfold for $\cdmh<10^{21}$ \pcs. For larger depths, changing $\chi$ only has a small effect. These dependencies on the physical parameters are consistent with the observations of \citet{sheffer08a}.

\begin{figure*}
\resizebox{\hsize}{!}{\includegraphics{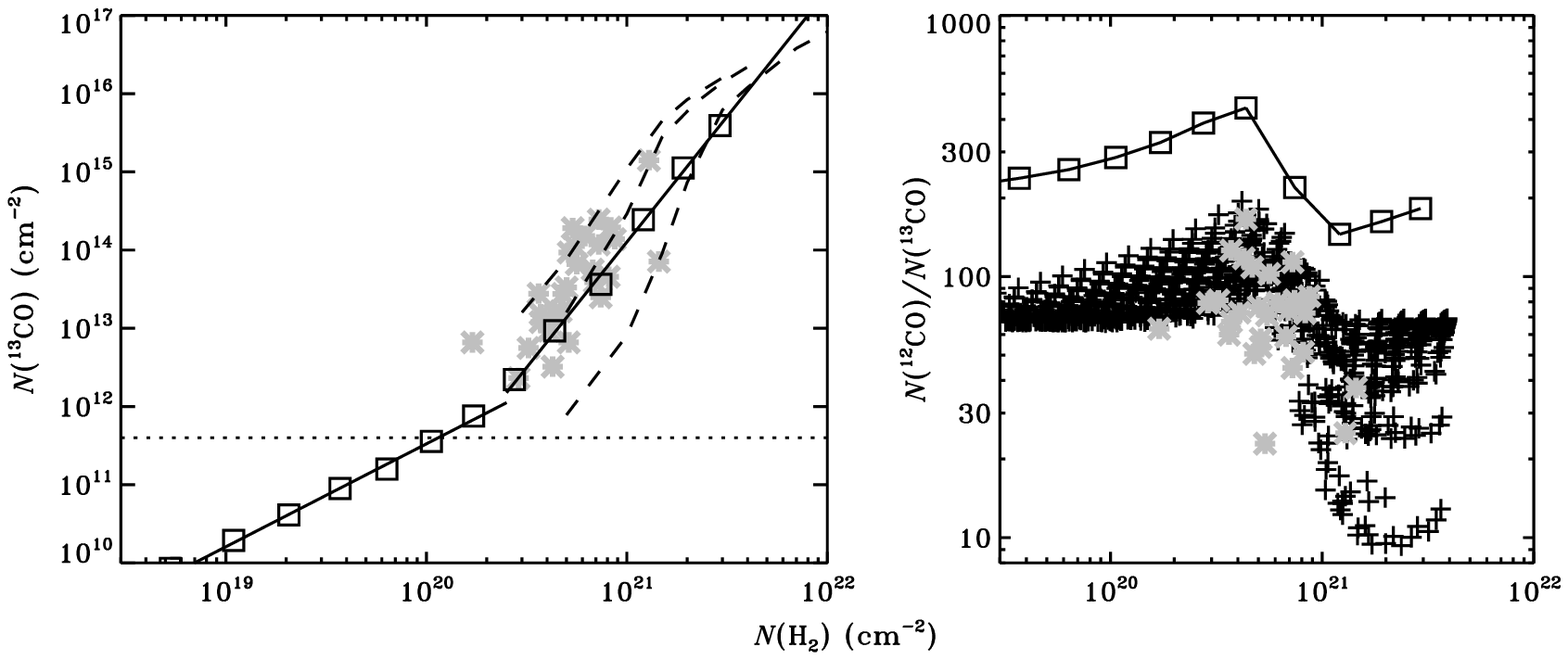}}
\caption{Column densities of \coc{13} (\emph{left}) and column density ratios of \coc{12} to \coc{13} (\emph{right}) plotted against the column density of \mh. Grey crosses: observations of \citet{sonnentrucker07a}, \citet{burgh07a} and \citet{sheffer07a}. Black squares and straight lines: 100-point means from our grid of translucent cloud models, offset by a factor of three in the right panel. Dotted line: median depth at which \coc{12} becomes self-shielding. Black plus signs: model data for all points with $\av\leq5$ mag. Dashed lines: data from the translucent cloud models of \vdb{} for $\chi=0.5$, 1 and 10 (\emph{left to right}).}
\label{fig:cd13co}
\end{figure*}

\onltab{10}{
\begin{table*}
\caption{Column densities (in \pcs) for our grid of translucent cloud models at $T_\el{gas}=15$ K. Results for $n_\el{H}=500$ and 700 \pcc{} are not shown.}
\label{tb:trans1}
\centering
\begin{tabular}{cccccccccccc}
\hline\hline
\rule{0pt}{1em}Model & $n_\el{H}$ & $\chi$ & $\av$ & $\cdmh$ & $\cdh$ & $N(\cocm{12})$ & $N(\coom{17})$ & $N(\coom{18})$ & $N(\cocm{13})$ & $N(\com{13}{17})$ & $N(\com{13}{18})$ \\
    & (\pcc)     &        & (mag) &         &        & & & & & & \\
\hline
\rule{0pt}{1em}A1$\alpha$ & \phantom{0}100 & \phantom{0}1 & 0.08 & 1.0(19) & 1.1(20) & 1.1(11) & 5.2(7)\phantom{0} & 2.0(8)\phantom{0} & 1.5(9)\phantom{0} & 7.7(5)\phantom{0} & 2.8(6)\phantom{0} \\
 & & & 0.14 & 3.0(19) & 1.7(20) & 4.6(11) & 2.2(8)\phantom{0} & 8.5(8)\phantom{0} & 6.7(9)\phantom{0} & 3.3(6)\phantom{0} & 1.2(7)\phantom{0} \\
 & & & 0.30 & 1.0(20) & 2.8(20) & 3.1(12) & 1.4(9)\phantom{0} & 5.5(9)\phantom{0} & 4.4(10) & 2.2(7)\phantom{0} & 8.1(7)\phantom{0} \\
 & & & 0.71 & 3.0(20) & 5.1(20) & 4.5(13) & 1.7(10) & 6.4(10) & 5.2(11) & 2.5(8)\phantom{0} & 9.2(8)\phantom{0} \\
 & & & 1.7\phantom{0} & 1.0(21) & 6.8(20) & 4.3(14) & 1.3(11) & 4.6(11) & 1.6(13) & 5.4(9)\phantom{0} & 1.8(10) \\
 & & & 4.2\phantom{0} & 3.0(21) & 6.9(20) & 1.6(16) & 7.4(12) & 2.6(13) & 1.4(15) & 6.6(11) & 2.4(12) \\
\rule{0pt}{1.2em}A1$\beta$ & \phantom{0}100 & \phantom{0}3 & 0.20 & 1.0(19) & 2.9(20) & 2.7(10) & 1.3(7)\phantom{0} & 5.0(7)\phantom{0} & 3.9(8)\phantom{0} & 1.9(5)\phantom{0} & 7.0(5)\phantom{0} \\
 & & & 0.29 & 3.0(19) & 3.9(20) & 1.3(11) & 6.1(7)\phantom{0} & 2.4(8)\phantom{0} & 1.9(9)\phantom{0} & 9.2(5)\phantom{0} & 3.4(6)\phantom{0} \\
 & & & 0.48 & 1.0(20) & 5.5(20) & 1.0(12) & 4.9(8)\phantom{0} & 1.9(9)\phantom{0} & 1.5(10) & 7.5(6)\phantom{0} & 2.7(7)\phantom{0} \\
 & & & 0.90 & 3.0(20) & 8.2(20) & 1.6(13) & 7.0(9)\phantom{0} & 2.6(10) & 2.1(11) & 1.0(8)\phantom{0} & 3.8(8)\phantom{0} \\
 & & & 1.9\phantom{0} & 1.0(21) & 1.0(21) & 1.9(14) & 6.4(10) & 2.3(11) & 6.2(12) & 2.2(9)\phantom{0} & 8.0(9)\phantom{0} \\
 & & & 4.5\phantom{0} & 3.0(21) & 1.0(21) & 1.1(16) & 5.0(12) & 1.8(13) & 9.7(14) & 4.7(11) & 1.7(12) \\
\rule{0pt}{1.2em}A1$\gamma$ & \phantom{0}100 & 10 & 0.46 & 1.0(19) & 7.0(20) & 1.2(10) & 5.5(6)\phantom{0} & 2.1(7)\phantom{0} & 1.6(8)\phantom{0} & 8.1(4)\phantom{0} & 3.0(5)\phantom{0} \\
 & & & 0.59 & 3.0(19) & 8.7(20) & 6.1(10) & 2.9(7)\phantom{0} & 1.1(8)\phantom{0} & 8.9(8)\phantom{0} & 4.4(5)\phantom{0} & 1.6(6)\phantom{0} \\
 & & & 0.81 & 1.0(20) & 1.1(21) & 5.7(11) & 2.7(8)\phantom{0} & 1.1(9)\phantom{0} & 8.5(9)\phantom{0} & 4.2(6)\phantom{0} & 1.5(7)\phantom{0} \\
 & & & 1.2\phantom{0} & 3.0(20) & 1.4(21) & 9.6(12) & 4.4(9)\phantom{0} & 1.7(10) & 1.4(11) & 6.7(7)\phantom{0} & 2.4(8)\phantom{0} \\
 & & & 2.3\phantom{0} & 1.0(21) & 1.6(21) & 1.0(14) & 3.9(10) & 1.4(11) & 3.4(12) & 1.3(9)\phantom{0} & 4.9(9)\phantom{0} \\
 & & & 4.8\phantom{0} & 3.0(21) & 1.6(21) & 7.3(15) & 3.5(12) & 1.3(13) & 7.1(14) & 3.5(11) & 1.3(12) \\
\hline
\rule{0pt}{1em}A2$\alpha$ & \phantom{0}300 & \phantom{0}1 & 0.04 & 1.0(19) & 4.4(19) & 4.3(11) & 2.1(8)\phantom{0} & 8.0(8)\phantom{0} & 6.2(9)\phantom{0} & 3.1(6)\phantom{0} & 1.1(7)\phantom{0} \\
 & & & 0.09 & 3.0(19) & 7.5(19) & 1.8(12) & 8.7(8)\phantom{0} & 3.4(9)\phantom{0} & 2.7(10) & 1.3(7)\phantom{0} & 4.9(7)\phantom{0} \\
 & & & 0.23 & 1.0(20) & 1.6(20) & 1.2(13) & 5.2(9)\phantom{0} & 2.0(10) & 1.6(11) & 7.8(7)\phantom{0} & 2.9(8)\phantom{0} \\
 & & & 0.60 & 3.0(20) & 3.5(20) & 1.7(14) & 5.3(10) & 1.9(11) & 1.6(12) & 7.4(8)\phantom{0} & 2.7(9)\phantom{0} \\
 & & & 1.5\phantom{0} & 1.0(21) & 4.5(20) & 1.4(15) & 3.6(11) & 1.1(12) & 8.1(13) & 2.1(10) & 6.6(10) \\
 & & & 4.1\phantom{0} & 3.0(21) & 4.5(20) & 1.5(17) & 6.8(13) & 2.5(14) & 2.8(15) & 1.1(12) & 4.2(12) \\
\rule{0pt}{1.2em}A2$\beta$ & \phantom{0}300 & \phantom{0}3 & 0.08 & 1.0(19) & 1.1(20) & 7.8(10) & 3.7(7)\phantom{0} & 1.4(8)\phantom{0} & 1.1(9)\phantom{0} & 5.6(5)\phantom{0} & 2.1(6)\phantom{0} \\
 & & & 0.14 & 3.0(19) & 1.7(20) & 3.7(11) & 1.8(8)\phantom{0} & 6.8(8)\phantom{0} & 5.4(9)\phantom{0} & 2.7(6)\phantom{0} & 9.9(6)\phantom{0} \\
 & & & 0.30 & 1.0(20) & 2.8(20) & 2.7(12) & 1.3(9)\phantom{0} & 4.8(9)\phantom{0} & 3.9(10) & 1.9(7)\phantom{0} & 7.1(7)\phantom{0} \\
 & & & 0.70 & 3.0(20) & 5.1(20) & 3.9(13) & 1.5(10) & 5.7(10) & 4.7(11) & 2.3(8)\phantom{0} & 8.3(8)\phantom{0} \\
 & & & 1.7\phantom{0} & 1.0(21) & 6.7(20) & 4.3(14) & 1.3(11) & 4.5(11) & 1.9(13) & 5.9(9)\phantom{0} & 2.0(10) \\
 & & & 4.2\phantom{0} & 3.0(21) & 6.8(20) & 9.8(16) & 4.3(13) & 1.5(14) & 2.3(15) & 9.8(11) & 3.6(12) \\
\rule{0pt}{1.2em}A2$\gamma$ & \phantom{0}300 & 10 & 0.21 & 1.0(19) & 3.2(20) & 1.9(10) & 8.9(6)\phantom{0} & 3.4(7)\phantom{0} & 2.7(8)\phantom{0} & 1.3(5)\phantom{0} & 4.8(5)\phantom{0} \\
 & & & 0.30 & 3.0(19) & 4.2(20) & 9.3(10) & 4.5(7)\phantom{0} & 1.7(8)\phantom{0} & 1.4(9)\phantom{0} & 6.8(5)\phantom{0} & 2.5(6)\phantom{0} \\
 & & & 0.50 & 1.0(20) & 5.9(20) & 8.3(11) & 4.0(8)\phantom{0} & 1.5(9)\phantom{0} & 1.2(10) & 6.1(6)\phantom{0} & 2.2(7)\phantom{0} \\
 & & & 0.92 & 3.0(20) & 8.6(20) & 1.4(13) & 6.0(9)\phantom{0} & 2.3(10) & 1.8(11) & 9.0(7)\phantom{0} & 3.3(8)\phantom{0} \\
 & & & 1.9\phantom{0} & 1.0(21) & 1.1(21) & 1.5(14) & 5.2(10) & 1.9(11) & 5.4(12) & 1.9(9)\phantom{0} & 7.0(9)\phantom{0} \\
 & & & 4.5\phantom{0} & 3.0(21) & 1.1(21) & 6.5(16) & 2.9(13) & 1.1(14) & 2.0(15) & 8.6(11) & 3.2(12) \\
\hline
\rule{0pt}{1em}A5$\alpha$ & 1000 & \phantom{0}1 & 0.02 & 1.0(19) & 1.9(19) & 2.5(12) & 1.1(9)\phantom{0} & 4.5(9)\phantom{0} & 3.5(10) & 1.7(7)\phantom{0} & 6.5(7)\phantom{0} \\
 & & & 0.06 & 3.0(19) & 4.0(19) & 1.0(13) & 4.4(9)\phantom{0} & 1.7(10) & 1.4(11) & 6.8(7)\phantom{0} & 2.5(8)\phantom{0} \\
 & & & 0.19 & 1.0(20) & 1.0(20) & 6.9(13) & 2.4(10) & 8.8(10) & 7.4(11) & 3.5(8)\phantom{0} & 1.3(9)\phantom{0} \\
 & & & 0.53 & 3.0(20) & 2.4(20) & 9.0(14) & 2.2(11) & 7.0(11) & 7.2(12) & 2.8(9)\phantom{0} & 1.0(10) \\
 & & & 1.5\phantom{0} & 1.0(21) & 3.0(20) & 1.4(16) & 1.9(12) & 5.6(12) & 2.4(14) & 3.4(10) & 1.1(11) \\
 & & & 4.0\phantom{0} & 3.0(21) & 3.0(20) & 5.9(16) & 2.1(13) & 7.3(13) & 8.8(14) & 3.0(11) & 1.1(12) \\
\rule{0pt}{1.2em}A5$\beta$ & 1000 & \phantom{0}3 & 0.04 & 1.0(19) & 4.0(19) & 4.5(11) & 2.1(8)\phantom{0} & 8.3(8)\phantom{0} & 6.6(9)\phantom{0} & 3.3(6)\phantom{0} & 1.2(7)\phantom{0} \\
 & & & 0.08 & 3.0(19) & 7.0(19) & 2.0(12) & 9.3(8)\phantom{0} & 3.6(9)\phantom{0} & 2.9(10) & 1.4(7)\phantom{0} & 5.3(7)\phantom{0} \\
 & & & 0.22 & 1.0(20) & 1.5(20) & 1.3(13) & 5.6(9)\phantom{0} & 2.1(10) & 1.7(11) & 8.5(7)\phantom{0} & 3.1(8)\phantom{0} \\
 & & & 0.60 & 3.0(20) & 3.4(20) & 1.8(14) & 5.6(10) & 2.0(11) & 1.7(12) & 8.0(8)\phantom{0} & 2.9(9)\phantom{0} \\
 & & & 1.5\phantom{0} & 1.0(21) & 4.3(20) & 1.8(15) & 4.5(11) & 1.4(12) & 9.3(13) & 2.2(10) & 6.7(10) \\
 & & & 4.1\phantom{0} & 3.0(21) & 4.3(20) & 4.7(16) & 1.8(13) & 6.4(13) & 7.7(14) & 2.8(11) & 9.8(11) \\
\rule{0pt}{1.2em}A5$\gamma$ & 1000 & 10 & 0.08 & 1.0(19) & 1.1(20) & 6.7(10) & 3.2(7)\phantom{0} & 1.2(8)\phantom{0} & 9.8(8)\phantom{0} & 4.8(5)\phantom{0} & 1.8(6)\phantom{0} \\
 & & & 0.14 & 3.0(19) & 1.7(20) & 3.4(11) & 1.6(8)\phantom{0} & 6.3(8)\phantom{0} & 5.0(9)\phantom{0} & 2.5(6)\phantom{0} & 9.1(6)\phantom{0} \\
 & & & 0.30 & 1.0(20) & 2.8(20) & 2.5(12) & 1.2(9)\phantom{0} & 4.6(9)\phantom{0} & 3.7(10) & 1.8(7)\phantom{0} & 6.8(7)\phantom{0} \\
 & & & 0.70 & 3.0(20) & 5.1(20) & 3.7(13) & 1.5(10) & 5.4(10) & 4.5(11) & 2.2(8)\phantom{0} & 8.0(8)\phantom{0} \\
 & & & 1.7\phantom{0} & 1.0(21) & 6.7(20) & 4.4(14) & 1.3(11) & 4.5(11) & 1.9(13) & 5.8(9)\phantom{0} & 2.0(10) \\
 & & & 4.2\phantom{0} & 3.0(21) & 6.7(20) & 4.0(16) & 1.6(13) & 5.6(13) & 6.8(14) & 2.6(11) & 9.2(11) \\
\hline
\end{tabular}
\end{table*}
}

\onltab{11}{
\begin{table*}
\caption{As \tb{trans1}, for $T_\el{gas}=30$ K.}
\label{tb:trans2}
\centering
\begin{tabular}{cccccccccccc}
\hline\hline
\rule{0pt}{1em}Model & $n_\el{H}$ & $\chi$ & $\av$ & $\cdmh$ & $\cdh$ & $N(\cocm{12})$ & $N(\coom{17})$ & $N(\coom{18})$ & $N(\cocm{13})$ & $N(\com{13}{17})$ & $N(\com{13}{18})$ \\
    & (\pcc)     &        & (mag) &         &        & & & & & & \\
\hline
\rule{0pt}{1em}B1$\alpha$ & \phantom{0}100 & \phantom{0}1 & 0.06 & 1.0(19) & 8.2(19) & 1.4(11) & 6.9(7)\phantom{0} & 2.7(8)\phantom{0} & 2.1(9)\phantom{0} & 1.0(6)\phantom{0} & 3.8(6)\phantom{0} \\
 & & & 0.12 & 3.0(19) & 1.2(20) & 6.2(11) & 3.0(8)\phantom{0} & 1.2(9)\phantom{0} & 9.1(9)\phantom{0} & 4.5(6)\phantom{0} & 1.7(7)\phantom{0} \\
 & & & 0.26 & 1.0(20) & 2.2(20) & 4.1(12) & 1.9(9)\phantom{0} & 7.3(9)\phantom{0} & 5.8(10) & 2.9(7)\phantom{0} & 1.1(8)\phantom{0} \\
 & & & 0.64 & 3.0(20) & 4.2(20) & 5.4(13) & 2.0(10) & 7.4(10) & 6.0(11) & 2.9(8)\phantom{0} & 1.1(9)\phantom{0} \\
 & & & 1.6\phantom{0} & 1.0(21) & 5.8(20) & 6.2(14) & 1.7(11) & 5.9(11) & 1.1(13) & 3.7(9)\phantom{0} & 1.3(10) \\
 & & & 4.2\phantom{0} & 3.0(21) & 5.9(20) & 3.4(16) & 1.6(13) & 5.7(13) & 1.3(15) & 6.1(11) & 2.2(12) \\
\rule{0pt}{1.2em}B1$\beta$ & \phantom{0}100 & \phantom{0}3 & 0.15 & 1.0(19) & 2.2(20) & 3.2(10) & 1.6(7)\phantom{0} & 5.9(7)\phantom{0} & 4.6(8)\phantom{0} & 2.3(5)\phantom{0} & 8.4(5)\phantom{0} \\
 & & & 0.23 & 3.0(19) & 3.0(20) & 1.5(11) & 7.4(7)\phantom{0} & 2.8(8)\phantom{0} & 2.2(9)\phantom{0} & 1.1(6)\phantom{0} & 4.1(6)\phantom{0} \\
 & & & 0.40 & 1.0(20) & 4.3(20) & 1.2(12) & 5.8(8)\phantom{0} & 2.2(9)\phantom{0} & 1.8(10) & 8.8(6)\phantom{0} & 3.2(7)\phantom{0} \\
 & & & 0.80 & 3.0(20) & 6.6(20) & 1.7(13) & 7.3(9)\phantom{0} & 2.7(10) & 2.2(11) & 1.1(8)\phantom{0} & 4.0(8)\phantom{0} \\
 & & & 1.8\phantom{0} & 1.0(21) & 7.8(20) & 1.0(14) & 3.7(10) & 1.3(11) & 2.0(12) & 8.1(8)\phantom{0} & 2.9(9)\phantom{0} \\
 & & & 4.3\phantom{0} & 3.0(21) & 7.9(20) & 2.0(16) & 9.4(12) & 3.4(13) & 7.8(14) & 3.8(11) & 1.4(12) \\
\rule{0pt}{1.2em}B1$\gamma$ & \phantom{0}100 & 10 & 0.37 & 1.0(19) & 5.6(20) & 1.1(10) & 5.3(6)\phantom{0} & 2.0(7)\phantom{0} & 1.6(8)\phantom{0} & 7.9(4)\phantom{0} & 2.9(5)\phantom{0} \\
 & & & 0.48 & 3.0(19) & 7.0(20) & 5.9(10) & 2.8(7)\phantom{0} & 1.1(8)\phantom{0} & 8.7(8)\phantom{0} & 4.3(5)\phantom{0} & 1.6(6)\phantom{0} \\
 & & & 0.70 & 1.0(20) & 9.0(20) & 5.8(11) & 2.8(8)\phantom{0} & 1.1(9)\phantom{0} & 8.7(9)\phantom{0} & 4.3(6)\phantom{0} & 1.6(7)\phantom{0} \\
 & & & 1.1\phantom{0} & 3.0(20) & 1.2(21) & 9.4(12) & 4.3(9)\phantom{0} & 1.6(10) & 1.3(11) & 6.5(7)\phantom{0} & 2.4(8)\phantom{0} \\
 & & & 2.2\phantom{0} & 1.0(21) & 1.4(21) & 1.5(14) & 5.4(10) & 2.0(11) & 2.5(12) & 1.0(9)\phantom{0} & 3.7(9)\phantom{0} \\
 & & & 4.7\phantom{0} & 3.0(21) & 1.4(21) & 1.5(16) & 6.9(12) & 2.5(13) & 5.9(14) & 2.8(11) & 1.0(12) \\
\hline
\rule{0pt}{1em}B2$\alpha$ & \phantom{0}300 & \phantom{0}1 & 0.03 & 1.0(19) & 3.2(19) & 6.4(11) & 3.0(8)\phantom{0} & 1.2(9)\phantom{0} & 9.3(9)\phantom{0} & 4.6(6)\phantom{0} & 1.7(7)\phantom{0} \\
 & & & 0.07 & 3.0(19) & 5.7(19) & 2.7(12) & 1.3(9)\phantom{0} & 4.9(9)\phantom{0} & 3.9(10) & 1.9(7)\phantom{0} & 7.1(7)\phantom{0} \\
 & & & 0.21 & 1.0(20) & 1.3(20) & 1.7(13) & 7.2(9)\phantom{0} & 2.7(10) & 2.2(11) & 1.1(8)\phantom{0} & 4.0(8)\phantom{0} \\
 & & & 0.56 & 3.0(20) & 2.9(20) & 2.3(14) & 6.7(10) & 2.4(11) & 2.1(12) & 9.3(8)\phantom{0} & 3.4(9)\phantom{0} \\
 & & & 1.5\phantom{0} & 1.0(21) & 3.8(20) & 2.1(15) & 4.8(11) & 1.5(12) & 5.3(13) & 1.3(10) & 3.9(10) \\
 & & & 4.0\phantom{0} & 3.0(21) & 3.9(20) & 3.4(17) & 1.4(14) & 5.1(14) & 6.0(15) & 2.3(12) & 8.6(12) \\
\rule{0pt}{1.2em}B2$\beta$ & \phantom{0}300 & \phantom{0}3 & 0.06 & 1.0(19) & 8.2(19) & 1.1(11) & 5.4(7)\phantom{0} & 2.1(8)\phantom{0} & 1.6(9)\phantom{0} & 8.1(5)\phantom{0} & 3.0(6)\phantom{0} \\
 & & & 0.12 & 3.0(19) & 1.2(20) & 5.4(11) & 2.6(8)\phantom{0} & 1.0(9)\phantom{0} & 7.9(9)\phantom{0} & 3.9(6)\phantom{0} & 1.4(7)\phantom{0} \\
 & & & 0.26 & 1.0(20) & 2.2(20) & 3.7(12) & 1.7(9)\phantom{0} & 6.7(9)\phantom{0} & 5.3(10) & 2.6(7)\phantom{0} & 9.8(7)\phantom{0} \\
 & & & 0.64 & 3.0(20) & 4.1(20) & 4.9(13) & 1.9(10) & 6.8(10) & 5.6(11) & 2.7(8)\phantom{0} & 1.0(9)\phantom{0} \\
 & & & 1.6\phantom{0} & 1.0(21) & 5.8(20) & 6.1(14) & 1.7(11) & 5.6(11) & 1.2(13) & 3.6(9)\phantom{0} & 1.2(10) \\
 & & & 4.2\phantom{0} & 3.0(21) & 5.8(20) & 2.3(17) & 9.2(13) & 3.4(14) & 4.6(15) & 1.8(12) & 6.6(12) \\
\rule{0pt}{1.2em}B2$\gamma$ & \phantom{0}300 & 10 & 0.16 & 1.0(19) & 2.4(20) & 2.2(10) & 1.1(7)\phantom{0} & 4.0(7)\phantom{0} & 3.2(8)\phantom{0} & 1.6(5)\phantom{0} & 5.8(5)\phantom{0} \\
 & & & 0.24 & 3.0(19) & 3.2(20) & 1.2(11) & 5.6(7)\phantom{0} & 2.2(8)\phantom{0} & 1.7(9)\phantom{0} & 8.5(5)\phantom{0} & 3.1(6)\phantom{0} \\
 & & & 0.42 & 1.0(20) & 4.7(20) & 1.0(12) & 4.9(8)\phantom{0} & 1.9(9)\phantom{0} & 1.5(10) & 7.5(6)\phantom{0} & 2.8(7)\phantom{0} \\
 & & & 0.82 & 3.0(20) & 7.0(20) & 1.5(13) & 6.5(9)\phantom{0} & 2.4(10) & 2.0(11) & 9.7(7)\phantom{0} & 3.5(8)\phantom{0} \\
 & & & 1.9\phantom{0} & 1.0(21) & 9.2(20) & 2.3(14) & 7.4(10) & 2.6(11) & 3.9(12) & 1.5(9)\phantom{0} & 5.2(9)\phantom{0} \\
 & & & 4.4\phantom{0} & 3.0(21) & 9.2(20) & 1.5(17) & 6.0(13) & 2.2(14) & 3.4(15) & 1.3(12) & 5.0(12) \\
\hline
\rule{0pt}{1em}B5$\alpha$ & 1000 & \phantom{0}1 & 0.02 & 1.0(19) & 1.4(19) & 3.6(12) & 1.7(9)\phantom{0} & 6.5(9)\phantom{0} & 5.1(10) & 2.5(7)\phantom{0} & 9.3(7)\phantom{0} \\
 & & & 0.06 & 3.0(19) & 3.1(19) & 1.5(13) & 6.3(9)\phantom{0} & 2.4(10) & 1.9(11) & 9.5(7)\phantom{0} & 3.5(8)\phantom{0} \\
 & & & 0.18 & 1.0(20) & 8.3(19) & 1.0(14) & 3.2(10) & 1.2(11) & 1.0(12) & 4.7(8)\phantom{0} & 1.7(9)\phantom{0} \\
 & & & 0.50 & 3.0(20) & 1.9(20) & 1.3(15) & 2.8(11) & 8.9(11) & 9.4(12) & 3.5(9)\phantom{0} & 1.3(10) \\
 & & & 1.4\phantom{0} & 1.0(21) & 2.6(20) & 3.2(16) & 3.0(12) & 8.8(12) & 6.7(14) & 6.1(10) & 1.9(11) \\
 & & & 3.9\phantom{0} & 3.0(21) & 2.6(20) & 5.5(17) & 1.7(14) & 6.4(14) & 8.5(15) & 2.6(12) & 9.7(12) \\
\rule{0pt}{1.2em}B5$\beta$ & 1000 & \phantom{0}3 & 0.03 & 1.0(19) & 3.0(19) & 6.9(11) & 3.3(8)\phantom{0} & 1.3(9)\phantom{0} & 1.0(10) & 5.0(6)\phantom{0} & 1.9(7)\phantom{0} \\
 & & & 0.07 & 3.0(19) & 5.4(19) & 3.0(12) & 1.4(9)\phantom{0} & 5.4(9)\phantom{0} & 4.3(10) & 2.1(7)\phantom{0} & 7.9(7)\phantom{0} \\
 & & & 0.20 & 1.0(20) & 1.2(20) & 1.9(13) & 7.9(9)\phantom{0} & 3.0(10) & 2.4(11) & 1.2(8)\phantom{0} & 4.4(8)\phantom{0} \\
 & & & 0.56 & 3.0(20) & 2.8(20) & 2.5(14) & 7.3(10) & 2.6(11) & 2.3(12) & 1.0(9)\phantom{0} & 3.7(9)\phantom{0} \\
 & & & 1.5\phantom{0} & 1.0(21) & 3.7(20) & 2.8(15) & 5.8(11) & 1.7(12) & 7.5(13) & 1.6(10) & 4.8(10) \\
 & & & 4.0\phantom{0} & 3.0(21) & 3.7(20) & 4.3(17) & 1.4(14) & 5.1(14) & 6.8(15) & 2.1(12) & 8.0(12) \\
\rule{0pt}{1.2em}B5$\gamma$ & 1000 & 10 & 0.06 & 1.0(19) & 8.2(19) & 1.0(11) & 4.8(7)\phantom{0} & 1.9(8)\phantom{0} & 1.5(9)\phantom{0} & 7.3(5)\phantom{0} & 2.7(6)\phantom{0} \\
 & & & 0.12 & 3.0(19) & 1.2(20) & 5.1(11) & 2.4(8)\phantom{0} & 9.4(8)\phantom{0} & 7.5(9)\phantom{0} & 3.7(6)\phantom{0} & 1.4(7)\phantom{0} \\
 & & & 0.26 & 1.0(20) & 2.2(20) & 3.6(12) & 1.7(9)\phantom{0} & 6.4(9)\phantom{0} & 5.2(10) & 2.6(7)\phantom{0} & 9.5(7)\phantom{0} \\
 & & & 0.64 & 3.0(20) & 4.1(20) & 4.7(13) & 1.8(10) & 6.6(10) & 5.5(11) & 2.6(8)\phantom{0} & 9.7(8)\phantom{0} \\
 & & & 1.6\phantom{0} & 1.0(21) & 5.8(20) & 6.0(14) & 1.7(11) & 5.5(11) & 1.2(13) & 3.6(9)\phantom{0} & 1.2(10) \\
 & & & 4.2\phantom{0} & 3.0(21) & 5.8(20) & 3.4(17) & 1.0(14) & 3.9(14) & 5.5(15) & 1.6(12) & 6.2(12) \\
\hline
\end{tabular}
\end{table*}
}

\onltab{12}{
\begin{table*}
\caption{As \tb{trans1}, for $T_\el{gas}=50$ K.}
\label{tb:trans3}
\centering
\begin{tabular}{cccccccccccc}
\hline\hline
\rule{0pt}{1em}Model & $n_\el{H}$ & $\chi$ & $\av$ & $\cdmh$ & $\cdh$ & $N(\cocm{12})$ & $N(\coom{17})$ & $N(\coom{18})$ & $N(\cocm{13})$ & $N(\com{13}{17})$ & $N(\com{13}{18})$ \\
    & (\pcc)     &        & (mag) &         &        & & & & & & \\
\hline
\rule{0pt}{1em}C1$\alpha$ & \phantom{0}100 & \phantom{0}1 & 0.05 & 1.0(19) & 6.5(19) & 1.9(11) & 9.3(7)\phantom{0} & 3.6(8)\phantom{0} & 2.8(9)\phantom{0} & 1.4(6)\phantom{0} & 5.1(6)\phantom{0} \\
 & & & 0.10 & 3.0(19) & 1.0(20) & 8.5(11) & 4.1(8)\phantom{0} & 1.6(9)\phantom{0} & 1.2(10) & 6.2(6)\phantom{0} & 2.3(7)\phantom{0} \\
 & & & 0.24 & 1.0(20) & 1.8(20) & 5.5(12) & 2.5(9)\phantom{0} & 9.6(9)\phantom{0} & 7.7(10) & 3.8(7)\phantom{0} & 1.4(8)\phantom{0} \\
 & & & 0.61 & 3.0(20) & 3.6(20) & 6.9(13) & 2.5(10) & 9.1(10) & 7.5(11) & 3.6(8)\phantom{0} & 1.3(9)\phantom{0} \\
 & & & 1.5\phantom{0} & 1.0(21) & 4.5(20) & 4.5(14) & 1.3(11) & 4.4(11) & 7.6(12) & 2.6(9)\phantom{0} & 8.8(9)\phantom{0} \\
 & & & 4.1\phantom{0} & 3.0(21) & 4.6(20) & 1.2(17) & 5.4(13) & 1.9(14) & 2.5(15) & 1.2(12) & 4.2(12) \\
\rule{0pt}{1.2em}C1$\beta$ & \phantom{0}100 & \phantom{0}3 & 0.12 & 1.0(19) & 1.7(20) & 4.0(10) & 1.9(7)\phantom{0} & 7.4(7)\phantom{0} & 5.8(8)\phantom{0} & 2.9(5)\phantom{0} & 1.0(6)\phantom{0} \\
 & & & 0.19 & 3.0(19) & 2.4(20) & 2.0(11) & 9.5(7)\phantom{0} & 3.6(8)\phantom{0} & 2.9(9)\phantom{0} & 1.4(6)\phantom{0} & 5.2(6)\phantom{0} \\
 & & & 0.36 & 1.0(20) & 3.6(20) & 1.5(12) & 7.3(8)\phantom{0} & 2.8(9)\phantom{0} & 2.2(10) & 1.1(7)\phantom{0} & 4.1(7)\phantom{0} \\
 & & & 0.74 & 3.0(20) & 5.8(20) & 2.0(13) & 8.6(9)\phantom{0} & 3.2(10) & 2.6(11) & 1.3(8)\phantom{0} & 4.7(8)\phantom{0} \\
 & & & 1.7\phantom{0} & 1.0(21) & 7.1(20) & 1.6(14) & 5.5(10) & 1.9(11) & 2.5(12) & 9.8(8)\phantom{0} & 3.5(9)\phantom{0} \\
 & & & 4.3\phantom{0} & 3.0(21) & 7.2(20) & 7.1(16) & 3.2(13) & 1.2(14) & 1.6(15) & 7.5(11) & 2.7(12) \\
\rule{0pt}{1.2em}C1$\gamma$ & \phantom{0}100 & 10 & 0.31 & 1.0(19) & 4.6(20) & 1.2(10) & 5.7(6)\phantom{0} & 2.2(7)\phantom{0} & 1.7(8)\phantom{0} & 8.4(4)\phantom{0} & 3.1(5)\phantom{0} \\
 & & & 0.41 & 3.0(19) & 6.0(20) & 6.5(10) & 3.1(7)\phantom{0} & 1.2(8)\phantom{0} & 9.6(8)\phantom{0} & 4.7(5)\phantom{0} & 1.7(6)\phantom{0} \\
 & & & 0.62 & 1.0(20) & 7.8(20) & 6.4(11) & 3.1(8)\phantom{0} & 1.2(9)\phantom{0} & 9.5(9)\phantom{0} & 4.7(6)\phantom{0} & 1.7(7)\phantom{0} \\
 & & & 1.0\phantom{0} & 3.0(20) & 1.0(21) & 9.8(12) & 4.5(9)\phantom{0} & 1.7(10) & 1.4(11) & 6.8(7)\phantom{0} & 2.5(8)\phantom{0} \\
 & & & 2.0\phantom{0} & 1.0(21) & 1.2(21) & 5.5(13) & 2.1(10) & 7.9(10) & 8.5(11) & 3.7(8)\phantom{0} & 1.3(9)\phantom{0} \\
 & & & 4.5\phantom{0} & 3.0(21) & 1.2(21) & 4.5(16) & 2.1(13) & 7.4(13) & 1.1(15) & 5.1(11) & 1.8(12) \\
\hline
\rule{0pt}{1em}C2$\alpha$ & \phantom{0}300 & \phantom{0}1 & 0.03 & 1.0(19) & 2.6(19) & 9.1(11) & 4.3(8)\phantom{0} & 1.7(9)\phantom{0} & 1.3(10) & 6.5(6)\phantom{0} & 2.4(7)\phantom{0} \\
 & & & 0.07 & 3.0(19) & 4.8(19) & 3.8(12) & 1.8(9)\phantom{0} & 6.8(9)\phantom{0} & 5.4(10) & 2.7(7)\phantom{0} & 9.9(7)\phantom{0} \\
 & & & 0.20 & 1.0(20) & 1.1(20) & 2.4(13) & 9.6(9)\phantom{0} & 3.6(10) & 2.9(11) & 1.4(8)\phantom{0} & 5.3(8)\phantom{0} \\
 & & & 0.54 & 3.0(20) & 2.6(20) & 3.0(14) & 8.5(10) & 3.0(11) & 2.6(12) & 1.2(9)\phantom{0} & 4.3(9)\phantom{0} \\
 & & & 1.5\phantom{0} & 1.0(21) & 3.4(20) & 3.2(15) & 6.3(11) & 1.9(12) & 5.5(13) & 1.2(10) & 3.7(10) \\
 & & & 4.0\phantom{0} & 3.0(21) & 3.5(20) & 3.9(17) & 1.5(14) & 5.5(14) & 6.2(15) & 2.3(12) & 8.5(12) \\
\rule{0pt}{1.2em}C2$\beta$ & \phantom{0}300 & \phantom{0}3 & 0.05 & 1.0(19) & 6.5(19) & 1.6(11) & 7.8(7)\phantom{0} & 3.0(8)\phantom{0} & 2.4(9)\phantom{0} & 1.2(6)\phantom{0} & 4.3(6)\phantom{0} \\
 & & & 0.10 & 3.0(19) & 1.0(20) & 7.7(11) & 3.7(8)\phantom{0} & 1.4(9)\phantom{0} & 1.1(10) & 5.6(6)\phantom{0} & 2.1(7)\phantom{0} \\
 & & & 0.24 & 1.0(20) & 1.8(20) & 5.1(12) & 2.4(9)\phantom{0} & 9.0(9)\phantom{0} & 7.2(10) & 3.6(7)\phantom{0} & 1.3(8)\phantom{0} \\
 & & & 0.61 & 3.0(20) & 3.6(20) & 6.5(13) & 2.3(10) & 8.6(10) & 7.1(11) & 3.4(8)\phantom{0} & 1.3(9)\phantom{0} \\
 & & & 1.5\phantom{0} & 1.0(21) & 4.5(20) & 3.8(14) & 1.1(11) & 3.8(11) & 6.2(12) & 2.1(9)\phantom{0} & 7.3(9)\phantom{0} \\
 & & & 4.1\phantom{0} & 3.0(21) & 4.5(20) & 2.8(17) & 1.1(14) & 3.9(14) & 4.6(15) & 1.7(12) & 6.4(12) \\
\rule{0pt}{1.2em}C2$\gamma$ & \phantom{0}300 & 10 & 0.13 & 1.0(19) & 1.9(20) & 2.9(10) & 1.4(7)\phantom{0} & 5.3(7)\phantom{0} & 4.2(8)\phantom{0} & 2.1(5)\phantom{0} & 7.6(5)\phantom{0} \\
 & & & 0.20 & 3.0(19) & 2.6(20) & 1.6(11) & 7.5(7)\phantom{0} & 2.9(8)\phantom{0} & 2.3(9)\phantom{0} & 1.1(6)\phantom{0} & 4.2(6)\phantom{0} \\
 & & & 0.37 & 1.0(20) & 3.9(20) & 1.3(12) & 6.3(8)\phantom{0} & 2.4(9)\phantom{0} & 1.9(10) & 9.6(6)\phantom{0} & 3.5(7)\phantom{0} \\
 & & & 0.76 & 3.0(20) & 6.1(20) & 1.8(13) & 7.7(9)\phantom{0} & 2.9(10) & 2.3(11) & 1.1(8)\phantom{0} & 4.2(8)\phantom{0} \\
 & & & 1.7\phantom{0} & 1.0(21) & 7.3(20) & 1.1(14) & 4.0(10) & 1.5(11) & 1.7(12) & 7.1(8)\phantom{0} & 2.6(9)\phantom{0} \\
 & & & 4.3\phantom{0} & 3.0(21) & 7.3(20) & 1.9(17) & 7.0(13) & 2.6(14) & 3.3(15) & 1.2(12) & 4.6(12) \\
\hline
\rule{0pt}{1em}C5$\alpha$ & 1000 & \phantom{0}1 & 0.02 & 1.0(19) & 1.2(19) & 5.0(12) & 2.2(9)\phantom{0} & 8.7(9)\phantom{0} & 6.8(10) & 3.4(7)\phantom{0} & 1.2(8)\phantom{0} \\
 & & & 0.05 & 3.0(19) & 2.7(19) & 2.1(13) & 8.3(9)\phantom{0} & 3.2(10) & 2.6(11) & 1.3(8)\phantom{0} & 4.6(8)\phantom{0} \\
 & & & 0.17 & 1.0(20) & 7.2(19) & 1.4(14) & 4.2(10) & 1.6(11) & 1.3(12) & 6.1(8)\phantom{0} & 2.2(9)\phantom{0} \\
 & & & 0.48 & 3.0(20) & 1.7(20) & 1.8(15) & 3.7(11) & 1.1(12) & 1.2(13) & 4.4(9)\phantom{0} & 1.6(10) \\
 & & & 1.4\phantom{0} & 1.0(21) & 2.0(20) & 2.3(16) & 2.3(12) & 6.6(12) & 4.2(14) & 4.2(10) & 1.3(11) \\
 & & & 3.9\phantom{0} & 3.0(21) & 2.0(20) & 5.4(17) & 1.7(14) & 6.2(14) & 8.0(15) & 2.5(12) & 9.3(12) \\
\rule{0pt}{1.2em}C5$\beta$ & 1000 & \phantom{0}3 & 0.03 & 1.0(19) & 2.4(19) & 1.0(12) & 4.7(8)\phantom{0} & 1.8(9)\phantom{0} & 1.5(10) & 7.2(6)\phantom{0} & 2.7(7)\phantom{0} \\
 & & & 0.07 & 3.0(19) & 4.5(19) & 4.2(12) & 1.9(9)\phantom{0} & 7.6(9)\phantom{0} & 6.0(10) & 3.0(7)\phantom{0} & 1.1(8)\phantom{0} \\
 & & & 0.19 & 1.0(20) & 1.1(20) & 2.7(13) & 1.1(10) & 4.0(10) & 3.3(11) & 1.6(8)\phantom{0} & 5.9(8)\phantom{0} \\
 & & & 0.54 & 3.0(20) & 2.5(20) & 3.4(14) & 9.4(10) & 3.2(11) & 2.9(12) & 1.3(9)\phantom{0} & 4.7(9)\phantom{0} \\
 & & & 1.5\phantom{0} & 1.0(21) & 3.3(20) & 4.0(15) & 7.2(11) & 2.1(12) & 7.1(13) & 1.4(10) & 4.2(10) \\
 & & & 4.0\phantom{0} & 3.0(21) & 3.4(20) & 4.4(17) & 1.4(14) & 5.1(14) & 6.7(15) & 2.0(12) & 7.7(12) \\
\rule{0pt}{1.2em}C5$\gamma$ & 1000 & 10 & 0.05 & 1.0(19) & 6.5(19) & 1.5(11) & 7.3(7)\phantom{0} & 2.8(8)\phantom{0} & 2.2(9)\phantom{0} & 1.1(6)\phantom{0} & 4.0(6)\phantom{0} \\
 & & & 0.10 & 3.0(19) & 1.0(20) & 7.4(11) & 3.5(8)\phantom{0} & 1.4(9)\phantom{0} & 1.1(10) & 5.4(6)\phantom{0} & 2.0(7)\phantom{0} \\
 & & & 0.24 & 1.0(20) & 1.8(20) & 5.0(12) & 2.3(9)\phantom{0} & 8.8(9)\phantom{0} & 7.1(10) & 3.5(7)\phantom{0} & 1.3(8)\phantom{0} \\
 & & & 0.61 & 3.0(20) & 3.6(20) & 6.4(13) & 2.3(10) & 8.4(10) & 7.0(11) & 3.4(8)\phantom{0} & 1.2(9)\phantom{0} \\
 & & & 1.5\phantom{0} & 1.0(21) & 4.5(20) & 3.5(14) & 1.0(11) & 3.5(11) & 5.7(12) & 1.9(9)\phantom{0} & 6.8(9)\phantom{0} \\
 & & & 4.1\phantom{0} & 3.0(21) & 4.5(20) & 3.3(17) & 9.8(13) & 3.7(14) & 5.1(15) & 1.5(12) & 5.7(12) \\
\hline
\end{tabular}
\end{table*}
}

\onltab{13}{
\begin{table*}
\caption{As \tb{trans1}, for $T_\el{gas}=100$ K.}
\label{tb:trans4}
\centering
\begin{tabular}{cccccccccccc}
\hline\hline
\rule{0pt}{1em}Model & $n_\el{H}$ & $\chi$ & $\av$ & $\cdmh$ & $\cdh$ & $N(\cocm{12})$ & $N(\coom{17})$ & $N(\coom{18})$ & $N(\cocm{13})$ & $N(\com{13}{17})$ & $N(\com{13}{18})$ \\
    & (\pcc)     &        & (mag) &         &        & & & & & & \\
\hline
\rule{0pt}{1em}D1$\alpha$ & \phantom{0}100 & \phantom{0}1 & 0.04 & 1.0(19) & 4.8(19) & 3.5(11) & 1.6(8)\phantom{0} & 6.4(8)\phantom{0} & 5.0(9)\phantom{0} & 2.5(6)\phantom{0} & 9.1(6)\phantom{0} \\
 & & & 0.09 & 3.0(19) & 7.8(19) & 1.5(12) & 7.1(8)\phantom{0} & 2.8(9)\phantom{0} & 2.2(10) & 1.1(7)\phantom{0} & 4.0(7)\phantom{0} \\
 & & & 0.22 & 1.0(20) & 1.6(20) & 9.5(12) & 4.2(9)\phantom{0} & 1.6(10) & 1.3(11) & 6.3(7)\phantom{0} & 2.3(8)\phantom{0} \\
 & & & 0.59 & 3.0(20) & 3.3(20) & 1.2(14) & 3.9(10) & 1.4(11) & 1.2(12) & 5.5(8)\phantom{0} & 2.0(9)\phantom{0} \\
 & & & 1.5\phantom{0} & 1.0(21) & 4.2(20) & 8.7(14) & 2.3(11) & 7.4(11) & 1.2(13) & 3.6(9)\phantom{0} & 1.2(10) \\
 & & & 4.0\phantom{0} & 3.0(21) & 4.2(20) & 2.7(17) & 1.2(14) & 4.3(14) & 4.1(15) & 1.8(12) & 6.6(12) \\
\rule{0pt}{1.2em}D1$\beta$ & \phantom{0}100 & \phantom{0}3 & 0.09 & 1.0(19) & 1.3(20) & 6.7(10) & 3.2(7)\phantom{0} & 1.2(8)\phantom{0} & 9.6(8)\phantom{0} & 4.8(5)\phantom{0} & 1.8(6)\phantom{0} \\
 & & & 0.15 & 3.0(19) & 1.8(20) & 3.4(11) & 1.6(8)\phantom{0} & 6.3(8)\phantom{0} & 5.0(9)\phantom{0} & 2.4(6)\phantom{0} & 9.0(6)\phantom{0} \\
 & & & 0.31 & 1.0(20) & 2.9(20) & 2.5(12) & 1.2(9)\phantom{0} & 4.5(9)\phantom{0} & 3.6(10) & 1.8(7)\phantom{0} & 6.6(7)\phantom{0} \\
 & & & 0.69 & 3.0(20) & 5.0(20) & 3.2(13) & 1.3(10) & 4.8(10) & 3.9(11) & 1.9(8)\phantom{0} & 6.9(8)\phantom{0} \\
 & & & 1.7\phantom{0} & 1.0(21) & 6.5(20) & 2.9(14) & 9.0(10) & 3.1(11) & 3.5(12) & 1.4(9)\phantom{0} & 4.9(9)\phantom{0} \\
 & & & 4.2\phantom{0} & 3.0(21) & 6.5(20) & 1.6(17) & 7.0(13) & 2.5(14) & 2.7(15) & 1.2(12) & 4.2(12) \\
\rule{0pt}{1.2em}D1$\gamma$ & \phantom{0}100 & 10 & 0.24 & 1.0(19) & 3.6(20) & 1.6(10) & 7.8(6)\phantom{0} & 3.0(7)\phantom{0} & 2.3(8)\phantom{0} & 1.1(5)\phantom{0} & 4.2(5)\phantom{0} \\
 & & & 0.33 & 3.0(19) & 4.7(20) & 9.4(10) & 4.5(7)\phantom{0} & 1.7(8)\phantom{0} & 1.4(9)\phantom{0} & 6.8(5)\phantom{0} & 2.5(6)\phantom{0} \\
 & & & 0.53 & 1.0(20) & 6.3(20) & 9.1(11) & 4.4(8)\phantom{0} & 1.7(9)\phantom{0} & 1.3(10) & 6.7(6)\phantom{0} & 2.4(7)\phantom{0} \\
 & & & 0.93 & 3.0(20) & 8.7(20) & 1.3(13) & 5.9(9)\phantom{0} & 2.2(10) & 1.8(11) & 8.9(7)\phantom{0} & 3.2(8)\phantom{0} \\
 & & & 1.9\phantom{0} & 1.0(21) & 1.0(21) & 1.1(14) & 3.9(10) & 1.4(11) & 1.4(12) & 6.0(8)\phantom{0} & 2.2(9)\phantom{0} \\
 & & & 4.5\phantom{0} & 3.0(21) & 1.1(21) & 1.0(17) & 4.5(13) & 1.6(14) & 1.8(15) & 7.7(11) & 2.8(12) \\
\hline
\rule{0pt}{1em}D2$\alpha$ & \phantom{0}300 & \phantom{0}1 & 0.03 & 1.0(19) & 2.0(19) & 1.6(12) & 7.6(8)\phantom{0} & 3.0(9)\phantom{0} & 2.3(10) & 1.2(7)\phantom{0} & 4.3(7)\phantom{0} \\
 & & & 0.06 & 3.0(19) & 4.0(19) & 6.7(12) & 3.0(9)\phantom{0} & 1.2(10) & 9.2(10) & 4.6(7)\phantom{0} & 1.7(8)\phantom{0} \\
 & & & 0.19 & 1.0(20) & 9.8(19) & 4.3(13) & 1.6(10) & 5.9(10) & 4.8(11) & 2.3(8)\phantom{0} & 8.6(8)\phantom{0} \\
 & & & 0.52 & 3.0(20) & 2.3(20) & 5.2(14) & 1.3(11) & 4.5(11) & 4.1(12) & 1.8(9)\phantom{0} & 6.4(9)\phantom{0} \\
 & & & 1.5\phantom{0} & 1.0(21) & 3.1(20) & 6.1(15) & 9.6(11) & 2.8(12) & 7.8(13) & 1.4(10) & 4.5(10) \\
 & & & 4.0\phantom{0} & 3.0(21) & 3.2(20) & 4.3(17) & 1.6(14) & 5.8(14) & 6.4(15) & 2.3(12) & 8.5(12) \\
\rule{0pt}{1.2em}D2$\beta$ & \phantom{0}300 & \phantom{0}3 & 0.04 & 1.0(19) & 4.8(19) & 3.1(11) & 1.5(8)\phantom{0} & 5.8(8)\phantom{0} & 4.6(9)\phantom{0} & 2.3(6)\phantom{0} & 8.3(6)\phantom{0} \\
 & & & 0.09 & 3.0(19) & 7.8(19) & 1.4(12) & 6.8(8)\phantom{0} & 2.6(9)\phantom{0} & 2.1(10) & 1.0(7)\phantom{0} & 3.8(7)\phantom{0} \\
 & & & 0.22 & 1.0(20) & 1.5(20) & 9.1(12) & 4.1(9)\phantom{0} & 1.5(10) & 1.2(11) & 6.1(7)\phantom{0} & 2.3(8)\phantom{0} \\
 & & & 0.59 & 3.0(20) & 3.2(20) & 1.1(14) & 3.8(10) & 1.4(11) & 1.1(12) & 5.4(8)\phantom{0} & 2.0(9)\phantom{0} \\
 & & & 1.5\phantom{0} & 1.0(21) & 4.2(20) & 6.6(14) & 1.7(11) & 5.8(11) & 8.1(12) & 2.7(9)\phantom{0} & 9.3(9)\phantom{0} \\
 & & & 4.0\phantom{0} & 3.0(21) & 4.2(20) & 3.2(17) & 1.2(14) & 4.3(14) & 4.8(15) & 1.8(12) & 6.4(12) \\
\rule{0pt}{1.2em}D2$\gamma$ & \phantom{0}300 & 10 & 0.10 & 1.0(19) & 1.4(20) & 5.2(10) & 2.5(7)\phantom{0} & 9.5(7)\phantom{0} & 7.5(8)\phantom{0} & 3.7(5)\phantom{0} & 1.4(6)\phantom{0} \\
 & & & 0.16 & 3.0(19) & 2.0(20) & 2.8(11) & 1.4(8)\phantom{0} & 5.2(8)\phantom{0} & 4.1(9)\phantom{0} & 2.0(6)\phantom{0} & 7.5(6)\phantom{0} \\
 & & & 0.32 & 1.0(20) & 3.1(20) & 2.2(12) & 1.0(9)\phantom{0} & 4.0(9)\phantom{0} & 3.2(10) & 1.6(7)\phantom{0} & 5.8(7)\phantom{0} \\
 & & & 0.71 & 3.0(20) & 5.2(20) & 2.8(13) & 1.2(10) & 4.3(10) & 3.5(11) & 1.7(8)\phantom{0} & 6.2(8)\phantom{0} \\
 & & & 1.7\phantom{0} & 1.0(21) & 6.6(20) & 2.0(14) & 6.4(10) & 2.3(11) & 2.3(12) & 9.7(8)\phantom{0} & 3.5(9)\phantom{0} \\
 & & & 4.2\phantom{0} & 3.0(21) & 6.7(20) & 2.2(17) & 8.1(13) & 3.0(14) & 3.5(15) & 1.2(12) & 4.6(12) \\
\hline
\rule{0pt}{1em}D5$\alpha$ & 1000 & \phantom{0}1 & 0.02 & 1.0(19) & 9.7(18) & 8.4(12) & 3.6(9)\phantom{0} & 1.4(10) & 1.1(11) & 5.4(7)\phantom{0} & 2.0(8)\phantom{0} \\
 & & & 0.05 & 3.0(19) & 2.3(19) & 3.6(13) & 1.3(10) & 5.1(10) & 4.1(11) & 2.0(8)\phantom{0} & 7.3(8)\phantom{0} \\
 & & & 0.17 & 1.0(20) & 6.3(19) & 2.4(14) & 6.6(10) & 2.4(11) & 2.1(12) & 9.2(8)\phantom{0} & 3.4(9)\phantom{0} \\
 & & & 0.47 & 3.0(20) & 1.4(20) & 3.2(15) & 5.8(11) & 1.7(12) & 1.9(13) & 6.6(9)\phantom{0} & 2.3(10) \\
 & & & 1.4\phantom{0} & 1.0(21) & 1.7(20) & 3.6(16) & 3.0(12) & 8.5(12) & 4.7(14) & 4.4(10) & 1.4(11) \\
 & & & 3.9\phantom{0} & 3.0(21) & 1.7(20) & 5.6(17) & 1.7(14) & 6.2(14) & 8.1(15) & 2.4(12) & 9.1(12) \\
\rule{0pt}{1.2em}D5$\beta$ & 1000 & \phantom{0}3 & 0.02 & 1.0(19) & 1.9(19) & 1.8(12) & 8.6(8)\phantom{0} & 3.4(9)\phantom{0} & 2.6(10) & 1.3(7)\phantom{0} & 4.8(7)\phantom{0} \\
 & & & 0.06 & 3.0(19) & 3.8(19) & 7.6(12) & 3.4(9)\phantom{0} & 1.3(10) & 1.0(11) & 5.1(7)\phantom{0} & 1.9(8)\phantom{0} \\
 & & & 0.19 & 1.0(20) & 9.4(19) & 4.9(13) & 1.8(10) & 6.6(10) & 5.4(11) & 2.6(8)\phantom{0} & 9.6(8)\phantom{0} \\
 & & & 0.52 & 3.0(20) & 2.2(20) & 5.9(14) & 1.5(11) & 5.0(11) & 4.6(12) & 2.0(9)\phantom{0} & 7.1(9)\phantom{0} \\
 & & & 1.5\phantom{0} & 1.0(21) & 3.0(20) & 7.3(15) & 1.0(12) & 3.0(12) & 9.1(13) & 1.5(10) & 4.8(10) \\
 & & & 4.0\phantom{0} & 3.0(21) & 3.0(20) & 4.6(17) & 1.4(14) & 5.1(14) & 6.8(15) & 2.0(12) & 7.5(12) \\
\rule{0pt}{1.2em}D5$\gamma$ & 1000 & 10 & 0.04 & 1.0(19) & 4.8(19) & 3.0(11) & 1.5(8)\phantom{0} & 5.6(8)\phantom{0} & 4.4(9)\phantom{0} & 2.2(6)\phantom{0} & 8.1(6)\phantom{0} \\
 & & & 0.09 & 3.0(19) & 7.8(19) & 1.4(12) & 6.7(8)\phantom{0} & 2.6(9)\phantom{0} & 2.0(10) & 1.0(7)\phantom{0} & 3.7(7)\phantom{0} \\
 & & & 0.22 & 1.0(20) & 1.5(20) & 9.0(12) & 4.0(9)\phantom{0} & 1.5(10) & 1.2(11) & 6.0(7)\phantom{0} & 2.2(8)\phantom{0} \\
 & & & 0.58 & 3.0(20) & 3.2(20) & 1.1(14) & 3.7(10) & 1.3(11) & 1.1(12) & 5.3(8)\phantom{0} & 1.9(9)\phantom{0} \\
 & & & 1.5\phantom{0} & 1.0(21) & 4.2(20) & 5.8(14) & 1.6(11) & 5.2(11) & 6.9(12) & 2.4(9)\phantom{0} & 8.3(9)\phantom{0} \\
 & & & 4.0\phantom{0} & 3.0(21) & 4.2(20) & 3.4(17) & 9.8(13) & 3.7(14) & 5.0(15) & 1.4(12) & 5.5(12) \\
\hline
\end{tabular}
\end{table*}
}

The full set of points already gives some indication that the data are not well represented by a global single-slope correlation. The right panel of \fig{co-vs-h2} shows the 100-point means of the full set, revealing the same two distinct regimes as found by \citet{sheffer08a}. Taking the uncertainties in both the observations and the models into account, the two sets of power-law fits (indicated by the black and grey lines) are identical. The break between the two slopes occurs at $\cdmh=\scim{2.5}{20}$ \pcs{} and is due to the switch from suprathermal to normal chemistry. It is clearly unrelated to the switch from UV photons to \element[+]{He} or H$_3^+$ as the main destroyer of \coc{12}, which does not occur until $\cdmh=1$--\scit{2}{21} \pcs. When we extend the grid to higher $\av$ (not plotted), we quickly reach the point where the \coc{12} abundance equals the elemental \element[][12]{C} abundance and the plot of $N(\cocm{12})$ versus $\cdmh$ continues with a slope of unity. The model results from \vdb{} (dashed lines, corresponding to $\chi=0.5$, 1 and 10) do not extend to low enough $\cdmh$ to show the suprathermal regime, but they do show at the upper end the transition to the regime where all gas-phase carbon is in CO.


\subsubsection{\coc{13}}
\label{subsec:trans13co}
With our model able to reproduce the observational relationship between \mh{} and \coc{12}, the next step is to look at the heavier isotopologues. \citet{sonnentrucker07a}, \citet{burgh07a} and \citet{sheffer07a} together presented a sample of 29 diffuse and translucent sources with derived column densities for both \coc{12} and \coc{13}. These data are plotted in the left panel of \fig{cd13co} (grey crosses) as $N(\cocm{13})$ versus $\cdmh$. The 100-point means from our model grid (black squares) agree to the observations to within a factor of three. The model results are again best fitted with two power laws, but the \coc{13} observations do not extend to low enough $\cdmh$ to confirm this.

The observations show considerable scatter in the ratio between $N(\cocm{12})$ and $N(\cocm{13})$, with some sources deviating by a factor of two either way from the local elemental \element[][12]{C}--\element[][13]{C} ratio of $69\pm6$ \citep{wilson99a}. It has long been assumed that the enhanced ratios are due to isotope-selective photodissociation and that the reduced ratios are due to isotope-exchange reactions. However, no answer has been found so far to the question of why one process dominates in some sources, and the other process in other sources \citep{langer80a,mccutcheon80a,sheffer92a,lambert94a,federman03a,sonnentrucker07a,sheffer07a}. Chemical models consistently show that isotope exchange is more efficient than selective photodissociation for observed cloud temperatures of up to $\sim$60 K. The models can easily reproduce $N(\cocm{12})/N(\cocm{13})$ ratios of less than 69, but they leave the ratios of more than 69 unexplained.

Our grid of models suggests that the answer lies in suprathermal chemistry. The right panel of \fig{cd13co} shows a plot of $N(\cocm{12})/N(\cocm{13})$ versus $\cdmh$ for the observations (grey) and for all depth steps at $\av\leq5$ from our models (black plus signs). Also shown are the 100-point means from the model, offset by a factor of three for clarity. The $N(\cocm{12})/N(\cocm{13})$ ratio initially increases from 69 to a mean of 151 at $\cdmh=\scim{4}{20}$ \pcs, then decreases to a mean of 51 at \scit{1}{21} \pcs, and gradually increases again for larger depths. The turnover from the initial rise to the rapid drop is due to the transition from suprathermal to normal chemistry.

At an Alfv\'en speed of $v_\el{A}=3.3$ km \ps, the effective temperature for \rx{imex1} is about 4000 K, so the forward and backward reactions are equally fast. Photodissociation is therefore the only active fractionation process and it enhances the amount of \coc{12} relative to \coc{13}. As we move beyond the depth to which the Alfv\'en waves reach ($\cdmh=\scim{4}{20}$ \pcs{} in our model), suprathermal chemistry is brought to a halt. The forward channel of \rx{imex1} becomes faster than the backward channel and isotope exchange now reverses the fractionation. When we reach a depth of $\cdmh=1$--\scit{2}{21} \pcs, the \element[+][13]{C} abundance drops too low to sustain further fractionation and the $N(\cocm{12})/N(\cocm{13})$ ratio gradually returns to the elemental ratio of 69.

The observations show a downward trend in the ratio between $N(\cocm{12})$ and $N(\cocm{13})$ in the $\cdmh$ range from \scit{2}{20} to \scit{2}{21} \pcs. This supports a rapid switch from selective photodissociation to isotope-exchange reactions being the dominant fractionation mechanism. A firm test of our model predictions requires observational data at both smaller and larger $\cdmh$ than are currently available.


\subsubsection{Other isotopologues}
\label{subsec:transother}
Moving to the next two isotopologues, \coo{17} and \coo{18}, the number of observed column densities goes down to three. \citet{lambert94a} determined an $N(\cocm{12})/N(\coom{18})$ ratio of $1550\pm440$ for the \zoph{} diffuse cloud, some three times larger than the elemental \element[][16]{O}--\element[][18]{O} ratio of $557\pm30$ \citep{wilson99a}. An even higher ratio of $3000\pm600$ was recorded by \citet{sheffer02a} for the X Per translucent cloud, along with an $N(\cocm{12})/N(\coom{17})$ ratio of $8700\pm3600$. The elemental \element[][16]{O}--\element[][18]{O} ratio is about four times smaller: $2000\pm200$.

The \mh{} column density towards X Per (HD 24534) is \scit{8.5}{20} \pcs{} \citep{sonnentrucker07a,sheffer07a}. For this value, our grid of models gives $N(\cocm{12})/N(\coom{18})=680$--3880 (median 1080) and $N(\cocm{12})/N(\coom{17})=2520$--10\,600 (median 3660), consistent with the observations. Out of the three parameters that we varied, $\chi$ affects the ratios most strongly. Both ratios decrease by a factor of two when $\chi$ is increased from 1 to 10. The onset of self-shielding of \coc{12}, and with it the onset of isotopic fractionation, occurs deeper into the cloud for a stronger UV field. Therefore, $N(\cocm{12})/N(\coom{18})$ and $N(\cocm{12})/N(\coom{17})$ are smaller at any given depth.

In order to further explore the behaviour of \coo{17} and \coo{18}, we take a more detailed look at our results for $n_\el{H}=300$ \pcc, $T_\el{gas}=50$ K and $\chi=1$ (model C2$\alpha$ from \tb{trans3}). This also allows for a comparison with \co{13}{17} and \co{13}{18}. Both of these have been detected in interstellar clouds \citep{langer80a,bensch01a}, but column densities have not yet been derived.

\begin{figure}
\resizebox{\hsize}{!}{\includegraphics{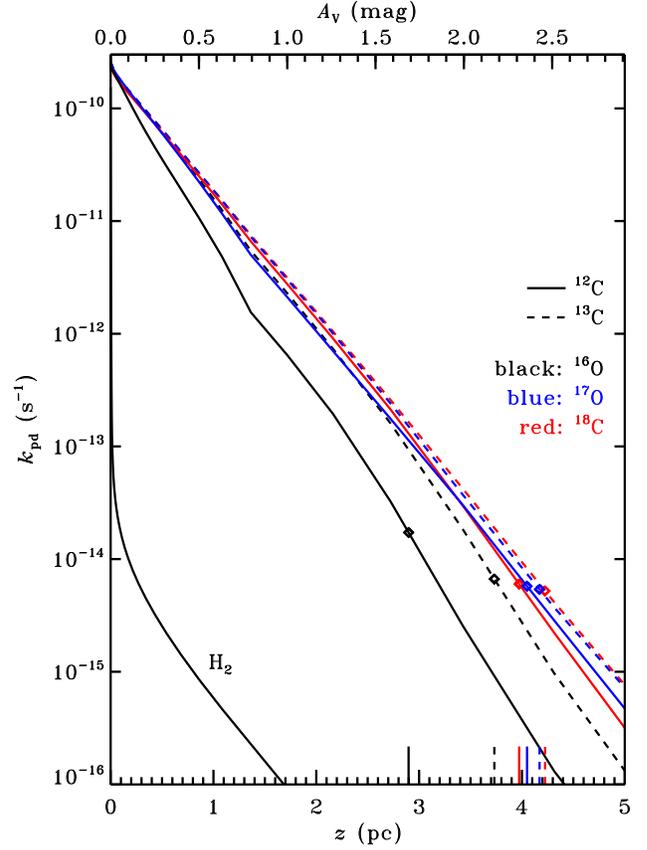}}
\caption{Photodissociation rates of \mh{} and the six CO isotopologues as a function of depth into translucent cloud model C2$\alpha$ with $n_\el{H}=300$ \pcc, $T_\el{gas}=50$ K and $\chi=1$. Attenuation by dust is included. The diamonds and corresponding markers on the bottom axis indicate the depth at which \element[+]{He} takes over from UV radiation as the dominant destroyer for each isotopologue.}
\label{fig:depth}
\end{figure}

The photodissociation rates of CO and \mh{} for this combination of parameters are plotted in \fig{depth} as a function of depth ($z$) into the cloud. The isotope-selective nature of the photodissociation is clearly visible. At a depth of 4.0 pc ($\av=2.3$ mag, $\cdmh=\scim{1.7}{21}$ \pcs), \coc{12} photodissociates about 7 times slower than \coc{13}, 15--18 times slower than \coo{17} and \coo{18}, and 23--26 times slower than \co{13}{17} and \co{13}{18}. At depths between 0.4 and 3.4 pc ($\av=0.3$--2.0 mag), the photodissociation of \coo{18} proceeds faster than that of \coo{17}. The lines of the latter lie closer to those of \coc{12} and are therefore more strongly shielded, as was the case for the diffuse \zoph{} cloud (\tb{zophrate}). As the CO column grows larger in our model cloud, \coo{18} becomes self-shielding and its rate drops below that of \coo{17} around 3.4 pc.

\begin{figure}
\resizebox{\hsize}{!}{\includegraphics{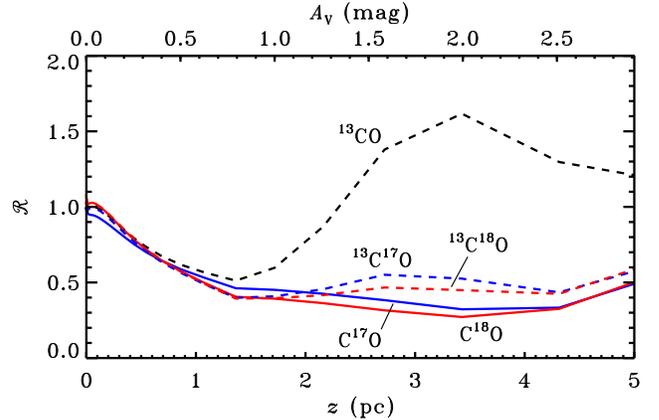}}
\caption{Cumulative column densities of the CO isotopologues normalised to \coc{12} and the elemental isotopic abundances (\eq{ncdr}) as a function of depth into translucent cloud model C2$\alpha$.}
\label{fig:ncdr}
\end{figure}

The extent to which each isotopologue is fractionated can easily be seen from the cumulative column density ratios when they are normalised to \coc{12} and the isotope ratios:
\begin{equation}
\label{eq:ncdr}
\mathcal{R}_i(z) = \frac{N_z(\com{x}{y})}{N_z(\cocm{12})} \frac{[\element[][12]{C}]}{[\element[][x]{C}]} \frac{[\element[][16]{O}]}{[\element[][y]{O}]}\,,
\end{equation}
with $[X]$ the elemental abundance of isotope $X$ and
\begin{equation}
\label{eq:ccd}
N_z(\com{x}{y})(z) = \int_0^z n(\com{x}{y})\el{d}z'\,
\end{equation}
the cumulative column density. A plot of $\mathcal{R}(\cocm{13})$ as function of depth (\fig{ncdr}) shows the same trends as the model data in \fig{cd13co}: the amount of \coc{13} is reduced ($\mathcal{R}<1$) up to a certain depth and is enhanced ($\mathcal{R}>1$) further in by isotope-exchange reactions. We only see $\mathcal{R}<1$ for \coo{17} and \coo{18} because \element[+][17]{O} and \element[+][18]{O} never become abundant enough to convert \coo{16} in a reaction similar to \rx{imex1}. Photodissociation ceases to be the dominant destruction mechanism for \coo{17} and \coo{18} around $z=4$ pc (\fig{depth}), so it can no longer cause substantial fractionation and $\mathcal{R}$ gradually returns to unity.

The remaining two isotopologues, \co{13}{17} and \co{13}{18}, experience the \element[][13]{C} and \element[][17]{O}/\element[][18]{O} fractionation simultaneously. Isotope exchange enhances their abundances relative to \coo{17} and \coo{18} in the same manner and at the same depths as it enhances the abundance of \coc{13} relative to \coc{12}. However, isotope-selective photodissociation is a stronger effect for the two heaviest isotopologues, so $\mathcal{R}$ remains less than unity. This result holds across our entire grid of translucent cloud models.


\subsection{Photon-dominated regions}
\label{subsec:pdr}
Photon-dominated regions (PDRs) form another type of objects where photodissociation is a key process in controlling the CO abundance. Their densities are higher ($n_\el{H}$ up to $\sim$\ten{6} \pcc) than those of diffuse and translucent clouds and they are exposed to stronger UV fields ($\chi$ up to $\sim$\ten{6}; \citealt{hollenbach97a}). Observational efforts have mostly gone into mapping the structure of PDRs, so CO column densities are tabulated only for a small number of sources. For most of these the column density of neutral carbon has also been determined. There are no direct measurements of \mh{} column densities; instead, $\cdmh$ is usually obtained from $N(\cocm{12})$ using a typical abundance ratio between the two species \citep[e.g.,][]{frerking82a}.

We present here the results from our combined photodissociation and chemistry model for a grid of physical parameters appropriate for PDRs: $T_\el{gas}=50$, 80 and 100 K, $n_\el{H}=10^3$, \ten{4}, \ten{5} and \ten{6} \pcc, and $\chi=10^3$, \ten{4} and \ten{5}. The excitation temperatures of CO and \mh{} are set equal to $T_\el{gas}$. Where necessary, we expand the number of rotational levels of CO beyond the default limit of $J''$=7, and likewise for \mh{} (see Sect.\ \ref{subsec:btexgrid}). UV pumping increases the population in the $v''$=0, $J''$>3 and $v''$>0 levels of \mh{}. However, this does not affect the CO photodissociation rates at these temperatures, so the level population of \mh{} is simply taken as fully thermal. The models are run to an $\av$ of 30 mag and are otherwise unchanged from the previous section.

\begin{figure*}
\resizebox{\hsize}{!}{\includegraphics{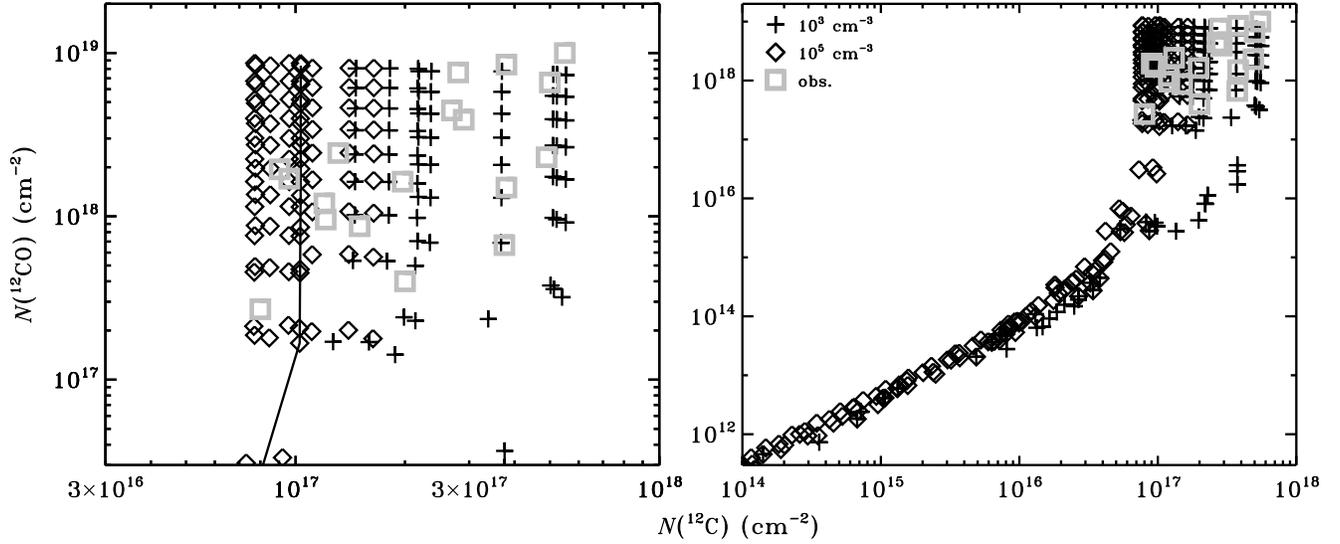}}
\caption{Column density of \coc{12} versus that of \element[][12]{C} on two different scales. Black: results from our grid of PDR models with $n_\el{H}=10^3$ \pcc{} (plus signs) and \ten{5} \pcc{} (diamonds). The line traces the results for different depth steps at $n_\el{H}=10^5$ \pcc{}, $T_\el{gas}=50$ K and $\chi=10^4$. Grey squares: observations of \citet{beuther00a}, \citet{kamegai03a}, \citet{kramer08a} and \citet{sun08a}.}
\label{fig:c-vs-co}
\end{figure*}

Figure \ref{fig:c-vs-co} shows the relationship between the column densities of \coc{12} and \element[][12]{C} for a large series of depths steps from our grid. The data for $n_\el{H}=10^4$ and \ten{6} \pcc{} are omitted for reasons of clarity. Overplotted are column densities determined towards a number of positions in different PDRs \citep{beuther00a,kamegai03a,kramer08a,sun08a}. Figure 5a of \citet{mookerjea06a} contains additional data for several dozen positions in the PDRs in \object{Cepheus B}. In none of these cases was the \coc{12} column density measured directly; instead, it was derived from the \coc{13} or \coo{18} column density using standard abundance ratios. The spread in both the model predictions and the observational data is large. The column of \element[][12]{C} for each individual model is nearly constant for $N(\cocm{12})>\scim{3}{17}$ \pcs, as shown by the solid line connecting the various depth steps for $n_\el{H}=10^5$ \pcc{}, $T_\el{gas}=50$ K and $\chi=10^4$. The observations seem to correspond more to the lower-density end of the model grid (the plus signs), but the sample is too small and is too scattered to be conclusive.

\begin{figure*}
\resizebox{\hsize}{!}{\includegraphics{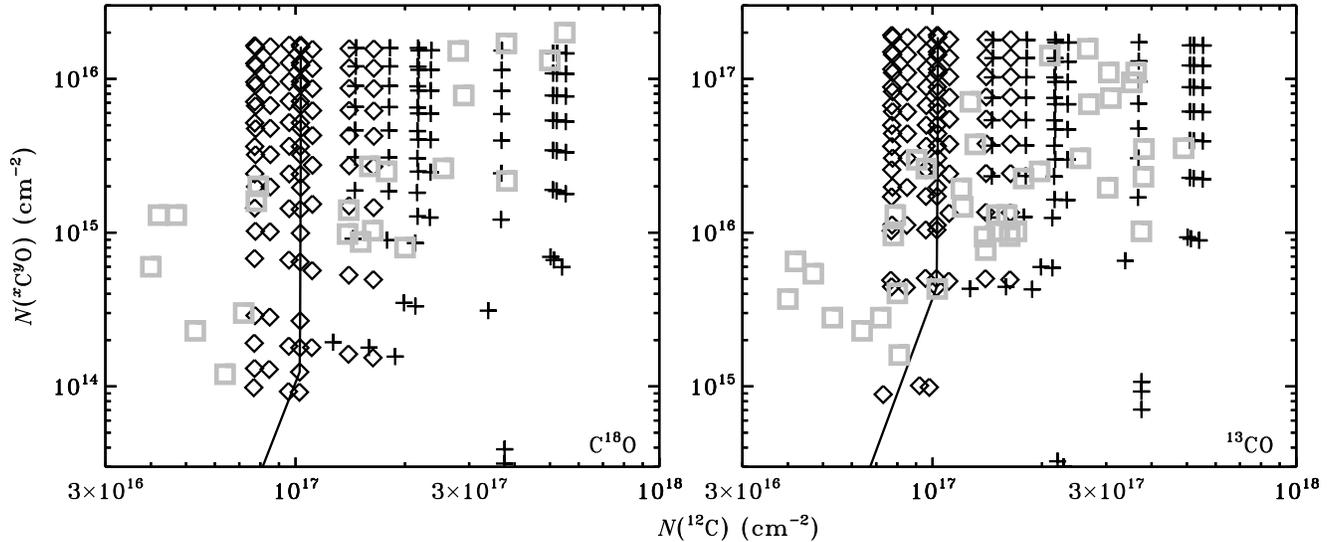}}
\caption{Column densities of \coo{18} (\emph{left}) and \coc{13} (\emph{right}) versus that of \element[][12]{C}. Symbols are as in \fig{c-vs-co}. See text for references to the observations.}
\label{fig:pdr-iso}
\end{figure*}

The onset of shielding is seen to occur around $N(\cocm{12})=10^{15}$ \pcs{} in the models. From that point onwards, the low-density models gain more atomic carbon than do the high-density models. For a given column of \coc{12}, the photodissociation rates (or, in other words, the \element[][12]{C} formation rates) are roughly independent of density, but the destruction of \element[][12]{C} occurs faster in higher densities. The same argument explained the density dependence of the $N(\cocm{12})/\cdmh$ ratio in the translucent clouds.

The two isotopologues for which column densities have been determined directly are \coo{18} and \coc{13} \citep{loren89a,gerin98a,plume99a,beuther00a,schneider03a,mookerjea06a,kramer08a,sun08a}. Again, both the observations and the models show a large spread in column densities (\fig{pdr-iso}). It seems the observations do not trace the high-extinction ($\av>10$ mag), high-density ($n_\el{H}>10^5$ \pcc) material shown in the upper left corner of each panel, but a more detailed study is required to draw any firm conclusions on this point. $N(\cocm{12})$, $N(\coom{18})$ and $N(\cocm{13})$ are mutually well correlated in both the observations and the models, but significant deviations can easily occur at specific depths due to the different photodissociation rates. When using standard abundance ratios to derive the column density of \coc{12} from that of another isotopologue, as was done for all cases in \fig{c-vs-co}, the result will generally be accurate to at best a factor of two.


\subsection{Circumstellar disks}
\label{subsec:disk}


\subsubsection{Model setup}
\label{subsec:diskmod}
Low-mass stars like our own Sun are formed through the gravitational collapse of a cold molecular cloud. As the young star grows at the centre, it gathers part of the remaining cloud material into a circumstellar disk. It is inside this disk that planets are formed, so its chemical composition is of great interest. The physical structure of the disk may be simplified as consisting of a cold, dense region near the midplane, covered by a warmer region of lower density. This surface layer intercepts the star's radiation and shows many similarities to common PDRs.

We adopt the standard disk model of \citet{dalessio99a}, whose chemistry has been studied extensively \citep{aikawa02a,vanzadelhoff03a,jonkheid04a}. The star, a T Tauri type, has a mass of 0.5 $M_\odot$, a radius of 2 $R_\odot$ and an effective temperature of 4000 K. It is surrounded by a disk of mass 0.07 $M_\odot$ and outer radius 400 AU. The disk is accreting onto the star at a constant rate of \ten{-8} $M_\odot$ yr$^{-1}$ and its viscosity is characterised by $\alpha=0.01$. We focus on a vertical slice through the disk at a radius of 105 AU, where the surface is located at a height of $z=120$ AU. This slice is irradiated from the top by a radiation field with an effective UV intensity of 516 times that of the interstellar radiation field. Following \citeauthor{vanzadelhoff03a}, we adopt two spectral shapes: the \citeauthor{draine78a} field and a 4000 K blackbody spectrum. The latter is very weak in the wavelength range where CO is photodissociated. We adopt again the chemical network from Sect. \ref{subsec:transmod}, and we set the gas temperature equal to the dust temperature from the \citeauthor{dalessio99a} model. Given the high densities (\scit{2}{4} \pcc{} at the surface and more than \ten{9} \pcc{} at the midplane), we also use $T_\el{gas}$ for $\texco$ and $\texmh$. As in the PDR models, UV pumping of \mh{} does not affect the CO photodissociation rates, so the \mh{} level populations are taken as fully thermal.

In reality, disks are not irradiated by the \citeauthor{draine78a} field or a pure blackbody. For example, T Tauri stars typically show UV emission in excess of these simple model spectra. Many of them also emit X-rays from accretion hotspots. As shown by \citet{vanzadelhoff03a}, the photochemistry of the disk can be sensitive to the details of the radiation field. Moreover, the excess emission can be concentrated in a few specific lines such as \ion{H}{i} Lyman-$\alpha$. CO cannot be dissociated by Ly-$\alpha$, but it can be by the \ion{H}{i} Ly-$\gamma$ line at 972.54 \AA{} through overlap with the $W0$ band at 972.70 \AA{} (No. 25 from \tb{moldata-ls}). A detailed treatment of these effects is beyond the scope of the current work.

\begin{figure*}
\resizebox{\hsize}{!}{\includegraphics{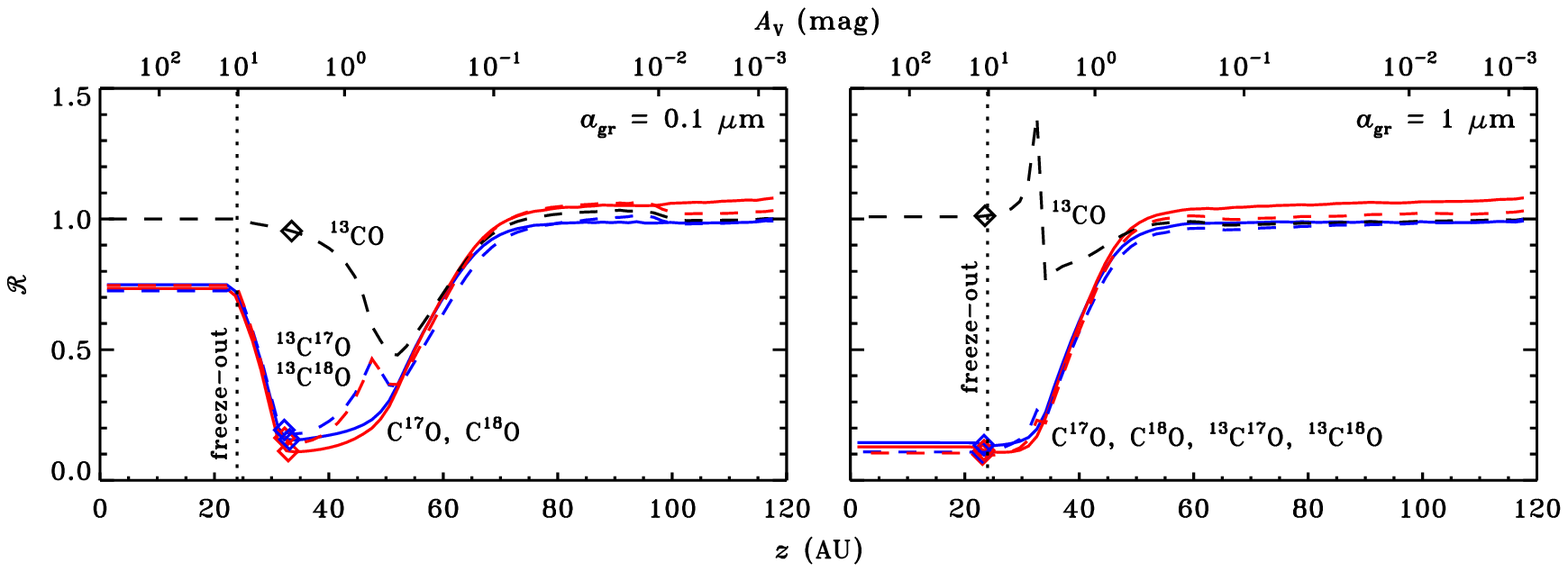}}
\caption{Cumulative column densities of the CO isotopologues normalised to \coc{12} and the elemental isotopic abundances (\eq{ncdr}) as a function of height at a radius of 105 AU in our disk model with the \citeauthor{draine78a} field and dust grains of 0.1 \micron{} (\emph{left}) or 1 \micron{} (\emph{right}). CO freezes out below 24 AU. The diamond symbols indicate where H$_3^+$ takes over from UV photons as the main destroyer of each isotopologue.}
\label{fig:diskncdr}
\end{figure*}


\subsubsection{Isotopologue ratios}
\label{subsec:diskiso}
The depth dependence of the photodissociation rates in the disk, illuminated by the \citeauthor{draine78a} field, is qualitatively the same as in the cloud and PDR models from Sects.\ \ref{subsec:trans} and \ref{subsec:pdr}. Several quantitative differences arise because of the different UV intensity and the higher densities. The profiles of the normalised column density ratios, $\mathcal{R}$ (\eq{ncdr}), do not show any enhancement in \coc{13}. Moving down from the surface, suprathermal chemistry initially prevents the formation of \coc{13} from \coc{12} and \element[+][13]{C}. The Alfv\'en waves penetrate to $\cdmh=\scim{4}{20}$ \pcs, which translates here to a height of about 50 AU. The absence of suprathermal chemistry below that point allows $\mathcal{R}(\cocm{13})$ to increase again. However, the \element[+][13]{C} abundance decreases at the same time, and $\mathcal{R}(\cocm{13})$ never gets above unity. This is consistent with the $N(\cocm{12})/N(\cocm{13})$ ratios of more than 69 determined for the disks around \object{HL Tau}, \object{AB Aur} and \object{VV CrA} \citep{brittain05a,schreyer08a,smith09a}.

In terms of local abundances, our results show a trend opposite to the model by \citet{woods09a}. Moving up vertically from the midplane at a given radius, their $n(\cocm{12})/n(\cocm{13})$ abundance ratio initially remains nearly constant, then drops to a lower value, and finally shows a shallow rise for the remainder of the column. We also find a constant ratio at first, but then we get a \emph{rise} (equivalent to $\mathcal{R}(\cocm{13})<1$) and a gradual decrease back to the elemental ratio of 69. The difference is due to the absence of suprathermal chemistry in the \citeauthor{woods09a} model. Suprathermal chemistry effectively shuts down the isotope-exchange reactions, leaving photodissociation (which increases $n(\cocm{12})/n(\cocm{13})$) as the only fractionation mechanism. Additional observations of \coc{12} and \coc{13} in circumstellar disks are required to ascertain whether isotope exchange or selective photodissociation is the key process in controlling the ratio of the abundances of these two species.

For \coo{17} and \coo{18}, $\mathcal{R}$ is $\sim$1 from the disk surface down to 70 AU. It then gradually drops to $\sim$0.1 due to their photodissociation rates being higher than that of \coc{12}. Shielding of \coo{17} and \coo{18} sets in around 32 AU, so that photodissociation no longer causes isotope fractionation and $\mathcal{R}$ increases again. However, it does not reach a value of unity. Below 24 AU, the temperature is low enough that most of the CO freezes out onto the dust grains. The tiny amount remaining in the gas phase does not add to the column density, so $\mathcal{R}$ is effectively frozen at 0.75 and 0.73 (\fig{diskncdr}). The corresponding column density ratios are $N(\cocm{12})/N(\coom{17})=2700$ and $N(\cocm{12})/N(\coom{18})=760$. Both values are consistent with the degree of fractionation determined for the disk around the T Tauri star VV CrA \citep{smith09a}.


\subsubsection{Model variations}
\label{subsec:diskvar}
If we illuminate the disk by a 4000 K blackbody (spectrum C from \citealt{vanzadelhoff03a}), it does not receive enough photons in the 912--1118 \AA{} range to cause substantial photodissociation of CO. Instead, the non-isotope-selective reactions with \element[+]{He} and H$_3^+$ are the main destruction mechanisms even at the top of our vertical slice at $R=105$ AU. The normalised column density ratios therefore remain unity at all heights. Adding the interstellar UV field at $\chi=1$ to the 4000 K blackbody spectrum allows CO to be photodissociated again in the surface layers. The isotope fractionation returns, but only partially, because the UV field is still weaker than in the original case. Freeze-out now locks $\mathcal{R}(\coom{17})$ and $\mathcal{R}(\coom{18})$ at 0.86 and 0.85 instead of $\sim$0.75. Clearly, some UV component is required in the irradiating spectrum to get isotope fractionation through the photodissociation of CO, but it does not have to be a very strong component: the interstellar UV field already has a significant effect in this model.

Another parameter to be varied is the grain size. The disk model so far contained small dust grains, with an average size of 0.1 \micron. Grain growth is an inevitable step towards planet formation and is indeed known to occur in disks \citep{vanboekel03a,przygodda03a,lommen07a}. Larger grains cause less extinction in the UV \citep{shen04a}, so CO can be photodissociated deeper into the disk. Grain growth also affects the \mh{} formation rate: the rate is inversely proportional to approximately the square of the grain radius \citep{jonkheid06a}. The right panel of \fig{diskncdr} shows the normalised vertical column density ratios at a radius of 105 AU for an average grain size of 1 \micron. The onset of fractionation occurs at a lower altitude than in the 0.1 \micron{} case: 55 AU instead of 75 AU. The slower formation rate of \mh{} has a direct influence on the formation of CO through the pathway starting with \rx{cp+h2}. There is less \mh{} and CO in the disk's surface layers when we increase the grain size, so we have to go down deeper before \coc{12} becomes self-shielding and isotope fractionation sets in. At the same time, the weaker continuum extinction allows the UV radiation to penetrate deeper, so the isotope-selective photodissociation goes on to lower altitudes. The spike in $\mathcal{R}(\cocm{13})$ to 1.4 at 30 AU is due to a narrow zone where the temperature is relatively low (35 K), the Alfv\'en waves do not reach, and the \element[+][13]{C} abundance is high enough to convert \coc{12} into \coc{13}. The model with 0.1 \micron{} grains also has a narrow zone with a high \element[+][13]{C} abundance and no suprathermal chemistry, but the kinetic temperature in this zone is 50 K. There is still some conversion of \coc{12} into \coc{13} at that temperature, but not enough to show as a spike in the normalised column density ratio.

The bigger dust grains also affect the fractionation of \coo{17} and \coo{18}. The point at which photodissociation ceases to be the main destruction mechanism now nearly coincides with the point where CO freezes out (\fig{diskncdr}). Both species have $\mathcal{R}\approx0.15$ when this happens, so the ratios are locked at that value. In the small-dust model, the ratios were locked at $\sim$0.75. Observations of face-on disks in principle probe CO all the way to the midplane, so our model predicts that grain growth from 0.1 to 1 \micron{} can increase the observed fractionation of \coo{17} and \coo{18} by a factor of five. The fractionation in the local abundances, which is relevant for incorporating material into comets, increases by the same amount from 0.1 to 1 \micron{} grains.

All the results so far in this section are for a T Tauri disk. Herbig Ae/Be stars are more massive and more luminous and tend to have warmer disks, where CO may not be frozen out at all. That would prevent $\mathcal{R}$ from being locked as it is in \fig{diskncdr}; instead, it would rapidly return to unity for all isotopologues. Observations probing the midplane of such a disk would not find any fractionation in \element[][17]{O} and \element[][18]{O} relative to \element[][16]{O}.


\subsubsection{Implications for meteorites}
\label{subsec:diskmet}
Regardless of the grain size and the absence or presence of freeze-out, the model shows \element[][17]{O} and \element[][18]{O} to be nearly equally fractionated: $\mathcal{R}$ has very similar values for \coo{17} and \coo{18}. The equal photodissociation rates implied by this result are partly due to our choice of molecular parameters. Spectroscopic data on \coo{17} are still very scarce, so we assume many of its oscillator strengths and predissociation probabilities to have the same value as for \coo{18} (Sects.\ \ref{subsec:osc} and \ref{subsec:eta}). Still, two important differences remain. First, the elemental abundance of \element[][18]{O} is higher than that of \element[][17]{O}, resulting in some self-shielding for \coo{18} (\fig{zoph4c}). Second, the shifts of the \coo{17} lines with respect to \coc{12} are generally somewhat smaller than those of the \coo{18} lines, so \coo{17} is more strongly shielded by \coc{12}. Both effects are minor, however, and the photodissociation rates end up being nearly the same.

According to the experiments of \citet{chakraborty08a}, the photodissociation rate due to the $E0$ and $E1$ bands may be up to $\sim$40\% higher for \coo{17} than for \coo{18}. If we mimic this by artificially increasing the $E0$ and $E1$ oscillator strengths of \coo{17} by 40\%, the overall change in $\mathcal{R}(\coom{17})$ is less than 10\% at any depth. When the two oscillator strengths are doubled from their standard value, $\mathcal{R}(\coom{17})$ changes by at most 20\%. In both cases, the change in $\mathcal{R}(\coom{17})$ is largest at the top of the disk and gradually decreases for lower altitudes; it is less than 1\% below 50 AU.

Observational evidence for equal fractionation of \element[][17]{O} and \element[][18]{O} in disks was recently obtained by \citet{smith09a} for the T Tauri star VV CrA. Their analysis points to isotope-selective photodissociation of CO as the most likely explanation, although they cautioned that more observations are needed to confirm that conclusion. Meanwhile, their observations are consistent with our model results.

A long-standing puzzle in our solar system involves the equal fractionation of \element[][17]{O} and \element[][18]{O} in the most refractory phases of meteorites \citep{clayton73a}. Current evidence points at isotope-selective photodissocation of CO in the solar nebula playing a key role \citep{clayton02a,lyons05a,lee08a}. It is generally accepted that the resulting atomic oxygen eventually makes it into the refractory material to produce the observed isotope ratios, but it remains unknown how this actually happens. Any additional mass-independent fractionation during that process must be limited, because photodissociation of CO already reproduces the observed isotope abundance ratios \citep{young07a}. One important requirement here is to have \coo{17} and \coo{18} photodissociate at the same rate, an assumption that had gone untested up to now. Our model confirms they do so in many different environments---including diffuse and translucent clouds, PDRs, and circumstellar disks---even when allowing for the isotope effects suggested by \cite{chakraborty08a}.

Equal (or nearly equal) \element[][17]{O} and \element[][18]{O} fractionation also occurs for other oxygen-containing molecules in the model, such as water. The equal fractionation is a clear sign of CO photodissociation playing a key role in the fractionation process. Other fractionation processes, such as isotope-exchange reactions, diffusion, freeze-out and evaporation, depend on the vibrational energies of the isotopologues and are therefore mass-dependent. Specifically, such processes would lead to \element[][17]{O} being 0.52 times as fractionated as \element[][18]{O} \citep{matsuhisa78a}. If mass-independent fractionation processes other than isotope-selective photodissociation play a role in circumstellar disks, they have not been discovered so far \citep{clayton02a,yurimoto04a,lyons05a}. Recently, \citet{kimura07a} did report on a set of chemical experiments yielding mass-independently fractionated silicates, but they were unable to identify the underlying reaction mechanism. Hence, it remains unknown if this would be relevant for disks. Self-shielding in O$_2$ \citep{thiemens83a} is unimportant because O$_2$ never becomes abundant enough to become a significant source of atomic O.


\section{Conclusions}
\label{sec:conc}
This paper presents an updated model, based on the method of \citet{vandishoeck88a}, for the photodissociation of carbon monoxide (CO) and its heavier isotopologues. It contains recent spectroscopic data from the literature and produces a photodissociation rate of \scit{2.6}{-10} \ps{} in the interstellar medium, 30\% higher than currently adopted values. Our model is the first to include \coo{17} and \co{13}{17} and we apply it to a broader range of model parameters than has been done before. The main results are as follows:

\begin{itemize}
\item[$\bullet$] Self-shielding is very important for \coc{12} and somewhat so for \coo{18} and \coc{13}. The rare isotopologues \co{13}{17} and \co{13}{18} have the highest photodissociation rate at most depths into a cloud or other object. The rates of \coo{17} and \coo{18} are very similar at all depths, but details in the line overlap and self-shielding can cause mutual differences of up to 30\% (Sects.\ \ref{subsec:depth} and \ref{subsec:transother}).

\item[$\bullet$] When coupled to a chemical network, the model reproduces column densities observed towards diffuse clouds and PDRs. It shows that the large spread in observed $N(\cocm{12})/N(\cocm{13})$ ratios may be due to a combination of isotope-selective photodissociation and suprathermal chemistry (Sect.\ \ref{subsec:trans13co}).

\item[$\bullet$] Photodissociation of CO is more strongly isotope-selective in cold gas than in warm gas (Sect.\ \ref{sec:temp}).

\item[$\bullet$] The results from the full computation are well approximated by a grid of pre-computed shielding functions, intended for easy use in various other models. Shielding functions are provided for a range of astrophysical parameters (Sect.\ \ref{sec:shield}).

\item[$\bullet$] Grain growth in circumstellar disks increases the vertical range in which CO can be photodissociated. If photodissociation is still important at the point where CO freezes out onto the grains, the observable gas-phase column density ratios $N(\cocm{12})/N(\coom{17})$ and $N(\cocm{12})/N(\coom{18})$ may become an order of magnitude larger than the initial isotopic abundance ratios (Sect.\ \ref{subsec:diskvar}).

\item[$\bullet$] Without a far-UV component in the irradiating spectrum, isotope fractionation through the photodissociation of CO is not possible. The interstellar radiation field can already cause substantial fractionation in disks (Sect.\ \ref{subsec:diskvar}).

\item[$\bullet$] The isotope-selective nature of the CO photodissociation results in mass-independent fractionation of \element[][17]{O} and \element[][18]{O}. Column density ratios computed for a circumstellar disk agree well with recent observations. Our model supports the hypothesis that the photodissociation of CO is responsible for the anomalous \element[][17]{O} and \element[][18]{O} abundances in meteorites (Sect.\ \ref{subsec:diskmet}).
\end{itemize}


\begin{acknowledgements}
We thank all the laboratory scientists for providing the much needed molecular data and discussions over the last twenty years, in particular Glenn Stark, Wim Ubachs and Fran{\c c}ois Rostas. Astrochemistry in Leiden is supported by a Spinoza Grant from the Netherlands Organization for Scientific Research (NWO) and a NOVA grant.
\end{acknowledgements}


\bibliographystyle{aa}
\bibliography{12129}


\longtabL{1}{
\begin{landscape}
\begin{longtable}{cc r@{.}l r@{.}l r@{ }l c r@{.}l c r@{.}l r@{.}l r@{.}l r@{.}l r@{.}l ll}
\caption{\label{tb:moldata-ls} New molecular data for \coc{12}.$^{\mathrm{a}}$} \\
\hline\hline
Band$^{\mathrm{b}}$ & ER90$^{\mathrm{b}}$ & \mc{$\lambda_0$} & \mc{$\nu_0$} & \mc{ID} & $v'$ & \mc{$\osc$} & $A_\el{tot}$ & \mc{$\eta$} & \mc{$\bvu$} & \mc{$\dvu$} & \mc{$\weu$} & \mc{$\wexeu$} & References & Notes \\
\#             & \#          & \mc{(\AA)} & \mc{(\pcm)} & \mc{ }                      &   & \mc{ }      & (\ps)        & \mc{ }   & \mc{(\pcm)} & \mc{(\pcm)} & \mc{(\pcm)} & \mc{(\pcm)} & & \\
\hline
\endfirsthead
\caption{continued.} \\
\hline\hline
Band$^{\mathrm{b}}$ & ER90$^{\mathrm{b}}$ & \mc{$\lambda_0$} & \mc{$\nu_0$} & \mc{ID} & $v'$ & \mc{$\osc$} & $A_\el{tot}$ & \mc{$\eta$} & \mc{$\bvu$} & \mc{$\dvu$} & \mc{$\weu$} & \mc{$\wexeu$} & References & Notes \\
\#             & \#          & \mc{(\AA)} & \mc{(\pcm)} & \mc{ }                      &   & \mc{ }      & (\ps)        & \mc{ }   & \mc{(\pcm)} & \mc{(\pcm)} & \mc{(\pcm)} & \mc{(\pcm)} & & \\
\hline
\endhead
\hline
\endfoot
\phantom{A0}1A & \phantom{A0}7A &  912&7037 & 109564&6 & &$^1\Pi$ & 0 & 3&4(-3) & 1(10) & 1&00 & 1&9226 & 5&9(-5) & 2169&8 & 13&29 & 1 & c \\
\phantom{B0}1B & \phantom{B0}7B &  913&4002 & 109481&0 & $(5p\sigma)$&$^1\Sigma^+$ & 1 & 1&7(-3) & 9(10) & 1&00 & 1&829 & 1&0(-5) & 2214&2 & 15&15 & 1,2 & d \\
\phantom{C0}1C & \phantom{C0}7C &  913&4255 & 109478&0 & $(5p\pi)$&$^1\Pi$ & 1 & 1&7(-3) & 1(10) & 1&00 & 1&9568 & 1&0(-4) & 2214&2 & 15&15 & & d \\
\phantom{D0}1D & \phantom{D0}7D &  913&6678 & 109449&0 & &$^1\Sigma^+$ & 2 & 2&7(-2) & 9(10) & 1&00 & 1&7781 & 5&4(-5) & 2169&8 & 13&29 & 1,2,3 & c \\
\phantom{A0}2A & \phantom{A0}8A &  915&7258 & 109203&0 & $(6p\pi)$&$^1\Pi$ & 0 & 2&0(-3) & 1(11) & 1&00 & 1&5796 & 6&737(-6) & 1562&8 & 13&93 & & e,f \\
\phantom{B0}2B & \phantom{B0}8B &  915&9708 & 109173&8 & $(6p\sigma)$&$^1\Sigma^+$ & 0 & 7&9(-3) & 1(11) & 1&00 & 1&685 & 1&0(-4) & 2214&2 & 15&15 & 3 & d \\
\phantom{0}3 & \phantom{A0}9A &  917&2721 & 109018&9 & &$^1\Pi$ & 2 & 2&3(-2) & 5(11) & 1&00 & 1&671 & 7&2(-5) & 2169&8 & 13&29 & 2 & c \\
\phantom{0}4 & \phantom{B0}9B &  919&2097 & 108789&1 & $(6s\sigma)$&$^1\Sigma^+$ & 0 & 2&8(-3) & 1(11) & 1&00 & 2&1357 & 4&633(-5) & 2214&2 & 15&15 & 4,5 & d \\
\phantom{0}5 & \phantom{C0}9C &  920&1410 & 108679&0 & $I'(5s\sigma)$&$^1\Sigma^+$ & 1 & 2&8(-3) & 1(11) & 1&00 & 1&9050 & 6&0(-6) & 2291&2 & 0&0 & & g \\
\phantom{0}6 & 10 &  922&7561 & 108371&0 & $(5d\sigma)$&$^1\Sigma^+$ & 0 & 6&3(-3) & 3(11) & 1&00 & 1&9675 & 6&318(-6) & 2214&2 & 15&15 & & d,h \\
\phantom{0}7 & 11 &  924&6309 & 108151&3 & &$^1\Sigma^+$ & 1 & 5&2(-3) & 1(11) & 1&00 & 1&8669 & 4&0(-5) & 2169&8 & 13&29 & & c \\
\phantom{0}8 & 12 &  925&8093 & 108013&6 & $W(3s\sigma)$&$^1\Pi$ & 3 & 2&0(-2) & 4(11) & 1&00 & 1&6506 & 1&1(-4) & 1745&1 & $-4$&22 & 6 & g,i \\
\phantom{0}9 & 13 &  928&6575 & 107682&3 & &$^1\Pi$ & 2 & 6&7(-3) & 4(10) & 1&00 & 1&936 & 3&1(-5) & 2169&8 & 13&29 & 1,2,3 & c \\
10 & 14 &  930&0611 & 107519&8 & &$^1\Pi$ & 2 & 6&3(-3) & 1(11) & 1&00 & 1&8202 & 2&6(-5) & 2169&8 & 13&29 & & c \\
11 & \phantom{A}15A &  931&0744 & 107402&8 & &$^1\Pi$ & 0 & 6&0(-3) & 1(11) & 1&00 & 1&650 & 1&0(-5) & 2169&8 & 13&29 & & c \\
12 & \phantom{B}15B &  931&6547 & 107335&9 & $(5p\pi)$&$^1\Pi$ & 0 & 1&2(-2) & 3(11) & 1&00 & 1&8682 & 4&3(-5) & 2214&2 & 15&15 & 2 & d,j \\
13 & \phantom{C}15C &  933&0583 & 107174&4 & $(5p\sigma)$&$^1\Sigma^+$ & 0 & 2&2(-2) & 3(10) & 1&00 & 2&128 & 1&0(-5) & 2214&2 & 15&15 & 1,2 & d,i \\
14 & 16 &  935&6638 & 106876&0 & &$^1\Sigma^+$  & 2 & 3&8(-3) & 3(11) & 1&00 & 1&95 & 0&0 & 2169&8 & 13&29 & 5 & c \\
15 & 17 &  939&9574 & 106387&8 & $I'(5s\sigma)$&$^1\Sigma^+$ & 0 & 2&1(-2) & 1(12) & 1&00 & 2&0365 & 8&8(-5) & 2291&2 & 0&0 & 4 & g,j \\
16 & 18 &  941&1685 & 106250&9 & $W(3s\sigma)$&$^1\Pi$ & 2 & 3&1(-2) & 1(11) & 1&00 & 1&6224 & $-1$&3(-5) & 1745&1 & $-4$&22 & 1,6 & g,i,k \\
17 & 19 &  946&2860 & 105676&3 & $(4d\sigma)$&$^1\Sigma^+$ & 0 & 7&6(-3) & 1(11) & 1&00 & 1&8952 & 1&73(-5) & 2214&2 & 15&15 & 1,2,3 & d \\
18 & 20 &  948&3860 & 105442&3 & $L(4p\pi)$&$^1\Pi$ & 1 & 2&8(-3) & 1(10) & 0&99 & 1&9606 & 1&0(-5) & 2170&5 & 0&0 & & g \\
19 & 21 &  950&0429 & 105258&4 & $H(4p\sigma)$&$^1\Sigma^+$ & 1 & 2&2(-2) & 1(12) & 1&00 & 1&9436 & 4&4(-5) & 2203&7 & 0&0 & 4 & g,j \\
20 & 22 &  956&2369 & 104576&6 & $W(3s\sigma)$&$^1\Pi$ & 1 & 1&6(-2) & 7(11) & 1&00 & 1&5696 & 5&8(-5) & 1745&1 & $-4$&22 & 6 & g \\
21 & 24 &  964&3973 & 103691&7 & $J(4s\sigma)$&$^1\Sigma^+$ & 1 & 2&8(-3) & 3(11) & 1&00 & 1&9202 & 9&0(-6) & 2235&7 & 0&0 & & g \\
22 & 25 &  968&3187 & 103271&8 & $L(4p\pi)$&$^1\Pi$ & 0 & 1&4(-2) & 2(9)\phantom{0} & 0&96 & 1&9598 & 7&06(-6) & 2170&5 & 0&0 & 7,8,9 & g,i,k \\
23 & 26 &  968&8815 & 103211&8 & $L'(3d\pi)$&$^1\Pi$ & 1 & 1&2(-2) & 2(11) & 1&00 & 1&754 & 1&0(-5) & 2214&2 & 15&15 & 1,2,3,9 & d,k \\
24 & 27 &  970&3588 & 103054&7 & $K(4p\sigma)$&$^1\Sigma^+$ & 0 & 3&4(-2) & 2(10) & 0&99 & 1&9166 & 5&99(-5) & 2203&7 & 0&0 & 1,2,3,9 & g \\
25 & 28 &  972&6996 & 102806&7 & $W(3s\sigma)$&$^1\Pi$ & 0 & 1&7(-2) & 1(10) & 0&97 & 1&5664 & 9&7(-5) & 1745&1 & $-4$&22 & 1,3,6 & g,i,k \\
26 & 29 &  977&3996 & 102312&3 & $W'(3s\sigma)$&$^3\Pi$ & 2 & 1&8(-3) & 4(11) & 1&00 & 1&5417 & 8&0(-6)  & 1562&8 & 13&93 & 1,2,3 & f,k \\
27 & 30 &  982&5914 & 101771&7 & $F(3d\sigma)$&$^1\Sigma^+$ & 1 & 4&8(-4) & 3(11) & 1&00 & 1&8506 & 1&4(-5) & 2030&0 & 0&0 & & g \\
28 & 31 &  985&6490 & 101456&0 & $J(4s\sigma)$&$^1\Sigma^+$ & 0 & 1&5(-2) & 1(12) & 1&00 & 1&9247 & 5&1(-5) & 2235&7 & 0&0 & 4 & g,j \\
29 & 32 &  989&7952 & 101031&0 & $G(3d\pi)$&$^1\Pi$ & 0 & 4&6(-4) & 1(11) & 1&00 & 1&9632 & 1&1(-5) & 2214&2 & 15&15 & & d \\
30 & 33 & 1002&5897 &  99741&7 & $F(3d\sigma)$&$^1\Sigma^+$ & 0 & 7&9(-3) & 3(11) & 1&00 & 1&8077 & 2&2(-4) & 2030&0 & 0&0 & & g \\
31 & 37 & 1051&7134 &  95082&9 & $E(3p\pi)$&$^1\Pi$ & 1 & 3&6(-3) & 6(9)\phantom{0} & 0&96 & 1&9280 & 6&64(-6) & 2238&8 & 42&51 & 10,11,12,13 & k \\
32 & 38 & 1063&0880 &  94065&6 & $C(3p\sigma)$&$^1\Sigma^+$ & 1 & 3&0(-3) & 2(9)\phantom{0} & 0&56 & 1&9238 & 6&25(-6) & 2175&9 & 14&76 & 14,15,16 & \\
33 & 39 & 1076&0796 &  92929&9 & $E(3p\pi)$&$^1\Pi$ & 0 & 6&8(-2) & 1(9)\phantom{0} & 0&80 & 1&9527 & 6&34(-6) & 2238&8 & 42&51 & 13,15,17,18 & k \\
\end{longtable}
\begin{list}{}{}
\item[References:] (1) $\lambda_0$ and $\nu_0$ from \citet{ubachs94a}; (2) $A_\el{tot}$ from \citet{ubachs94a}; (3) $\bvu$ and $\dvu$ from \citet{ubachs94a}; (4) $\lambda_0$ and $\nu_0$ from \vdb; (5) $\bvu$ and $\dvu$ from \vdb; (6) $\osc$ and $A_\el{tot}$ from \citet{eidelsberg06a}; (7) $\lambda_0$, $\nu_0$, $\bvu$ and $\dvu$ from \citet{cacciani02a}; (8) $A_\el{tot}$ from \citet{drabbels93b}; (9) $\osc$ from \citet{eidelsberg04a}; (10) $\lambda_0$, $\nu_0$, $\bvu$ and $\dvu$ from \citet{ubachs00a}; (11) $\osc$ from \citet{eidelsberg06a}; (12) $A_\el{tot}$ from \citet{ubachs00a}; (13) $\weu$, $\wexeu$ from \citet{kepa88a}; (14) $\lambda_0$, $\nu_0$, $A_\el{tot}$, $\bvu$ and $\dvu$ from \citet{cacciani01a}; (15) $\osc$ from \citet{federman01a}; (16) $\weu$, $\wexeu$ from \citet{tilford68a}; (17) $\lambda_0$, $\nu_0$, $\bvu$ and $\dvu$ from \citet{cacciani04a}; (18) $A_\el{tot}$ from \citet{cacciani98a}
\item[Notes:]
\item[$^{\mathrm{a}}$] $\lambda_0$, $\nu_0$, ID and $v'$ from \citet{eidelsberg92a}, $\osc$, $A_\el{tot}$, $\bvu$ and $\dvu$ from \er, unless noted otherwise. All $\eta$ are recomputed from the $\osc$ and $A_\el{tot}$. The notation $a(b)$ in this and following tables means \scit{a}{b}. Many values are rounded off from higher-precision values in the references.
\item[$^{\mathrm{b}}$] The numbering follows \vdb. Their bands 1 and 2 are split into four and two components. The corresponding \er{} indices are also given.
\item[$^{\mathrm{c}}$] $\weu$ and $\wexeu$ from the CO ground state \citep{guelachvili83a}. $\weyeu$ and $\wezeu$ (not listed) are included in the model.
\item[$^{\mathrm{d}}$] $\weu$ and $\wexeu$ from the \cop{} $X\,^2\Sigma^+$ state \citep{haridass00a}.
\item[$^{\mathrm{e}}$] $\bvu$ and $\dvu$ from the \cop{} $A\,^2\Pi$ state \citep{haridass00a}.
\item[$^{\mathrm{f}}$] $\weu$ and $\wexeu$ from the \cop{} $A\,^2\Pi$ state \citep{haridass00a}. $\weyeu$ (not listed) is included in the model.
\item[$^{\mathrm{g}}$] Vibrational constants derived from the different $\nu_0$ in one of six vibrational series: bands 30--27, 28--21, 25--20--16--8, 24--19, 22--18 or 15--5.
\item[$^{\mathrm{h}}$] $\bvu$ and $\dvu$ from the \cop{} $X\,^2\Sigma^+$ state \citep{haridass00a}.
\item[$^{\mathrm{i}}$] $A_\el{tot}$ depends on parity and/or rotational level (see \tb{parrota}). $A_\el{tot}$ and $\eta$ are listed here for $J'$=0 and $f$ parity.
\item[$^{\mathrm{j}}$] $\bvu$ and $\dvu$ computed from the \coo{18} values of \er.
\item[$^{\mathrm{k}}$] $\bvu$ and $\dvu$ depend on parity (see \tb{parbc}); values are listed here for $f$ parity.
\end{list}
\end{landscape}
}


\end{document}